\documentclass[11pt]{article}
\usepackage{color,amsfonts,amsmath,amssymb,mathrsfs,verbatim,graphicx}

\def\nailz{\mbox{\LARGE{$\intercal$}}{\mbox{\hspace{-9.279pt}\raisebox{-2.6pt}{$\color{white}\bigvee$}}}  }
\def\naila{\mbox{\raisebox{+1.3pt}{\LARGE{$\intercal$}}}
   {\mbox{\hspace{-9.352pt}\raisebox{-1.2pt}{$\color{white}\bigvee$}}}   \mbox{\hspace{-11pt}\raisebox{-1.5pt}{$-\!\!-$}} }
\def\nailb{\mbox{\raisebox{-0.1pt}{\LARGE{$\intercal$}}}
   {\mbox{\hspace{-9.352pt}\raisebox{-2.6pt}{$\color{white}\bigvee$}}}   \mbox{\hspace{-11pt}\raisebox{-1.5pt}{$-\!\!-$}} }
 \def\nailc{\mbox{\raisebox{-1.5pt}{\LARGE{$\intercal$}}}
   {\mbox{\hspace{-9.352pt}\raisebox{-4pt}{$\color{white}\bigvee$}}}   \mbox{\hspace{-11pt}\raisebox{-1.5pt}{$-\!\!-$}} }  
   \def\naild{\mbox{\raisebox{-2.9pt}{\LARGE{$\intercal$}}}
   {\mbox{\hspace*{-9.352pt}\raisebox{-5.4pt}{$\color{white}\bigvee$}}}   \mbox{\hspace{-11pt}\raisebox{-1.5pt}{$-\!\!-$}} }

\def\bx{\mbox{\boldmath $x$}}
\def\bdx{\mbox{\boldmath $\delta x$}}
\def\bxf{\mbox{\boldmath $\delta x$}_4}
\def\bxn{\mbox{\boldmath $\delta x$}_n}
\def\bxnf{\mbox{\boldmath $\delta x$}_{n-4}}

\def\Rcon{R_{\mathrm{con}}}
\def\Hint{H_{\mathrm{int}}}
\def\stot{\sigma_{\mathrm{tot}}}
\def\bp{\mbox{\boldmath $p$}}

\def\bk{\mbox{\boldmath $k$}}
\def\bPP{\mbox{\boldmath $P$}}

\def\rrr{{\mathbb{R}}}
\def\ccc{{\mathbb{C}}}

\def\ooo{{\mathbb{O}}}

\def\hkc{\mbox{h}_k\ccc}
\def\htwc{\mbox{h}_2\ccc}

\def\hpc{\mbox{h}_p\ccc}

\def\sltc{\mbox{SL}(2,\ccc)}
\def\slthc{\mbox{SL}(3,\ccc)}
\def\sltho{\mbox{SL}(3,\ooo)}
\def\slpc{\mbox{SL}(p,\ccc)}
\def\slptc{\mbox{SL}(k,\ccc)_{\! D}}
\def\sukd{\mbox{SU}(k)_{\! D}}
\def\uod{\mbox{U}(1)_{\! D}}

\def\esi{\mbox{E}_6}
\def\ese{\mbox{E}_7}
\def\ee{\mbox{E}_8}

\def\bone{\mbox{\boldmath $1$}}

\def\soot{\mbox{SO}^+(1,3)}
\def\sootn{\mbox{SO}^+(1,n-1)}

\def\bv{\mbox{\boldmath $v$}}
\def\bvf{\bv_4}

\def\sutw{\mbox{SU}(2)}
\def\suth{\mbox{SU}(3)}

\def\uo{\mbox{U}(1)}
\def\hG{\hat{G}}

\def\Lsl{L\!\!\!\!\!\!\;{\mbox{\scriptsize${\diagup}$}}}

\def\mcSL{{\mathcal S}_{\mbox{\tiny ${\!\mathcal L}$}}}
\def\mcM{{\mathcal M}}
\def\lag{\mathcal{L}}

\def\Amin{A_{\mbox{-}}}
\def\Apls{A_{\mbox{\tiny{+}}}}

\def\pAm{p_{\mbox{\tiny$\! A$}_{\mbox{-}}}}

\def\rAm{\rho_{\mbox{\tiny$\! A$}_{\mbox{-}}}}

\def\pAMp{p_{\mbox{\tiny$\! AM_{\;\!\!+}$}}}
\def\rAMp{\rho_{\mbox{\tiny$\! AM_{\;\!\!+}$}}}

\def\pvac{p_{\mbox{\tiny$V$}}}
\def\rvac{\rho_{\mbox{\tiny$V$}}}
\def\wvac{w_{\mbox{\tiny$V$}}}

\def\rSM{\rho_{\mbox{\tiny$\!S\!M$}}}

\def\rDM{\rho_{\mbox{\tiny$\!D\!M$}}}
\def\rDE{\rho_{\mbox{\tiny$\!D\!E$}}}
\def\rDS{\rho_{\mbox{\tiny$\!D\!S$}}}

\def\bphiD{\phi_{\!\:\! D}}
\def\bpsiD{\psi_{\!\:\! D}}

\def\AmaxD{A^{\mbox{\raisebox{+3pt}{\tiny\boldmath$\!\!\!\circ$}}}_{\! D}}

\def\Mplu{M_{\mbox{\tiny{+}}}}

\def\setb{\setlength{\baselineskip}{0.625\baselineskip}}

\begin{document} 

{\setlength{\baselineskip}{0.625\baselineskip}

\begin{center}
   
           {\LARGE {\bf Generalised Proper Time and}}  
                 
              \vspace{10pt}          
              
            {\LARGE {\bf the Universal Bootstrap}}  

   \vspace{22pt}
 
 \mbox {{\Large David J. Jackson}} \\ 

    \vspace{10pt}  
 
  {david.jackson.th@gmail.com}  \\

   \vspace{10pt}
 
   {August 8, 2023} 

 \vspace{28pt}

{\bf  Abstract}

 
\end{center}

 The generalisation of proper time, as an alternative to models with extra dimensions of space, has been proposed as the source of the elementary structures of matter. Direct connections with the Standard Model of particle physics together with dark sector candidates can be attained in this manner, through the additional components and symmetry breaking pattern identified on extracting the substructure of the local 4-dimensional spacetime form.
   In this paper we focus more on the extended and global properties of the theory, both in relation to amalgamating gravity with quantum theory and to broader cosmological issues. Contrast will be made with both early $S$-matrix bootstrap models and more recent cosmological bootstrap techniques. With time playing a central role in the new theory we describe how a `universal bootstrap' can be constructed and elaborate upon the implications.

\vspace{5pt}


\tableofcontents 



\pagebreak


\section{Introduction}
\label{boot1}

  \begin{quotation}
             \flushright{\noindent In the broadest sense, bootstrap  philosophy asserts that \\
                      ``nature is  as  it is because this is the only possible  \\
                                             nature consistent with itself.''}
             \vspace{-3pt}
            \flushright{ Geoffrey F. Chew (\cite{Chew68} opening, 1968)}
  \end{quotation}
\vspace{10pt}

   The motivation for `bootstrap' models
   stems from the attraction of deducing physical quantities, as empirically observed, from a minimal set of reasonable and general constraints that define the theory. Such internally self-consistent schemes are named after the analogy with a 
     self-supporting individual, originally associated with the fictional character Baron von M\"{u}nchhausen, apparently capable of levitating himself into the air by pulling on his own bootstraps, and hence floating free of any external impetus.  
  This type of approach was first attempted to describe the strong interaction properties and scattering of the growing number of hadronic states as observed in the 1950s and 1960s and
   termed the `$S$-matrix bootstrap'~\cite{Chew61,Chew62,Venez,Chew71}. More recently similar ideas have been employed to deduce  spatial correlations in the distribution of galaxies, as can be observed today, from general constraints on inflation in the very early universe 
   using the `cosmological bootstrap'~\cite{ABLP,Baum19,Baum20,Baum22}.
  
      In this paper we propose a self-contained `universal bootstrap' and argue how even the consistency constraints \textit{themselves} are contained \textit{within} this bootstrap in a self-sufficient manner. We shall describe how such a construction follows inevitably for a theory based upon generalised proper time, with the flow of time considered the simple basic entity. 
      As a unification scheme the approach of generalised proper time has achieved a series of successes in accommodating the Standard Model of particle physics~\cite{TimeE}, 
      providing a framework for unifying general relativity with quantum theory~\cite{QGrav}, and
          implying dark energy and dark matter candidates~(\cite{Basis} section~4). These are the traditional targets for any proposed unification scheme~(\cite{Short} section~1).
      
      However, for most unification schemes the basic foundations are generally simply \textit{posited} and justified largely through the explanatory power of the physics that can be derived \textit{from} the theory in comparison with empirical observations. 
     This leaves an inherent, and seemingly irreducible, element of incompleteness in that
     an \textit{underlying} justification for the origin of the basic entities or mathematical foundations of the theory is not provided.
       We argue that it may be unnecessarily pessimistic to leave this as the ultimate fate of unification, and describe how it is possible for the theory of generalised proper time to `pull itself up by its own bootstraps' in a \textit{fully}
        self-sufficient manner as a consequence of the fundamental role played by time.
        The corresponding  self-contained and complete unification motivates the term `universal bootstrap'. This structure could be considered a realisation of the `complete bootstrap' envisioned in~\cite{Chew68,Chew70} as enabled by subsequent scientific advances.

          The paper is organised as outlined in the following. 
          In the next section we first review the original $S$-matrix bootstrap for strong hadronic interactions and discuss its scope and limitations as well as the ongoing influences stemming from this programme. 
          One influence in particular has been upon   the cosmological bootstrap in aiming to relate the large-scale structure of  spatial galactic correlations to the primordial particle content of the universe.
            In both cases we    cite several key references motivating and developing these bootstrap ideas.

           In the three subsections of     section~\ref{boot3} we  review and further elaborate upon the novel approach to unification of generalised proper time.
           The underlying motivation for this approach 
           will be described in subsection~\ref{boot31}, where the 
          direct connections identified with the Standard Model of particle physics will also be summarised.      
            \mbox{The construction} of the extended 4-dimensional spacetime  manifold
            from this local basis will be reviewed in 
            subsection~\ref{boot32}, where 
                the means of accommodating quantum phenomena in a manner fully consistent with general relativity within this framework will be described and further clarified, 
                     with emphasis upon the reconceptualisation of particle interaction processes and the quantum field theory calculation limit.  Contrast will be made with elements of the   $S$-matrix bootstrap,  in particular in providing a source of specific matter fields and their  interactions and in the ability to identify a foundation for quantum theory itself. 
                      In subsection~\ref{boot33}, on returning to the local symmetry breaking structure of the theory, we review how the mathematical possibility of alternative branches for
               generalised proper time implies a basis for a significant dark sector, and describe how the general properties are  consistent with a source for both dark energy and dark matter candidates.  
               Now in contrast with elements of the cosmological bootstrap,
        we then  consider  
           broad cosmological questions on the global scale regarding 
               how a basis for large-scale structure formation in the very early universe, and potentially for inflation itself, can be identified for this theory, implying further possible means for empirical tests.

             With the successes of the theory, in accounting for the above elementary structures of physics, hinging upon the analysis of generalised proper time, in section~\ref{boot4} we turn to the underlying origin of the flow of time itself  
        and the implications that follow.
        A key observation is  the manner in which  time not only plays a fundamental role in physics, but  how the flow of time is also an irreducible element behind everything that we can
         perceive in the world generally.  This  leads to the proposal of a corresponding intrinsic  source for  temporal flow   rooted in neural structures of the brain which, as we review in subsection~\ref{boot41}, is supported by contemporary findings in neuroscience.
              These observations regarding the elementary and ubiquitous nature of time, and as similarly
               argued for  perception of space,
                will lead directly and inevitably to the construction of the self-contained system of the universal bootstrap in subsection~\ref{boot42}.
               As well as presenting support for this system we also describe how the $S$-matrix and cosmological bootstraps fit into this picture.
                While raising opportunities for further investigation, the immediate benefits of the universal bootstrap in addressing several outstanding foundational questions regarding the physical universe will be expounded in subsection~\ref{boot43}. There we also elaborate further on the
                  implications of this framework for
                  particle physics, cosmology and for other fields, together with their mutual interconnections.

                 In section~\ref{boot5} we review the  approach to these elementary foundational questions suggested by a range of other theories. 
                  We argue how generalised proper time and the 
                   universal bootstrap present a robust, complete and fully scientific account of `why there is something rather than nothing', without the residual loose ends typically remaining for  other approaches.
                 \mbox{The successes} of the unified physical theory of generalised proper time in relation to
                  the Standard Model of 
                  particle physics  and cosmology will be summarised in section~\ref{boot6}. 
                 There we also conclude 
                 with further general remarks on the nature of the theory, together 
         with analogies and a summary of the universal bootstrap picture, which provides a very firm basis for this
        unified theory in physics.


\pagebreak

\section{Review of $S$-Matrix and Cosmological Bootstraps}
\label{boot2}

    By the 1960s, as the number of strongly interacting hadronic states discovered in particle physics experiments continued to climb, it was unclear whether any model in which some states were considered more elementary than others could be sustained. Hence, in contrast to for example the Sakata model with
    a select `aristocracy' of the proton, neutron and $\Lambda$ baryons taken as elementary, Geoffrey Chew and others proposed a `nuclear democracy'. Under this democratic approach \textit{all} hadronic states were considered \textit{equally} elementary, with the basis for strong interactions between them determined by the properties of the scattering amplitude, that is, the $S$-matrix~\cite{Chew61,Chew62}.

   Scattering amplitudes for hadronic states were to be determined from basic consistency principles in accord with relativity and quantum theory, including Lorentz invariance, locality, quantum superposition, unitarity and analyticity, as sufficient constraints. With the probability of a transition from a given initial state to a given final state determined by the squared modulus of the complex amplitude, the unitarity condition $SS^{\dagger} =\bone$ is physically interpreted as the conservation of a total probability of unity. That the transition amplitude should vary smoothly as a function of the momentum variables is expressed mathematically by the analyticity of the $S$-matrix. 
   
    The broad picture based upon such constraints provided a non-perturbative alternative to the standard field theory perturbative approach that employs the details of proposed Lagrangian terms and the properties of Feynman diagrams, and which becomes particularly problematic for strong coupling.
     This `$S$-matrix bootstrap' was then the first example of such a framework, with laws of physics determined purely through mutual consistency, as if the theory were `pulled up by its own bootstraps', and with the $S$-matrix considered a basic entity in strong interaction processes as pictured in figure~\ref{bubble}.
     
     
\begin{figure}[htbp]  
\centering
\leavevmode
\includegraphics[width=14.3cm]{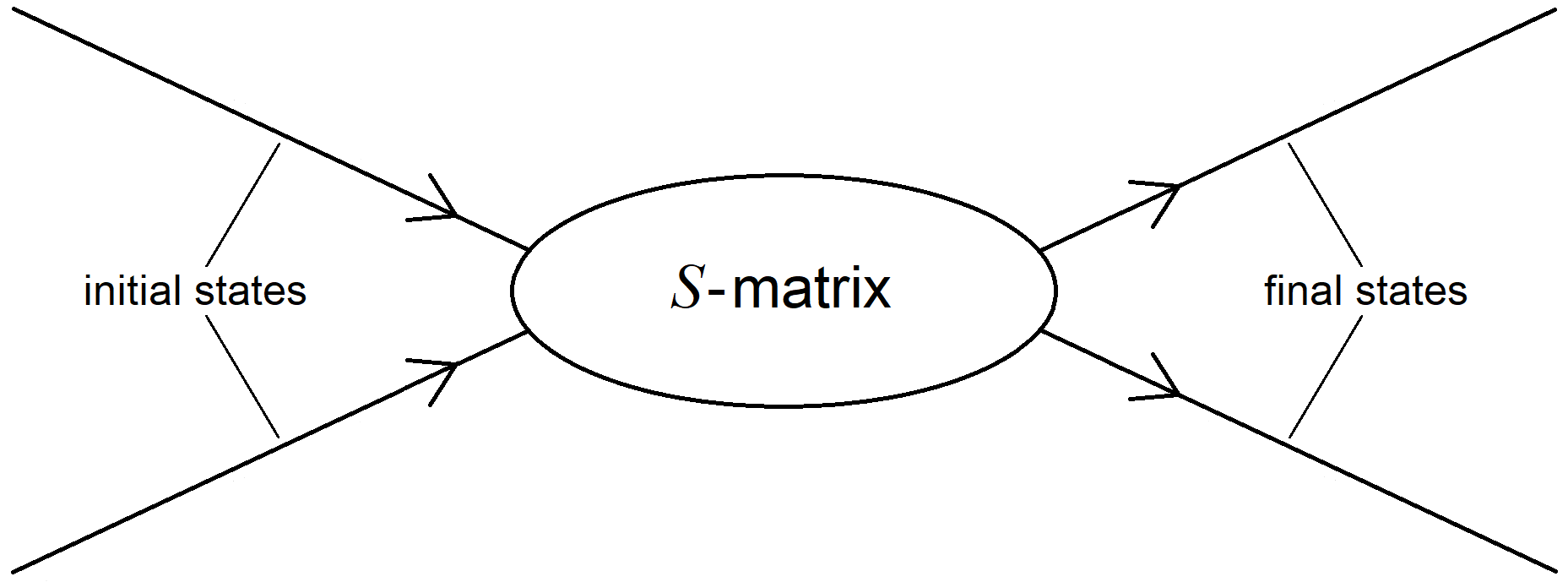}
\caption{\setb
   For the $S$-matrix bootstrap any strong interaction scattering process connecting initial `in' states and final `out' states can be depicted
    in a `bubble diagram'. The bubble represents the $S$-matrix amplitude as a basic entity, consistent with constraints such as unitarity and analyticity but independent of any underlying field theory.
     A further constraint concerned the relativistic property of `crossing symmetry', with the scattering amplitude required to be invariant under the interchange of an `in' and `out' state while replacing them with their respective antiparticle partners, as could be applied in the above diagram (see for example
      \protect\cite{Venez} figure~1).
   }
\label{bubble}
\end{figure}

   Given the above basic constraints, together potentially with supplementary principles such as isotopic spin and baryon number conservation, it was conceivable that there may be only one possible unique $S$-matrix structure that might then be determined and empirically verified. More generally, the ultimate ambition of the  bootstrap programme  would be to uniquely determine the laws of nature through self-consistency arguments alone.
   
     With the basis for strong interactions in the analytically continued $S$-matrix as a function of linear momentum variables,  bound states and resonances  could be identified with energy poles on the real axis. On extending the analyticity of the
      \mbox{$S$-matrix}  as a function also of angular momentum corresponding `Regge poles'  could be identified tracing out families of particles with uniform increments in spin and mass  along `Regge trajectories'  (\cite{Chew62} figure~1, \cite{Venez} figure~3). 
      With the $S$-matrix approach aiming to treat Regge poles as a primary feature, rather than a derived aspect of the theory, it could be proposed that all strongly interacting particles, potentially infinite in number, could be associated with Regge poles and the spectrum correspondingly determined.

     In this picture all hadrons are considered composite as bound states of each other, with none more elementary than any other. The possibly infinite variety of particles mutually strongly interact on the scale of order $10^{-15}\,$m, with the bound states or resonances exhibiting the properties of their constituents.     
        With strong interactions between the hadron states hence determined by reasonable general assumptions concerning the $S$-matrix mutual collisions between hadrons, or their decays, could only generate further hadrons in a democratic manner with all such states having an equally elementary status. Hadronic interactions hence merely transmute these states in a perpetually `cyclic' manner without any underpinning lowest level, furthering the analogy with the self-levitating `bootstrap' picture.
     
     Since all hadrons are reciprocally interdependent under the self-consistent bootstrap, for any one strongly interacting particle to exist and be completely understood the entire collection must exist and be understood  within a mutually supporting framework. Hence, even though all physical consequences should follow uniquely from self-consistency, such an `All or Nothing' character makes detailed empirical predictions difficult. Nevertheless, while such a simultaneous complete solution for $S$-matrix theory would be far too complicated, some empirical consequences can be deduced. In particular, the behaviour of Regge poles and trajectories, and anticipated cross-sections, are in general distinct from the expectations of
      models based on elementary particle states. However, the predictions of Regge theory did not prove reliable and ultimately only ever provided a heuristic phenomenological guide.

      It could also not be seen how electromagnetic and weak interactions, together with lepton and photon states, could emerge from the bootstrap principle. Further, the $S$-matrix approach seemed necessarily only partial in that quantum mechanical principles were presupposed, including the need for a classical concept of `measurement' as requires the classical application of electromagnetism~\cite{Chew71,Chew68}. 
       Hence it seemed inevitable that a truly inclusive `complete bootstrap' would remain elusive. 
       
       Moreover,  the success of the quark model and quantum chromodynamics (QCD) 
       led to the inevitable    
       decline of interest in the \mbox{$S$-matrix} bootstrap  as a theory of the strong interaction itself  through the 1970s and beyond~(\cite{Chew70}, \cite{Kragh} chapter 6 `The Rise and Fall of the Bootstrap Programme'). 
       Ironically, Gell-Mann had been in part influenced by the bootstrap idea when in 1964 he  proposed the quark model~\cite{GellM},  with all hadronic states now `equally non-elementary' in being composed of two or three quark components, which subsequently provided the elementary matter states 
          for QCD~\cite{GrossW,Polit}.

          On the other hand, given that the perturbative expansion approach of QCD with Feynman diagrams remains problematic owing to the large coupling of strong interactions, non-perturbative $S$-matrix bootstrap techniques have again played a role in recent years (see for example~\cite{White,BernDK,Alme}). 
       Further, dating back to the late 1960s, the insights of the consistency conditions for the $S$-matrix bootstrap programme became highly influential for the early development of string theory~\cite{Schw,CERNV}, which itself has had a highly significant lasting impact.
       As opposed to string-like structures on the hadronic strong interaction scale, the fundamental strings of string theory are far smaller, of order the Planck length at $10^{-35}\,$m, 
       as motivated  to  incorporate  a quantum theory of gravity alongside all the internal gauge interactions
       in what remains  a leading candidate for  a unifying framework.

       There has also been a resurgence of interest in bootstrap ideas in a cosmological context.
        Observations of the large-scale structure of the universe, with vast clusters of galaxies and the cosmic web  together with the voids in between, raise the question concerning the composition and nature of the very early universe that could have acted as the source seeding these formations. With galaxies tending to initially form under the pull of gravity in regions of over-density, small fluctuations in the primordial matter distribution and their spatial correlations would become imprinted in the large-scale structure at later times.
         The study of cosmological spatial correlators over very large distances, as observed by present and future galaxy surveys and also in the photons from the cosmic microwave background (CMB), hence encode the properties of the primordial physics.
         
          Running the clock backwards from the CMB era, where the observed temperature variations are of order one part in $10^5$, the primordial fluctuations are deduced to correspond to almost Gaussian and scale-invariant initial conditions -- such as may have been generated by a pattern of quantum fluctuations in the creation and properties of particle states as  stretched to very large distances by a primordial inflationary phase prior to the hot Big Bang. The perturbations, imprinted upon the final spacelike boundary of such an inflationary era,  would have then grown as seen in the CMB and ultimately in large-scale galactic structures. Observations of the spatial correlations then provide insight into the physics of the inflationary phase.

          The measured 2-point spatial correlation function is consistent with the production
     of  particle pairs  as fluctuating into existence  out of the quantum vacuum and  torn apart by the inflationary expansion in the very early universe  to become frozen into the spacetime geometry, seeding the eventual gravitational formation of galaxies and galactic clusters and their 2-point spatial correlations.     
        An inflationary epoch, with approximately constant energy density and nearly time-translation invariant dynamics, can generate the required scale invariant 2-point correlations. 
        However, these are largely fixed by symmetry constraints and reveal little else about inflationary dynamics. Large 2-point correlations can also be directly generated by production of  the  inflaton state itself. 
        
                On the other hand,  if the initial particle pair consists of new massive states that can decay, into the inflaton or other states,
                 during the inflationary epoch then 3-point and 4-point correlations can also be generated and in principle observed at late times, encoding information about the microscopic dynamics of the inflationary process.  
           That such higher-point correlations   
              are particularly difficult to calculate explicitly using time-dependent perturbation theory  motivates the  ongoing attempts to employ self-consistency bootstrap techniques.

              The `cosmological bootstrap' aims to take contemporary cosmological observations of 2-point, and in principle higher-point, correlations over extremely large distances to unravel the nature of the primordial particle content and interactions of the very early universe directly through consistency with basic physical principles and constraints alone~\cite{ABLP,Baum19,Baum20,Baum22}.
               In addition to locality, causality and unitarity, these principles include the isometries of inflationary de Sitter spacetime, with the dilation symmetry of the latter closely associated with the scale invariance. Collectively the principles tightly  constrain the  cosmological correlators that inflation can produce.
       
       While these principles could in general be imposed upon a Lagrangian constructed in terms of specific local fields, the cosmological bootstrap bypasses any model based upon a Lagrangian, equations of motion and Feynman diagrams.
     By sidestepping the need for explicit particle dynamics and establishing a direct link between  the basic principles and the observables the non-perturbative cosmological bootstrap approach can provide direct insight into hidden structure at a deeper microscopic level while avoiding model-dependence,  as represented in figure~\ref{cosmo}.

\begin{figure}[htbp]  
\centering
\leavevmode
\includegraphics[width=14.1cm]{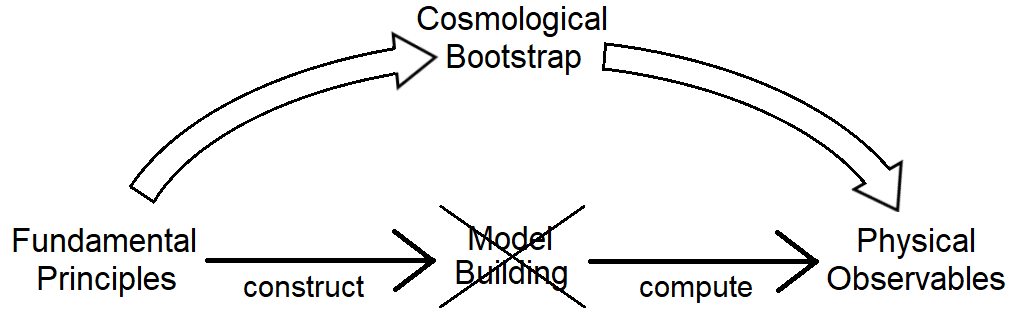}
\caption{\setb
   In the cosmological bootstrap a direct connection is established from the basic principles, including the constraints of causality, unitarity and the symmetries of the spacetime, to cosmological observables, such as large-scale spatial correlations in galactic structure. By circumventing the assumptions and structures of specific models the calculations can be greatly simplified and the link between the fundamental principles and the physical observations clarified. 
   }
\label{cosmo}
\end{figure}

    Pairs of extremely massive particles,  up to of order $10^{14}\,$GeV in mass, can be spontaneously produced out of the vacuum during  the rapid inflationary expansion, attaining energies far higher than can be achieved with terrestrial particle accelerators. These particle pairs created at spacetime points may very quickly decay leaving  remnants  of the event, with 
    such   inflationary particle production in the very early universe effectively acting as a giant cosmological collider~\cite{ArkMa}. These fluctuations and the correlations generated by the remnants might hence allow the probing of new physics beyond the Standard Model at far higher energies than possible in the analogous laboratory high energy physics (HEP) experiments.  
    
      The aim is then to utilise observable higher-point correlations to infer the nature of these primordial events, to determine what kinds of states are `colliding' and their scattering properties. With the early universe seen as an extremely high-energy collider experiment, the cosmological correlator functions contain the scattering amplitudes of this experiment, encoding the physics of the very early universe and 
      establishing a link with the $S$-matrix bootstrap.

      The correlator functions, expressed in momentum space, are largely constrained or even fixed by the combined role of symmetries and the singularities in the $S$-matrix scattering amplitude structure they contain. The challenge is to determine the correlations in the physically observable region via analytic continuation from the singular points, or poles, in the unphysical negative energy region that act as boundary conditions, with
      the singularities playing a role analogous to the Regge poles described earlier in this section.  
          Higher-point correlation functions can be related to, and built up from, simpler lower-point functions  in a manner similar to the relation between scattering amplitudes in a flat spacetime as constrained by unitarity and the optical theorem.
          With `cutting rules' for cosmological correlators identified similarly as for the `Cutkosky rules' for $S$-matrix calculations with Feynman diagrams, the correlators can be bootstrapped from information about their singularities~\cite{Baum21}.
      
       With the  space of possibilities constrained in this manner, signatures for new massive particles created during inflation can be classified. Since all states up to the scale of $10^{14}\,$GeV can be excited during inflation, including a Standard Model contribution that is suppressed since the direct coupling of a single such field to the inflaton is typically prohibited  
 by gauge symmetry (\cite{ArkMa} section~1, see also~\cite{ChenWX}), there is a significant potential to discover new physics.
      There is hence an interplay between particle physics and cosmology in which the physics of scattering amplitudes can provide new insights into large-scale cosmological structure, while cosmological 
      observations may reveal novel microscopic particle phenomena.

        Through the possibility of accounting for cosmological correlators by quantum fluctuations during inflation,
     employing the symmetries of the very early universe de Sitter spacetime, the cosmological bootstrap 
        could also provide further evidence for such an inflationary era itself.
     This would add to 
      the theoretical successes of inflation in addressing the horizon, flatness and other `problems' that have been known for some time~\cite{Guth,Linde,Achu}.
     Alternative early universe scenarios such as `bounce' or `cyclic' cosmological models would imply different early universe spacetime symmetries and different correlators.
     The specific mechanism for inflation is also still a mystery, as is the nature of the
      inflaton particle itself as  typically associated with a scalar field and as a source of primordial fluctuations. Whether the inflaton is elementary or composite, and whether it may relate to string theory or other higher-dimensional theories, supersymmetric models or even Higgs physics might also be probed by the cosmological bootstrap.
     
     These non-perturbative bootstrap methods might also be sensitive to a primordial role for a quantum gravity theory. All inflationary models essentially involve two massless particle states -- the postulated scalar inflaton   and presumed spin-2 graviton, and hence it is desirable to classify all possible interactions of these states. There are tight consistency constraints on the graviton, which should couple to all states, including the inflaton, equally as an expression of the equivalence principle. As well as making a small direct contribution to the 2-point correlation function, the graviton also mediates exchanges playing a role in the inflationary origin of higher-point correlations.
     While these effects may all be small, through the cosmological bootstrap calculations cosmological observations might also then provide hints of the quantum  properties of gravity.
     
    Collectively, the cosmological correlators as observable today can be traced back to the future spacelike boundary  of the approximately de Sitter spacetime inflationary era, without reference to the interior of that spacetime region itself.  That is, the temporal dynamics of the primordial 4-dimensional spacetime translates into 3-dimensional spatial features on that final spacelike boundary, with the spatial correlators after inflation encoding a `memory' of the physics during inflation. This relation between the full temporal evolution and its complete transposition into spatial patterns is a key aspect of the cosmological bootstrap, allowing the equations to be solved for higher-point correlators without explicit reference to evolution in time.

      In this `time without time' approach, time-dependent effects emerge in solutions obtained from time-independent bootstrap constraints, allowing in principle the inference of the mass and spin of particle states in the inflationary background from spatial signals alone. The emergence of a temporal structure of the physics in spacetime from static spatial correlations across great distances is analogous to a `holographic' principle~\cite{ABLP,Baum22}. In this manner the 
      cosmological bootstrap and modern-day amplitude programme have helped reveal hidden mathematical structures. These relate to new mathematical objects such as the `amplituhedron'~\cite{ArkTr}, in which the conventional formalism of field theory and Feynman diagrams is also bypassed in obtaining a new understanding of scattering amplitudes. In that approach scattering amplitudes, for a non-trivial interacting 
       quantum field theory in 4-dimensional spacetime, are proposed to be directly computed as the `volume' of the amplituhedron. 
      
       The cosmological spatial correlations and the bootstrap equations might then be considered to be encoded in such an amplituhedron, as a  geometric object that does not possess an explicit time variable. Under this reconceptualisation of particle collisions within an atemporal geometry the familiar primordial cosmological description \textit{in} time can be considered an apparent illusion.
        Given this `timeless' foundation, 
        with the `time evolution' of particle interactions encoded in the momentum dependence of inflationary spacelike boundary conditions and boundary observables computed without reference to the spacetime interior, `time' might be considered an emergent entity and itself `bootstrapped'.

      The above then presents a rather different picture compared with the new theory of interest to be reviewed  in the following section. As will be clear from subsection~\ref{boot31}, here `time' is considered \textit{the} basic fundamental entity of the theory. 
       Further, while the $S$-matrix and cosmological bootstraps exhibit 
     an inclusive and broad scale scope the underlying basic principles and  symmetry properties still need to be posited.  
       However, for the new approach
      the principal constraints on the physics  can themselves be \textit{derived} directly from the elementary properties of generalised proper time itself, from the symmetry breaking structure implied in the necessary extraction of the substructure of the local \mbox{4-dimensional} spacetime form.
        The initial motivation for employing generalised proper time and a broad range of successes are presented in the following section, with a new source of primordial fluctuations and potentially inflation itself proposed in subsection~\ref{boot33}, before returning to consider a possible `bootstrap' construction for the new theory of universal extent in section~\ref{boot4}.


\section{Generalised Proper Time}
\label{boot3}

\subsection{Motivation and Connections with the Standard Model}
\label{boot31}

   A characteristic feature of the bootstrap approach, as reviewed in the previous section, is the adoption of a minimal set of basic constraints and the avoidance of postulating any specific underlying fields, interactions or additional states. In this sense of removing model-dependence the bootstrap is a `subtractive' approach,
    in sharp contrast with most models that typically introduce new elementary entities or terms in a Lagrangian for example. However, such an `additive' approach, towards a more elaborate foundation, runs against the spirit of unification, for which a minimal and simple starting point would ideally be desired. 
   
     Studies with groups of volunteers set various problem solving tasks  confirm that subtractive and simple solutions tend to get overlooked in favour of approaches that add new elements or a degree of complication,  as a consequence of cognitive, cultural, social and other factors~\cite{Adams}. Reasons for missing a `subtractive' solution include the assumption that simple ideas may already have been tried or the potential seeming initial implausibility of such solutions.  On the other hand reasons for seeking an `additive' solution include the general availability of more options and the tendency of explicit extensions to provide a more tangible contribution, particularly valued in an academic setting, without rolling back on earlier contributions.
     
      Regarding unification schemes in physics, models that posit \textit{extra} spatial dimensions provide an archetypal example of an explicitly \textit{additive} solution.
      With proper time expressed as the continuum:
\begin{equation}
    \label{srrr}
        s \in\rrr
\end{equation}              
       an infinitesimal interval of proper time   can be expressed in
    a local inertial reference frame with coordinates $\{x^0,x^1,x^2,x^3\}\in\rrr^4$
       at the level of the local \mbox{4-dimensional} spacetime structure   in the form:
\begin{equation}
  \label{sxxxx}
   (\delta s)^2 \, = \, (\delta x^0)^2 -  (\delta x^1)^2 
                          -  (\delta x^2)^2 -  (\delta x^3)^2 
\end{equation}         
  The Lorentz symmetry $\soot$, acting on the infinitesimal intervals of the local spacetime coordinates
    $\{\delta x^0,\delta  x^1,\delta x^2,\delta x^3\}$, leaves $\delta s$ invariant -- motivating the 
      name \textit{proper} time.
       The approach of extra spatial dimensions then involves the \textit{addition} of a further $(n-4)$ spacelike
        components $\{\delta x^4,\delta  x^5,\ldots,\delta x^{n-1}\}$ under the quadratic $n$-dimensional
         spacetime form: 
 \begin{equation}
  \label{sxn}
   (\delta s)^2 \, =  \, (\delta x^0)^2 -  (\delta x^1)^2 
                          -  (\delta x^2)^2 -  (\delta x^3)^2 
                           -  (\delta x^4)^2 \ldots -  (\delta x^{n-1})^2 
\end{equation}         
   The proper time interval $\delta s$ is then invariant under the augmented Lorentz group $\sootn$ acting upon the full set of $n$ components $\{\delta x^a; a=0,\ldots,n-1\}$.
   
    Given that the goal is to account for the elementary structure and properties of matter \textit{in} 4-dimensional spacetime it might seem necessary and inevitable that \textit{something} should be \textit{added} to the original spacetime, and hence the \textit{addition} of extra spatial components over and above the 4-dimensional spacetime base would seem to be a plausible approach consistent with this requirement. 
     However, direct or significant connections with the Standard Model  have not been achieved with such approaches. Rather, while typically considering the global spacetime picture, significant geometric assumptions regarding the extended higher-dimensional spacetime together with the addition of further fields, entities or other structures  are needed to obtain any features resembling the familiar multiplet properties of empirical particle physics in the original \mbox{4-dimensional} spacetime. 
     
      By contrast, the new unification scheme described in this paper proposes an explicitly \textit{subtractive} solution, starting out by \textit{removing all} spatial dimensions at the local level and stripping equation~\ref{sxxxx} down to a basic interval of time alone, that is simply:
\begin{equation}
  \label{sint}
    \delta s \in \rrr
\end{equation}    
   This hence assumes only the existence of the temporal continuum, as modelled by the
    real numbers in equation~\ref{srrr}, as the basic fundamental entity.
     Without adding \textit{anything} we simply observe that the \textit{intrinsic} properties of the real continuum imply that the infinitesimal interval in equation~\ref{sint} contains the inherent arithmetic substructure of
      $p^{\mathrm{th}}$-order homogeneous polynomial forms as can be expressed
        (with each coefficient
     $\alpha_{abc\ldots} = -1,0$ or $+1$, and adopting the summation convention over repeated upper and lower indices): 
\begin{equation}
 \label{salpha}
  (\delta s)^p  \, = \, \alpha_{abc\ldots}\delta x^a 
                            \delta x^b \delta x^c \ldots
\end{equation}
  The continuous transformations of a symmetry group $\hG$ act upon the full set of $n$ components
   $\{\delta x^a\} \equiv \bxn \in \rrr^n$ leaving the left-hand side invariant, with $\delta s \in \rrr$ now considered `generalised proper time'. Equation~\ref{salpha} subsumes the 4-dimensional form of  equation~\ref{sxxxx}  and takes the place of the restricted quadratic form of equation~\ref{sxn}.

    Equation~\ref{salpha} can hence be written in a form explicitly isolating the quadratic local 4-dimensional spacetime substructure with $\{\delta x^0,\delta  x^1,\delta x^2,\delta x^3\} \equiv \bxf \in \rrr^4$ the  external spacetime components
        and $\bxnf \in \rrr^{n-4}$  the residual components,  as subsets of the original full set $\bxn \in \rrr^n$.
 That is, with  $(\bxnf)^{p-2}$ and $ (\bxn)^{p}$ as shorthand representing the appropriate polynomial expressions  as consistent with equation~\ref{salpha} (and $\eta_{ab} = \mbox{diag}(+1,-1,-1,-1)$ the local Lorentz metric):

  \vspace{-15pt}
{{\setlength{\baselineskip}{1.5\baselineskip}
\begin{eqnarray}
   \label{sbreak}
  (\delta s)^p   =  \overbrace{ \alpha_{abc\ldots}\delta x^a 
                            \delta x^b \delta x^c \ldots}^{\stackrel
                            {\mbox{\raisebox{0pt}{\small{generalised proper time}}}}
                            {\mbox{\raisebox{-9pt}{\small{$a,b,c,\ldots\, =\, 0,\ldots,n-1$}}}}
                                          } 
                                          &  = &
                         \overbrace{( \eta_{ab}\delta x^a \delta x^b )}^
                         {\stackrel
                         {\mbox{\raisebox{0pt}{\small{4D spacetime}}}}
                         { \mbox{\raisebox{-9pt}{\small{$a,b\, =\, 0,1,2,3$}}}}
                         }  
	  \!\!  \times   \overbrace{(\bxnf)^{p-2}  \,\;\,
    +  \,\;\,  (\bxn)^{p}}^
         {\stackrel
               {\mbox{\raisebox{+3pt}{\small{basis for matter}}}} 
               {\mbox{\raisebox{+1pt}{\small{$\;\, \delta x^c;\, c\ge 4\qquad \;\;
            \delta x^a;\,  a\ge 0 \!\!$}}}}
               }       \,  \\        
   \label{gbreak}          
       \underbrace{\qquad\qquad\!\!\hG\qquad\qquad\!\!}_{\mbox{\small{full symmetry}}} 
                                   \;\;\,  &     \supset &  \;
                \underbrace{\quad\! \mbox{Lorentz} \quad\!}_{\mbox{\small{external}}}
          \;  \times \:
           \underbrace{\qquad\qquad\;\;\, G \qquad\qquad\;\;\,}_{\mbox{\small{internal}}}
            \quad\,
\end{eqnarray}
\par}}

\noindent
  The full symmetry $\hG$ of equation~\ref{salpha} is correspondingly fully broken
   in extracting 
    the external Lorentz symmetry of 4-dimensional spacetime as indicated in equation~\ref{gbreak}. 
    The residual components  in equation~\ref{sbreak},  in general transforming under \textit{both} the external Lorentz symmetry and the remaining internal symmetry $G$ in equation~\ref{gbreak}, provide the basis for the local structure of matter fields  as coupled to gauge fields associated \mbox{with $G$}. 
  The \textit{necessity} of breaking the full symmetry $\hG$ to construct 4-dimensional spacetime itself implies that the local symmetry for any physics in that spacetime is that of the group product
    $\mbox{Lorentz}\times G$ on the right-hand side of equation~\ref{gbreak}, as consistent with the requirements of the Coleman-Mandula theorem applied for the quantum field theory limit described in the following subsection.
  
  The progression in local forms for time from the simple basis in equation~\ref{sint}, via the specific quadratic form of equation~\ref{sxxxx}, to the general form for proper time in equation~\ref{salpha} and the resulting symmetry breaking of equations~\ref{sbreak} and \ref{gbreak} is pictured in figure~\ref{sgener}. 
  Given that for the Standard Model of particle physics we are interested in the local symmetry and local interaction properties of particle states this approach provides a reasonable basis for identifying the elementary structures of matter.
\vspace{-4pt}
\begin{figure}[htbp]  
\centering
\leavevmode
\includegraphics[width=14.4cm]{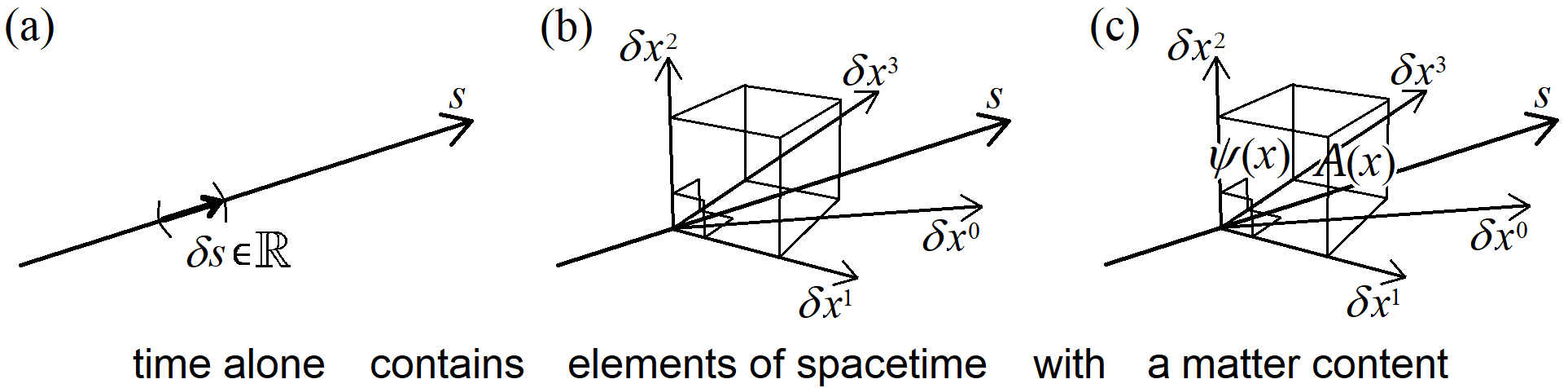}
\vspace{-29pt}    
\caption{\setb
   The progression in forms for proper time. (a) On assuming the temporal continuum $s \in \rrr$ of equation~\ref{srrr} intervals of arbitrarily short duration as expressed in equation~\ref{sint} can be identified. (b) The specific quadratic \textit{arithmetic substructure} of such infinitesimal intervals in equation~\ref{sxxxx}, with four components 
     $\{\delta x^0,\delta x^1,\delta x^2,\delta x^3\}$ and a Lorentz symmetry, has a \textit{geometric interpretation} as a local basis for a \mbox{4-dimensional} spacetime with its causal structure. (c) Generalised proper time in equation~\ref{salpha} then provides the basis not only for the external spacetime but also a matter content, including $\psi(x)$ matter fields associated with the residual components in equation~\ref{sbreak} and gauge fields $A(x)$ associated with the residual internal symmetry $G$ of equation~\ref{gbreak} (where $x$ represents coordinates in an extended 4-dimensional spacetime). 
   }
\label{sgener}
\end{figure}     


   A non-trivial direct augmentation from the 4-dimensional spacetime form for proper time in equation~\ref{sxxxx} can be achieved on first writing this quadratic expression in four components in a matrix form 
    as   $(\delta s)^2 = \det (\bxf)$ on employing the determinant of the $2\times 2$ complex Hermitian matrix $\bxf \in \htwc$, upon which the symmetry group $\sltc$ acts as the double cover of the Lorentz group
     (\cite{TimeE} equation~44, \cite{Basis} equation~4). 
  This structure can then be embedded in turn within a cubic determinant form over nine components
   with an $\slthc$ symmetry, a cubic 27-component form with an $\esi \equiv \sltho$ symmetry and through to a quartic \mbox{56-component} form for generalised proper time with an $\ese$ symmetry 
    (\cite{TimeE} equations 45 and 46, \cite{Basis} equations~5--7 and references therein).

     On extracting the local 4-dimensional spacetime substructure the transformation properties of these 56 components,  fragmented under the breaking of the full \mbox{$\hG = \ese$} symmetry as described for equations~\ref{sbreak} and \ref{gbreak}, reveal a series of Standard Model properties. These include a Lorentz spinor structure, gauge $\suth_c$ colour singlets and triplets and $\uo_Q$ electromagnetic fractional charges, 
 as well as significant features of the electroweak sector including a basis for a left-right asymmetry. As a source for the structure of matter these components can then be closely correlated with a generation of leptons and quarks in the Standard Model.
  Elements of electroweak symmetry breaking can be shown to arise from the impingement of some of the generators of the internal symmetry $G$ in equation~\ref{gbreak} upon some of the external 4-dimensional spacetime components of $\bxf$ in equation~\ref{sbreak}. These external components are then at the heart of both the overall symmetry breaking structure and also the electroweak symmetry breaking, with the scalar function
    $\mbox{det}(\bxf)$ associated with the Standard Model Higgs field (\cite{TimeE} section~4 and figure~4 discussion therein).

   The need to recover the full Standard Model, including three generations of leptons and quarks with a full Lorentz spinor structure and complete electroweak sector, together with the properties of known mathematical structures, leads to the theoretical prediction of an octonion-rich $\ee$-type symmetry for the ultimate form of generalised proper time, potentially of octic order and over 248 components
    (\cite{TimeE} section~5, \cite{Basis} equation~8, \cite{Octo} equation~9).
    Further ongoing developments  in relating algebraic structures associated with $\ee$ and the octonions to the Standard Model 
      (including for example~\cite{GordSM,AranA,KVSin,Wilson,ChesRM,FurH,Todo} and references therein) may provide further input for deducing this proposed mathematical structure in the context of the conceptual framework of the new theory.
    
  The possible form an $\ee$ symmetry for generalised proper time could take, as also consistent with the nature of the embedding of Standard Model properties in the mathematical structure of the $\esi$ and $\ese$ levels, implies a source for new physics in closely connected Higgs and neutrino sectors~\cite{BSM,Gener}.
   The scalar Higgs itself, associated with $\mbox{det}(\bxf)$ over the external spacetime components as noted above, can also be interpreted as effectively a composite of right-handed neutrino $\nu_R$ degrees of freedom, while a possible leptogenesis origin for
    the matter-antimatter asymmetry  is conceivable through $C\!P$-violating decays of the  remaining heavy $\nu_R$ physical states in the early universe; as collectively guided by existing models (\cite{Gener} subsection~1.1 and section~4 and references therein).

   These `Standard Model and beyond' properties hence arise \textit{very directly} for this `subtractive' approach of generalised proper time as introduced for \mbox{equations~\ref{sint}--\ref{gbreak}} and figure~\ref{sgener} through unique  mathematical structures.
     On the other hand, models based upon the `additive' approach with extra dimensions of space, which have led to further models with yet further additional structures involving for example assumptions about not only the higher-dimensional bulk  geometry but also  supersymmetry and even string theory frameworks, have not made clear or convincing connection with explicit structures of the Standard Model, but rather may rely upon anthropic selection from a vast  landscape or multiverse of possible worlds.
     
      Further, unlike the posited mathematical structures of these elaborate additive approaches, for which there remains no tangible empirical evidence, the new theory is based purely on the continuous flow of time, as represented in figure~\ref{sgener}(a) and as an intimately familiar element of \textit{all} our observations in the world. While perhaps sounding initially implausible  the theory of generalised proper time, as represented in figures~\ref{sgener}(b) and \ref{sgener}(c), is fully consistent with the ideal of unification in keeping the basis for the theory `as simple as possible, but not simpler' (paraphrasing Einstein, see also \cite{Struct} section~5). This underlying simplicity together with the direct connections established with empirical data make a compelling case for this approach.

       The  new theory is based in part on a reassessment of the possible options for generalising beyond the geometry of 4-dimensional spacetime to incorporate a matter content. Generalised proper time can then be conceived in a similar spirit to the unified theories of circa the 1920s which were based upon generalising from the metric structure of general relativity, such as provided the original motivation for  Kaluza-Klein theory with one extra dimension of space (\cite{Gener} subsections~1.2 and 5.1). 
        However, here by contrast for the `subtractive' approach of generalised proper time all spatial dimensions are initially removed, as described for equations~\ref{sint} and \ref{salpha}, which amounts also to \textit{removing} the assumption that an augmentation from the 4-dimensional spacetime form  of equation~\ref{sxxxx} should take the \textit{quadratic} form of equation~\ref{sxn}.
  After all, the whole purpose is to seek a source for matter, \textit{not} a source for further dimensions of space which are not in themselves needed for anything, and hence the \textit{quadratic} `spatial' restriction can readily be dropped.
  
    There is also an analogy with the progression from special relativity to general relativity itself, with the \textit{dropping} of the restriction to a global flat 4-dimensional Minkowski spacetime in favour of a more general 
 curved pseudo-Riemannian manifold. While the immediate mathematics of differential geometry is somewhat more complicated, this came with the enormous benefit of leading directly to a theory of gravity, with general relativity continuing to prove  highly successful empirically.
   With the curved geometry described by the Einstein tensor $G^{\mu\nu}(x)$, as a function of the metric
    $g_{\mu\nu}(x)$, 
  this success has been achieved through the Einstein field equation that directly relates this curvature 
   to the energy-momentum tensor $T^{\mu\nu}(x)$ via  a  normalisation constant $-\kappa$ factor
    (with $\mu,\nu$ general coordinate indices in 4-dimensional spacetime):
 \vspace{-3pt}
\begin{equation}
  \label{Eineq}
      G^{\mu\nu} \; = \; -\kappa T^{\mu\nu}
\end{equation}
 \vspace{-15pt}

  For the present theory on \textit{dropping} the quadratic form restriction for proper time, as imposed for models with extra spatial dimensions, the immediate mathematics for generalised proper time, in leading directly to symmetry groups such as $\esi$, $\ese$ and potentially $\ee$, is also more complicated but has the great benefit of directly accounting for esoteric properties of the Standard Model multiplet structure in the patterns of symmetry breaking over the local 4-dimensional spacetime base. The resulting elementary matter content of the theory, arising from this local generalisation from the 4-dimensional spacetime form,  then matches that empirically desired as underlying $T^{\mu\nu}(x)$ on the right-hand side of equation~\ref{Eineq}, very much as was sought for the early unified field theories in the 1920s by generalising general relativity.

  While this theory is conservative, in being based on the flow of time alone in equations~\ref{srrr} and \ref{sint}, and simple, employing the direct arithmetic generalisation of equation~\ref{salpha}, a further desired feature for unification is uniqueness. Compared with models of extra spatial dimensions which begin with a \textit{choice} of the number of dimensions and the properties of a \textit{global} geometric structure
  together typically with a mechanism to `compactify' the extra dimensions,  
   for generalised proper time in beginning with a \textit{single} temporal component and the \textit{local} form of equation~\ref{salpha} there is considerably less room for ambiguity in directly deducing the elementary local structure of matter.
  However, while the exceptional Lie groups $\esi$, $\ese$ and $\ee$, and their smallest non-trivial representations in 27, 56 and 248 dimensions respectively, are unique mathematical structures, the general expression of equation~\ref{salpha} can take other explicit forms that also accommodate the substructure of the local 4-dimensional spacetime form in equation~\ref{sxxxx} via equation~\ref{sbreak}. This apparent \textit{small} degree of ambiguity will  though itself have a significant beneficial role  given the need to also account for a significant dark sector, as we shall describe in subsection~\ref{boot33}.
  
       While these visible and dark sectors of matter are identified from the symmetry breaking in the local structure of the theory, the incorporation of general relativity and the Einstein field equation~\ref{Eineq} requires the construction of an extended global \mbox{4-dimensional} spacetime. This single
       external  spacetime together with its smooth geometry 
        can be built up by the contiguous and continuous juxtaposition of the local spacetime elements of figure~\ref{sgener}(c). In this case there is an \textit{enormous} degree of ambiguity in the possible interconnections of the underlying matter content contributions for the right-hand side of equation~\ref{Eineq}, even for any given external geometric structure on the left-hand side and from within the visible matter sector alone. The benefit here, however, will be in providing a basis for amalgamating the geometric structure of  classical general relativity  with the indeterminacy and other characteristic phenomena of quantum theory, as we describe in the following subsection.


\vspace{-3pt}
\subsection{General Relativity with a Quantum Theory Limit}
\label{boot32}
 
   In the previous subsection we reviewed how the non-trivial multiplet structures of the Standard Model can derive from the direct symmetry breaking structures of generalised proper time, without needing to artificially postulate these esoteric features. This motivates in turn the aim to \textit{derive} a formalism for the quantum phenomena of the corresponding particle states, within a framework consistent with general relativity, in a similarly direct manner without the need to impose a set of postulates for quantum theory by hand. In this subsection we describe how this might be achieved.
   
    It is a fairly common view that quantum mechanics should encompass \textit{all} physical phenomena, with the apparent classical world \textit{emergent} for sufficiently large mass or macroscopic distance scales. However, from all of our practical experience in the laboratory and through observations in general quantum mechanics only covers a quite specific aspect of physical phenomena as applied for processes on a limited scale \textit{within} a classically defined framework that is needed \textit{a priori} to make sense of the observations. This dualistic structure could also imply the inevitable incompleteness  of quantum mechanics as a description of nature and the need to understand quantum phenomena within a broader theoretical framework, as will be the view taken here.
   
    Here we are not setting out by imposing any quantisation conditions, and in particular the theory will \textit{not} involve any quantisation of the gravitational field or the 4-dimensional spacetime geometry itself. There is after all no evidence to suggest that gravity \textit{should} be quantised and in fact we do not know of any underlying reasons why \textit{anything} should be quantised (see for example~\cite{Hugg,Kief}).
    Further, gravity is simply \textit{different} from the other forces of nature in that it is 
    based upon \textit{external} spacetime curvature rather than an \textit{internal} gauge symmetry and  applies universally rather than in a manner determined by particle type, while also being far weaker. It is also technically much more difficult to `quantise' gravity.  
        However, these observations do \textit{not} mean giving up on a unified theory of `quantum gravity'. On the contrary, not quantising gravity and rather going the other way and giving the conception and formalism of classical general relativity the upper hand will be central to explaining the origin of the \textit{quantisation} for all non-gravitational elementary fields in spacetime. Such a unification scheme might then rather be classified as an approach of `gravitising quantum theory' (see also section~\ref{boot5}).
      
      The focus will be upon particle and quantum physics events as observed in laboratory experiments. We first note that  \textit{no} empirical observations, either from individual events or experiments or when taken collectively, indicate  that classical general relativity must break down in this environment. In violating Bell's inequalities correlations between observables in EPR-type (Einstein-Podolsky-Rosen) experiments are consistent with the pragmatic calculations of quantum theory, without the theory giving any definitive unambiguous picture of `what is actually physically happening' between and in the measurements. No equivalent of Bell's theorem has been formulated for general relativity that could indicate any necessary inconsistency between what is \textit{observed} in these experiments and the continuous spacetime geometry of solutions for
       the Einstein field equation~\ref{Eineq}.

        However, general relativity \textit{is} inconsistent with the quantum mechanical \textit{account} of such elementary events in terms a wavefunction undergoing a continuous evolution followed by a discrete collapse corresponding to a measurement outcome. This suggests 
        it should be worth
         attempting to reconceptualise and reformulate observations of laboratory quantum phenomena in manner consistent with, and subsumed within, classical general relativity and equation~\ref{Eineq}. Pursuing this strategy, within the context of generalised proper time, a framework for amalgamating general relativity with a quantum theory limit can indeed be constructed, while also giving an explicit picture of `what is actually physically happening' in particle processes, as we describe below.

          The starting point for generalised proper time is at the very local level of \mbox{4-dimensional} spacetime, as described for equations~\ref{sxxxx} and \ref{sint}--\ref{sbreak} and figure~\ref{sgener}.
           While this limiting infinitesimal local structure incorporates the geometry of a 4-dimensional spacetime inertial frame, in continuously piecing such elements together a generally curved spacetime, consistent with the pseudo-Riemannian manifold, equivalence principle and causal structure of general relativity, can be constructed (\cite{Tflow} section~3).
      The extended external geometry, described by the Einstein tensor $G^{\mu\nu}(x)$ of equation~\ref{Eineq}, will be interdependently related to both the matter field components
       $\psi(x)$, associated with the residual $\{\delta x^a\}$ components in equation~\ref{sbreak},  and
      internal gauge fields $A(x)$, associated with the internal symmetry $G$ in equation~\ref{gbreak} (see for example~\cite{KKone}), in the 4-dimensional spacetime as represented at the local level in figure~\ref{sgener}(c).
      
       This will generate a vast number of possible solutions for building a consistent 4-dimensional spacetime with an apparent matter content. In particular there will be a local \textit{degeneracy} of matter field states 
        $\{A,\psi \}$ underlying the \textit{same} local smooth external spacetime geometry. This local ambiguity in how the matter components can fit together at a local level, in constructing the continuous extended manifold with a    
      matter content, underlies the indeterminacy and probabilistic nature of the quantum phenomena
      of matter. A basis for reconceptualising quantum theory, with classical general relativity remaining essentially intact, can then be identified, and with an energy-momentum tensor $T^{\mu\nu}(x)$ \textit{defined} in turn in terms of the underlying   matter field structure $f^{\mu\nu}(A,\psi)$:    
  \vspace{-3pt}
\begin{equation}
  \label{GRQT}
      \mbox{\raisebox{9pt}{$\overbrace{{\mbox{\raisebox{-10pt}{$G^{\mu\nu}  \; = \!\!\! \!\!\!   \!\!\!
           \underbrace{\, f^{\mu\nu}(A,\psi) \,}_{\;\;\;\;\; \mbox{{Quantum Theory}}}   
    \!\!\!\!\! \!\! \!\!\! =: \,  -\kappa T^{\mu\nu}$}}}}^{\mbox{{General Relativity}}}$}}
\end{equation}     

   While the Einstein tensor $G^{\mu\nu}(x)$ describes ten of the twenty degrees of freedom of the full Riemann curvature tensor in general relativity, its significance is in a large part due to the key property of the contracted Bianchi identity with the vanishing of   
    $\nabla_{\!\mu}G^{\mu\nu} = 0$ (where $\nabla_{\!\mu}$ is the  covariant derivative,  \cite{Unifi} section~3.3). Hence on defining the  energy-momentum tensor through equation~\ref{GRQT} the
    corresponding identity  
     $\nabla_{\!\mu}T^{\mu\nu} = 0$ in turn  \textit{implies} the conservation of energy-momentum in the flat spacetime approximation (\cite{Unifi} section~5.2 opening). 
     
     The interactions of the `matter field' components $\{A,\psi \}$ in equation~\ref{GRQT},  interpreted as composing the energy-momentum $T^{\mu\nu}(x)$, are in general more readily and clearly observable than the small deviations from a flat  spacetime geometry, 
       as described by $G^{\mu\nu}(x)$ as a function of the metric field  $g_{\mu\nu}(x)$. Hence a distribution of matter together with the symmetry of a system  can be used collectively as `boundary conditions' to solve the second order differential equations for $g_{\mu\nu}(x)$ to determine a complete description of the 4-dimensional spacetime  geometry. This is then very similar to the employment of equation~\ref{Eineq} in  general relativity in practice to yield full 4-dimensional spacetime geometries, such as for example the Schwarzschild solution given a central mass and spherical symmetry as the boundary conditions (\cite{Unifi} equation~5.49).

   A full 4-dimensional spacetime solution for equation~\ref{GRQT}, consistent with the Einstein field equation~\ref{Eineq}, contains an everywhere flowing progression in time from which it is constructed (locally as described for figure~\ref{sgener}) and which parametrises an ordered progression in local degenerate matter field exchanges, paralleling a description in terms of the evolution of states in quantum theory. In this manner the `block' and `dynamic' conceptions of the physical world are brought together. The whole structure is analogous to the `fixed' geological feature of a vast meandering river that contains an everywhere `flowing' progression of water, and is key to the establishment of a framework for amalgamating general relativity with a quantum theory limit  (\cite{Tflow} sections~6 and 7).

    Constraints on the possible local exchanges between matter field components  $\{A,\psi \}$ are determined by the structure of generalised proper time itself, in place of any postulated Lagrangian as tailor-made to match observations. With time the fundamental entity not only are the infinitesimal intervals of equation~\ref{salpha} invariant under a full symmetry $\hG$, the property of invariance will also necessarily apply for arbitrary finite temporal  intervals 
    expressed as an integral sum over infinitesimal elements  of the $n$-component form for
    generalised proper time  (\cite{Basis} equation~23 and discussion):  
\begin{equation}
     \label{extgpt}
     S_{n} \; = \;    \int_{n}  \delta s      \; = \;
          \int_{n}   \left( \alpha_{abc\ldots} \frac{\delta x^{a}}{ \delta s}
         \frac{\delta x^{b}}{ \delta s}\frac{\delta x^{c}}{ \delta s}\ldots 
                                                          \right)^{\!\frac{1}{p}}  \delta s
 \end{equation}  
          The required  symmetry breaking,
      in selecting four preferred components as local external spacetime coordinate intervals
         $\{\delta x^0,\delta x^1,\delta x^2,\delta x^3 \}$ on projecting out the local 4-dimensional spacetime geometric
          form as described for equations~\ref{sbreak} and \ref{gbreak},   implies significant constraints on the physics in the extended spacetime.
          
   It is convenient to rewrite the general expression for infinitesimal intervals in equation~\ref{salpha} in terms of the generally finite  components
 $v^a := \frac{\delta x^a}{\delta s}   {\big{\vert}}_{\mbox {\tiny $\delta s \! \to \! 0$}}$ 
   of an $n$-vector $\bv_n \in \rrr^n$  (more closely associated with the matter fields $\psi(x)$
   introduced in figure~\ref{sgener}(c) and  incorporated 
    into equation~\ref{GRQT})  and to
  employ the notation:
\begin{equation}
  \label{lpvn}
  L_p(\bv_n)_{\hat{G}} 
  \; := \; \alpha_{abc\ldots} \frac{\delta x^a}{\delta s}
   \frac{\delta x^b}{\delta s}\frac{\delta x^c}{\delta s}\ldots
   \Big\vert_{\delta s \to 0} \; = \;
    \alpha_{abc\ldots}v^a v^b v^c \ldots \; = \; 1
\end{equation}
  incorporating the homogeneous power $p$, dimension $n$ and full symmetry
   $\hat{G}$ for an explicit instantiation of the general form for proper time  (\cite{Gener} equation~13).
      The form of equation~\ref{lpvn} can then be substituted into equation~\ref{extgpt}.

     In building up the extended 4-dimensional spacetime \textit{from} the  elements of  time  that compose $S_n$ in equation~\ref{extgpt}, rather than introducing an action $\mcSL$
     based on a Lagrangian $\lag$ and
      defined \textit{within} a pre-existing spacetime,  the overall construction is inevitably somewhat different to that of an action principle. However, the consequences of breaking the symmetry of equation~\ref{extgpt}, listed  in terms of `geometric' and `matter' parts in equation~\ref{sconst}, do collectively exhibit a correspondence with the Lagrangian terms of the Standard Model as summarised in
       equation~\ref{asmlag}, as indicated
       in general terms below (see for example \cite{Lang} for the standard  terms and \cite{QGrav} equations~36--44 for more details on the correspondence identified):

\vspace{-35pt}   
{{\setlength{\baselineskip}{2.8\baselineskip} 
\begin{eqnarray}
   \label{sconst}
 \!\!\!\!\!  \overbrace{\tilde{I} = \int \big( R + FF \big) d^4 x \, ;}^{\mbox{{geometric}}}
                                      \,\,\,\,    &   & \!\!\!\!
    \raisebox{14pt}{$\overbrace{\raisebox{-14pt}{$\Lsl_p(\bv_n)_{\mathrm{Lorentz}\times G} 
                                                 \, = \,  1; \quad
                          \gamma^{\mu}\! D_{\mu} \, 
   \Lsl_p(\bv_n)_{\mathrm{Lorentz}\times G} \, = \,  0 $} {\color{white}.}\!
                                   }^{\mbox{{matter}}}$} 
                                    \qquad  
	  \\
  \setlength{\unitlength}{25pt}
\begin{picture}(0.0,0.0)(0.0,0.0)
  {\Large
	 \put(1.9,1.47){\line(1 ,-2){0.5}}
	 \put(3.1,1.47){\line(1,-1){1.0}}
	 \put(6.9,1.47){\line(-1 ,-2){0.5}}
	 \put(7.9,1.47){\line(4,-1){4.1}}
	 \put(10.6,1.47){\line(-2 ,-1){2.07}}
	 \put(11.7,1.47){\line(-1,-1){1.04}}	 
   }       
 \end{picture}	             
   \label{asmlag}          
         \mcSL = \int   \Big(  \!\!\!\!\! \underbrace{  
                     \;\; R\;\; }_{\stackrel{\mbox{\small{scalar}}}
                                                                                 {\mbox{\small{curvature}}}}   
                  +  \underbrace{  \;\; FF \;\; }_{\stackrel{\mbox{\small{gauge}}}
                                                                                 {\mbox{\small{kinetic}}}}                                                                
            \!\!   &  + &     \!\!  
                  \underbrace{ Y \bar{f} \Phi f }_{\stackrel{\mbox{\small{fermion}}}
                                                                                 {\mbox{\small{mass}}}}                                                                
   \; +\; \underbrace{ \bar{f} \gamma^{\mu} \! D_{\mu} f }_{\stackrel{\mbox{\small{fermion}}}
                                                                                 {\mbox{\small{kinetic}}}}
    \; + \;  \underbrace{ (D_{\mu}\Phi)^2  - V(\Phi)  }_{\stackrel{\mbox{\small{Higgs}}}
                                                                          {\mbox{\small{sector}}}} \,  \Big) d^4 x
                                                               \quad
\end{eqnarray}  
  \par}}

 \vspace{5pt}

   The `geometric' part of the symmetry breaking in equation~\ref{sconst} corresponds to an effective action integral $\tilde{I}$ over a perturbation from the 4-dimensional scalar curvature $R(x)$ of the Einstein-Hilbert action in general relativity that incorporates also a gauge kinetic term quadratic in the gauge field strength tensor  $F(x)$
    (\cite{KKone} equation~91).
    \mbox{The `matter' part} consists of the fragmented terms of 
equation~\ref{lpvn}, as denoted by `$\Lsl$', as  partitioned  under the $\mbox{Lorentz} \times G$
  broken symmetry of 
  equation~\ref{gbreak} into sums of individually 
invariant parts,
 together with the vanishing of the corresponding gauge covariant derivative $D_{\mu}$
  as can be contracted with the 
   Dirac gamma matrices to form scalar $\gamma^{\mu}\! D_{\mu}  \Lsl$
 terms  (\cite{QGrav} equations~43 and 44). 
 
 With the fragmented terms in equation~\ref{sconst} containing factors 
 interpreted as Yukawa couplings $Y$
  and Higgs components $\Phi$ in addition to lepton and quark components $f$,
  as briefly reviewed in the previous subsection, 
   some of the $\Lsl$  terms correspond to Standard Model fermion mass terms
    as indicated by the correspondence with equation~\ref{asmlag}.
    The scalar Higgs field is associated with  $\mbox{det}(\bxf)$, as discussed following figure~\ref{sgener},  and hence  with the external
   Lorentz \mbox{4-vector} scalar magnitude $\vert \bvf \vert$. 
   This quantity parametrises the symmetry breaking projection itself, 
   over the local \mbox{4-dimensional} spacetime,
    attaining a stable vacuum value $\vert \bvf(x) \vert = h_0$  in the very early universe
       and taking the place of the Higgs potential $V(\Phi)$.  Similarly the fragmented $\gamma^{\mu}\! D_{\mu} \,  \Lsl$ terms  
 correspond to fermion kinetic terms, which incorporate interactions with the internal gauge fields, and also take the place of the Higgs model in generating  electroweak gauge boson masses as also alluded to in
 the discussion following figure~\ref{sgener}.

     Given the direct consequences of the symmetry breaking of the full $\hG= \esi$, $\ese$ and proposed 
      $\ee$ forms for generalised proper time, as  summarised in the previous subsection, the terms arising from equations~\ref{sconst} can be seen and further anticipated  to closely match the field content and terms of the Standard Model Lagrangian of equation~\ref{asmlag}. The constraints of this familiar Lagrangian, which contains many terms when written out in full, do not then need to be imposed by hand but rather these Standard Model properties are intrinsically identified in the symmetry breaking of generalised proper time.
      
       This also implies that some degree of deviation from the structure of the Standard Model can arise.  For example the geometric part of equation~\ref{sconst} contains $FF$ terms but \textit{not} $F^{\,\ast}\!F$ terms, where ${}^{\,\ast}\!F$ is the dual field strength tensor,  potentially accounting for the `strong $C\! P$ problem' of the Standard Model (see \cite{QGrav} equation~94 discussion and references therein). New physics is also implied in the Higgs and neutrino sectors, as briefly reviewed in the previous subsection in relation to a possible origin for the matter-antimatter asymmetry in the early universe, and with the potential for yet further  physics beyond the Standard Model that could also have observable implications.

    The action principle itself has a long history of employing paths or integrals of quantities over a time variable, as can be interpreted as a `principle of extremal time' (\cite{Basis} section~5).
    In the present theory with proper time $S_n$ on the left-hand side of equation~\ref{extgpt} considered the primary basic entity it is necessarily the case that 
     $\Delta S_n = 0$ for any variation in the components  \textit{through which it is expressed}
      on the right-hand side. This underlying construction then not only differs from and takes the place of an `action' but also explains \textit{why} the `stationary action' principle works, with for example $\Delta \mcSL = 0$ with respect to field component variations on the right-hand side of equation~\ref{asmlag} generating the standard field equations of motion.
      The structure of this physical realisation for finite intervals of generalised proper time, incorporating both the external geometry and matter field components as mutually constrained through equations~\ref{GRQT}--\ref{sconst}, is central to the consistent amalgamation of classical general relativity with quantum theory within the present framework.

     The notation
   $\Rcon$, for `Redescription constraints', is  employed for the collection of  terms deriving from equations~\ref{sconst}  as representing the allowed field exchanges.
    Any given local spacetime geometry in equation~\ref{GRQT}
     (such as represented by $g_{\mu\nu}(x)$ in figure~\ref{sinter} below) can in general be constructed in terms of a degenerate set of possible underlying matter fields $\{A(x),\psi(x)\}$. Local interchanges between the matter fields are then freely permitted provided they are consistent with the  $\Rcon$ constraints deriving from equations~\ref{sconst} and leave the local geometry invariant.

      Under the approximation of a nearly flat spacetime, as appropriate for laboratory experiments, a linearised approximation for general relativity can be employed to describe the external geometry on the left-hand side of equation~\ref{GRQT} (\cite{QGrav} section~5). In this case the matter field contributions can be expanded in terms of
    $e^{\pm ik\cdot x}$    Fourier modes, with wave 4-vector $k$, allowing a simple expression for field exchanges leaving terms of $\Rcon$ invariant. For example  the   $\gamma^{\mu}\! D_{\mu}   \Lsl$  
    expression in equations~\ref{sconst}, with a gauge covariant derivative, will generate terms with factors of the form $\Rcon \sim \overline{\psi} \gamma^{\mu} \! A_{\mu} \psi$ coupling a spinor matter field $\psi(x)$ with a gauge field $A(x)$. 
     This permits an exchange of the form (see \cite{QGrav} equations~62--67 for a more complete example):   
 \vspace{-2pt}
{\setlength{\baselineskip}{0.0\baselineskip}
\begin{eqnarray}
     \mbox{\raisebox{-17pt}{$\Rcon$}} &  \mbox{\raisebox{-17pt}{$\sim$}} &  
	    \mbox{\raisebox{-17pt}{$\left(\overline{\psi}(k_2)e^{+ik_2\cdot x}\right)   
	\gamma^{\mu} \! A_{\mu}
	 \left(\psi(k_1) e^{+ik_1\cdot x}\right)$}}  
	  \nonumber  \\ 
\setlength{\unitlength}{25pt}
\begin{picture}(0.0,0.0)(0.0,0.0)
  {\Large
	 \put(3.35,0.52){\vector(3 ,-2){1.2}}
	 \put(6.55,0.52){\vector(-3,-2){1.2}}
   }
\end{picture}		
       &   &   \label{slide}   \\ 
		& = & \qquad\quad \!\!
    \overline{\psi} \left(\gamma^{\mu}\! A_{\mu}(k)e^{+ik\cdot x}\right)   
  \! \psi    \quad   \mbox{provided} \quad  k_1 + k_2 \; = \; k   
   \nonumber
\end{eqnarray}    
\par}

 \vspace{-2pt}
      
      With the spinor field $\psi(x) \sim e^{+ik_1\cdot x}$  associated with a lepton or quark field and 
    the gauge field  $A(x) \sim e^{+ik\cdot x}$ with a component of the Standard Model gauge symmetry 
      \mbox{$\suth_c \times \sutw_L \times \uo_Y$}, as deriving from equations~\ref{sbreak} and \ref{gbreak} respectively, the possibility of such an exchange can be correlated with a possible  three-prong vertex in a Feynman diagram as associated with 
      $ \bar{f} \gamma^{\mu} \! D_{\mu} f $ terms in equation~\ref{asmlag}
        (\cite{QGrav} figure~3). 
      This intrinsic mathematical degeneracy, in the sliding transfer of $e^{\pm ik\cdot x}$ modes between field contributions as permitted by the $\Rcon$ terms deriving from equation~\ref{sconst}, parallels the somewhat mechanical mutual action of field component  annihilation $a(\bp)$  and creation 
       $a^{\dagger}(\bp)$ operators, featuring in the expansion of interaction Hamiltonian $\Hint$ terms associated with equation~\ref{asmlag}, in canonical quantum field theory calculations
       (see for example~\cite{Unifi} section~10.3). Here such interaction $\Hint$ terms will be interpreted as   \textit{modelling} the underlying role of the constraint $\Rcon$ terms.

      For particle interaction processes between initial and final states observed in high energy physics  experiments the physical particle states are here described more precisely by real wave-packet 
      functions (\cite{QGrav} equation~72 and discussion). In the adopted flat spacetime approximation real Gaussian wave-packets naturally resolve into matching
       $e^{+ik\cdot x}$ and $e^{-ik\cdot x}$ complex sub-component Fourier mode contributions, over a Fourier transform Gaussian distribution in the wave 4-vector $k$.
      An interaction probability $P_{fi}$, between the initial $i$ and final $f$ states, will then be proportional to the product of the degeneracy $D_+$, that is 
         the number of possible ways,  for exchanging the   $e^{+ik\cdot x}$ 
 field components with the degeneracy   $D_-$ for exchanging the   $e^{-ik\cdot x}$ 
 field components in the manner described for equation~\ref{slide}.
  A possible contribution to this calculation for a particular nested time sequence of field exchanges for
   a wave 4-vector component contribution for a specific process is    pictured in figure~\ref{sinter}.  
   
   \vspace{8pt}
   
\begin{figure}[htbp]  
\centering
\leavevmode
\includegraphics[width=14.4cm]{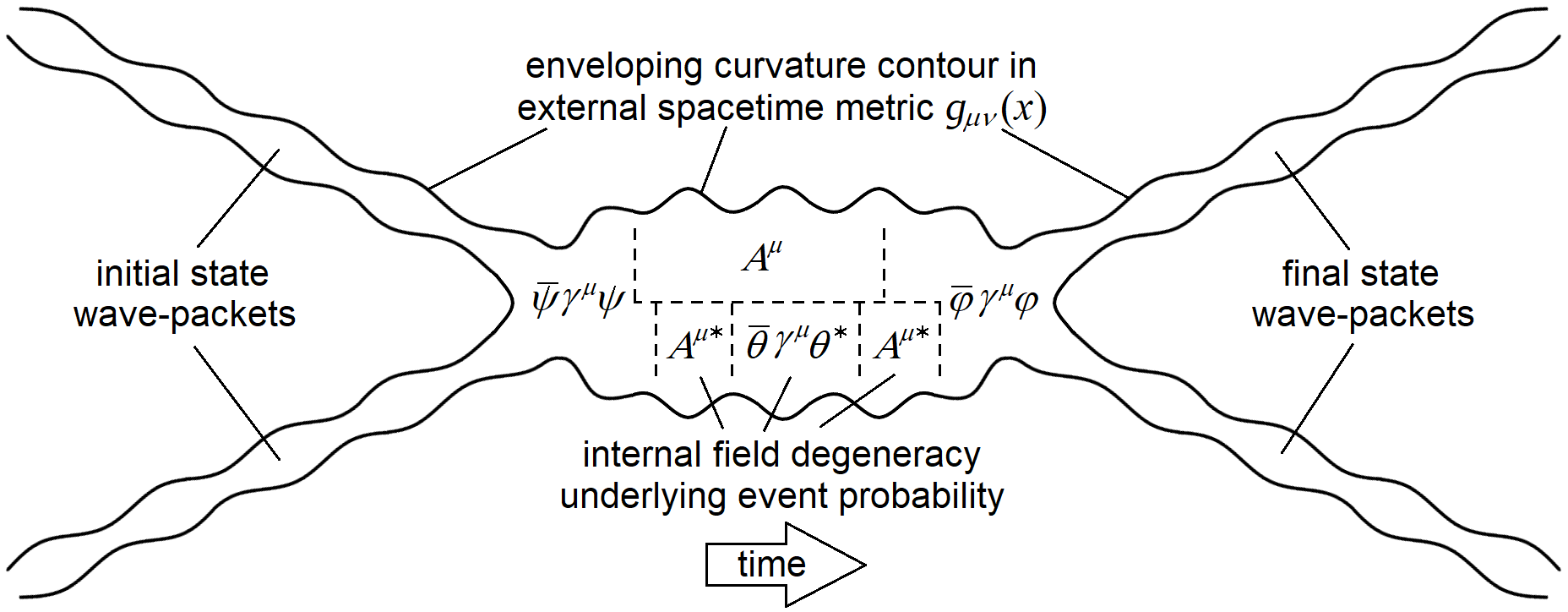}
\vspace{-20pt}
\caption{\setb
   Interaction between initial and final real particle states, as for example representing the process
    $e^+e^- \to \mu^+\mu^-$, described by wave-packets in the matter fields such as $\psi(x)$ or
     $\varphi(x)$  \textit{and} also in the gravitational field $g_{\mu\nu}(x)$.  The process is mediated by a hybrid sequence of $e^{+ik\cdot x}$ exchanges (such as described for equation~\protect\ref{slide}), shown above the dashed partition, 
     together with counterpart exchanges in $e^{-ik\cdot x}$  components, shown below the dashed partition (as indicated by the `$\ast$' superscripts),  
      for each Fourier contribution, while in general including further possible intermediate matter field components such as depicted for $\theta(x)$ as well as for the
       gauge fields $A_{\mu}(x)$  
       (see also \cite{Unifi} figure~11.7 and discussion). 
     The overall enveloping continuous gravitational field (represented by the outer contour) is consistent with general relativity through equation~\ref{GRQT}, while the degeneracy in possible local field exchanges through the constraints of equation~\protect\ref{sconst} underlie the quantum indeterminacy of such events.
   }
\label{sinter}
\end{figure}     

\pagebreak

    Calculations in quantum field theory (QFT) with wave-packet particle states are equivalent to those that take the single central wave-vector value and instead assume a simple plane-wave state  (\cite{QGrav} equation~71 and discussion). Hence a further correspondence with the standard formalism can be sought under this simplifying plane wave approximation.

      With each matter field exchange moderated by an $\Rcon$ term deriving from equation~\ref{sconst}, as described for equation~\ref{slide}, a formal correspondence can be identified between a nested time sequence of both $e^{+ik\cdot x}$ and  $e^{-ik\cdot x}$ field exchanges, such as pictured in the centre of figure~\ref{sinter}, and a contribution to the imaginary part of the forward scattering amplitude $\mbox{Im}(\mcM_{ii})$ as based on a Lagrangian and $\Hint$ terms associated with equation~\ref{asmlag} in QFT. In turn, for a given initial state, the real number $\mbox{Im}(\mcM_{ii})$ is directly related to the total interaction cross-section $\stot$ by the optical theorem. That is:
 \begin{equation}
      P_{fi} \, \propto \,  \overbrace{D_+ \, D_-  \, \sim \, 
	     \mbox{Im}({\mathcal M}_{ii})}^{\mbox{formal similarity}}
	     \!\!\!\!\!\!\!\!\!\! \!\!\!\!\! \!\!\!\!\!\!\!\!\! 
	      \underbrace{ \qquad  \qquad \,\, \propto  \,  \stot \, }_{\mbox{optical theorem}} 
		     \propto  \,   \sum_f  \vert\mcM_{fi}\vert^2   
    \label{pddddm}
\end{equation}

      Each of the correspondences, indicated by the over and under braces, holds at every given order of a perturbative expansion, establishing an overall explicit connection between the degeneracy count described for figure~\ref{sinter} and calculations in QFT. The `formal similarity' identified in equation~\ref{pddddm} involves a correspondence between the construction of the probability degeneracy count and the
       application of `cutting rules' as consistent with unitarity and the optical theorem in QFT (see \cite{Unifi} figure~11.8 and discussion).
       With the `$\sim$' in equation~\ref{pddddm} indicating that the left-hand side provides a \textit{contribution} to the total cross-section on the right-hand side, there remains a need to disentangle the transition amplitude $\mcM_{fi}$ for a specific final state $f$, and to take into account the time-ordering structure, in order to fully determine the connection between the event probability  $P_{fi}$  on the  left-hand side and the QFT calculation for the given final state
   (see discussion of \cite{QGrav} equation~70 and \cite{Unifi} equation~11.46).

        Reading equation~\ref{pddddm} from left-to-right corresponds to a re-expressing of the degeneracy count probability $P_{fi}$ calculation through a complexified structure to a transition amplitude
         $\mcM_{fi}$ linking the initial $i$ and final $f$ states, as the non-trivial part of the $S$-matrix
          (that is $S= \bone + iT$ with
           $S_{fi} =  \langle f \vert S \vert i \rangle$  and
           $\mcM_{fi} \propto  \langle f \vert T \vert i \rangle$, 
           \cite{Unifi} equations~10.4--10.6). 
       In QFT such an initial state $\vert i \rangle$ undergoes a unitary $U(+\infty,-\infty)$  evolution,
       from an initial time $t = -\infty$ to a final time $t = +\infty$,  
        according to interactions described by $\Hint$ terms.
      Given an initial state normalisation $\langle i \vert i \rangle = 1$ such an evolution leaves invariant:
\begin{equation}
   \label{uprob}
       \sum_f  \vert S_{fi}\vert^2   \, = \, 
       \sum_f \vert \langle f \vert U(t,-\infty) \vert i \rangle \vert^2 
           \, \mbox{\raisebox{-3pt}{\LARGE{$\vert$}}}_{t \to +\infty}
            \, = \, 1
\end{equation}
   with the identity with unity on the right-hand side also holding for all intermediate time $t$ values.
    Probability conservation with necessarily $\sum_f P_{fi} = 1$, normalised taking into account all permitted field exchange sequences,  is then \textit{modelled} through the properties of  complex amplitudes and the unitarity of the $S$-matrix with $SS^{\dagger} =\bone$
     (\cite{QGrav}~equations~11--13).

  As a key constraint of the $S$-matrix bootstrap,
    as noted in the opening of section~\ref{boot2}, unitarity is a highly restrictive demand in itself.
 Compared with the \mbox{$S$-matrix} bubble diagram in figure~\ref{bubble} in addition to the `bootstrap' constraints an explicit source of internal interactions is  here identified in figure~\ref{sinter} through the constraints of generalised proper time. That is, equations~\ref{extgpt} and \ref{lpvn}, through the symmetry breaking projection over the local external 4-dimensional spacetime, take the place of an explicit Lagrangian as described for equations~\ref{sconst}~and~\ref{asmlag}.

     Rather than \textit{postulating} the `squared amplitude' definition of an event probability,  as for quantum theory and the case of $\vert\mcM_{fi}\vert^2$, such a  calculation hence here \textit{originates} from  the local construction of 4-dimensional spacetime itself, as described for equation~\ref{GRQT}. Here the degeneracy count on the left-hand side of equation~\ref{pddddm} itself defines an essentially \textit{classical} probability, corresponding to the relative number of ways  that spacetime itself can be constructed as effectively 
     composed of matter fields at the microscopic level.
     
     The sequences of matter field exchanges, contributing to $D_+$ and $D_-$, while being indeterministic, follow a strictly causal order in time, as implied in the discussion of the `river' analogy shortly after equation~\ref{GRQT} and indicated by the `time' arrow in figure~\ref{sinter}, with the constraints on these exchanges determined by the generalised local form for the flow of proper time itself via   equations~\ref{extgpt}--\ref{sconst}.
      With classical general relativity incorporated through the Einstein equation bookending equation~\ref{GRQT}, apparent quantum and particle phenomena are enveloped within `microscopic' solutions for a single smooth 4-dimensional spacetime geometry, on the scale of systems probed in the laboratory, with gravity hence playing an essential and irreducible role. 
     
      By the equivalence principle, as for general relativity, any arbitrarily small region of the 4-dimensional spacetime can be treated as a  local inertial frame, with the metric structure $\eta_{ab}$  and the external Lorentz symmetry in the first term and factor on the right-hand side of equations~\ref{sbreak} and \ref{gbreak} respectively. For macroscopic purposes this Lorentz symmetry from the symmetry breaking structure can also be taken to be globally respected under the flat spacetime approximation being employed in the environment of laboratory phenomena.  
      The Lorentz transformation properties of the initial and final states, that is the particle spin structure, will restrict the possible field interactions and determine the form of the overall event likelihood as a function of geometric angular distributions in the local rest frame, for processes    such as described in figure~\ref{sinter}.   
   Consistency with the corresponding transition amplitude calculations through equation~\ref{pddddm} should then imply the reproduction of the implications of QFT calculations incorporating non-trivial particle spin  (such as \cite{Unifi}  equations~10.10 and 10.11 for figure~10.2 therein, as might correspond to figure~\ref{sinter} herein). In solutions for the external spacetime geometry particle spin might here also be locally associated with non-zero torsion components (this is an open question as discussed in \cite{QGrav}).

   As noted immediately after equation~\ref{GRQT}
    the geometric contracted Bianchi identity  implies  the conservation of energy-momentum with 
     $\partial_{\mu}T^{\mu\nu} = 0$ in the flat spacetime limit (see also \cite{QGrav} equation~45), which also implies the conservation of angular momentum (see for example \cite{WeinGR} sections~2.9, 5.3 and 7.6).  These conservation properties 
       apply   in particular through all  interactions as accommodated in the central part of  equation~\ref{GRQT}.
      All such interactions are strictly local, in deriving from the degeneracy of matter field states underlying a common external local 4-dimensional spacetime geometry such as depicted in the central part of figure~\ref{sinter}.
      
      \pagebreak
     
      Collectively the local conservation of 4-momentum, with $\sum_i P^{\mu}_i=\sum_f P^{\mu}_f$ when summing over initial $i$ or final $f$ state particles, as well as the independent conservation of the wave 4-vector components $\sum_i k^{\mu}_i = \sum_f k^{\mu}_f$ through exchanges as described for equation~\ref{slide}, together with the kinematic and wave properties of particles described by real wave-packet propagation,  imply a universal relation for all such particle states propagating between interactions of the 4-vector form (\cite{QGrav} subsection~6.1, equation~85):
\begin{equation}
         P^{\mu} = \hbar k^{\mu}
  \label{phbarku}		  
\end{equation}	      
     This fixed relation here  introduces  and \textit{defines} Planck's constant $\hbar$ as the universal normalisation 
        between the 4-momentum $P^{\mu}$ and wave 4-vector $k^{\mu}$ for an elementary propagating particle state, where $k^{\mu}$ corresponds to the central peak of the Fourier transformed  Gaussian wave-packet. Equation~\ref{phbarku} incorporates the de Broglie relations $E=\hbar \omega$, relating the particle energy $E=P^0$ to the angular frequency $\omega = k^0$, and  
          $\bPP = \hbar \bk$, for the spatial 3-vector components,  as also apply in the non-relativistic limit
          (\cite{QGrav} equations~1 and 2).
          
     For a propagating particle described by a Gaussian wave-packet the basic Fourier transform properties imply that the spread $\Delta k^a$ for any $a=1,2,3$ component of the wave 3-vector $\bk$   
      and the spread
      in the corresponding spatial location component $\Delta x^a$   are related as
       $\Delta k^a \Delta x^a  \simeq 1$ (no sum over $a$) as an intrinsic `uncertainty relation' for
        a Gaussian wave-packet.
         The spatial components of equation~\ref{phbarku} then imply (for each $a=1,2,3$):
 \begin{equation}
     \label{HUP}
           \Delta P^a \Delta x^a  \simeq \hbar   
 \end{equation}      
       This can be considered the \textit{origin} of the `Heisenberg uncertainty relations', which in quantum mechanics are interpreted as expressing the minimal statistical spread in expectation values for a combined measurement of the momentum and position of a particle (see also for example \cite{Barl} section~2). That is, equation~\ref{HUP} sets an intrinsic limit on how sharply these quantities can be simultaneously defined. 
       Similarly, the spread in angular frequency $\Delta \omega$ and that in location in  time 
        $\Delta t$  is related as $\Delta w \Delta t  \simeq 1$ for a Gaussian wave-packet, implying via the first component of equation~\ref{phbarku} the uncertainty relation $\Delta E \Delta t  \simeq \hbar$. These are lower bounds in the sense that the Fourier transform properties for non-Gaussian wave-packet shapes imply values on the right-hand side greater than Planck's constant.

        The uncertainty relations are then primarily a property of the wave-like propagation structure, which via the universal constraint of equation~\ref{phbarku} are transferred onto a property of particle-like states as seen in experiments. While agreement can be sought with the predictions of quantum mechanics, the aim is ultimately to account for these laboratory phenomena directly rather than necessarily to fully reproduce the formalism of quantum mechanics, which can be seen as a means of modelling particle phenomena in a suitable limit.  After all, unresolved interpretational issues remain for both the 
         Heisenberg uncertainty relations and the general formalism of quantum mechanics itself~\cite{Schl,CarrS}.

     There are a wide range of existing schemes for quantisation, including the   historical approaches of matrix mechanics and wave mechanics as well as more modern techniques in non-relativistic  quantum mechanics (see for example~\cite{AliEn}) and 
     the half-a-dozen or more schemes such as canonical quantisation and the path integral approach
      for quantum field theory in the relativistic case (see for example~\cite{Kaku} section~3.1).
      This need for a variety of schemes to fit different situations, with no one in particular preferred overall, itself suggests that an underlying conceptual basis for the  origin of quantum phenomena is still missing.
      
      The construction here, centring on equations~\ref{GRQT}, \ref{sconst}, \ref{slide}, \ref{pddddm} and \ref{phbarku}, essentially amounts to a new scheme of `quantisation' in itself. 
        In contrast with the above existing schemes
         here there is no \textit{imposition} of any quantisation
       \textit{postulates}, but rather the underlying source of these phenomena is intrinsically identified in the theory.
         A formalism of QFT is \textit{derived} as a consequence of constructing extended
            4-dimensional spacetime itself, in the special relativistic limit, from the symmetry breaking projection out of local elements of generalised proper time, with non-relativistic quantum mechanics seen as a further limiting approximation.
       
       Existing quantum theories can then be interpreted as 
         mathematical models  for calculating and relating observables. 
        Such a perspective dates back to early quantum theory and the matrix mechanics of Heisenberg in the 1920s (see for example his lecture in~\cite{Bethe}), and also led to the original proposal of $S$-matrix theory  by Heisenberg in 1943  in  dealing only with directly measurable quantities (\cite{Venez}, \cite{Kragh} chapter~6).
         Rather than merely linking observables
      an underlying \textit{explanation} of quantum phenomena is here provided consistently in a
       full mathematical theory based upon generalised proper time, with this new framework correspondingly providing a realistic account of `what is actually physically happening' in quantum and particle interaction processes, such as pictured in figure~\ref{sinter}.

      Reading equation~\ref{pddddm} from right-to-left this approach then takes apart the mathematical structure of QFT cross-section calculations and provides an underlying reconceptualisation of HEP processes. This is consistent with QFT, as employed for the Standard Model, considered as 
      an \textit{effective} theory, and for example provides a means of accounting for the infinities 
      that arise in QFT calculations 
    and the need for renormalisation involving an element of calibration against the data, which again themselves provide a strong hint of the need for such a new conceptual basis. 
      Infinities in QFT calculations are generated by loops in Feynman diagrams that can be interpreted in terms of the exchange of intermediate `virtual particle' states each with unlimited \mbox{4-momentum}. This is permitted in loops since unlike real particles the virtual states
       are `off-mass-shell' and
      while 4-momentum is conserved at each Feynman vertex there is an ambiguity in how it
       is shared between such states.

       Even for QFT itself these virtual states can be considered an illusory figment of the mathematical structure     (\cite{Unifi} discussion of equation 11.43).
       With this mathematics of the effective QFT seen in turn as derived in the limit of a more fundamental theory the notion of `virtual' particle states becomes even yet less `real'. The physics in reality derives from the relative degeneracy count of matter fields underlying the local external geometry as for figure~\ref{sinter}, rather than the  apparent exchange of virtual particle states.
      The theoretical difficulties concerning the QFT calculation of vacuum energy, as associated with the fleeting production of pairs of virtual particle states, further  suggests that QFT is not providing the full picture (as will be discussed further in the following subsection in relation to equation~\ref{HUPE}).

      While the underlying field degeneracy is central to calculating \textit{interaction} probabilities, as pictured in the centre of figure~\ref{sinter} and reviewed for equation~\ref{pddddm}, degeneracy of the kind described for equation~\ref{slide}  applies \textit{everywhere} for the construction of the extended
       4-dimensional spacetime through equation~\ref{GRQT}. This will then in particular also apply for the apparently \textit{freely propagating} real particle wave-packet states, such as for the initial and final states pictured to the left and right in figure~\ref{sinter}, since the corresponding field cannot be fully dissociated from  mutual interactions with other matter fields. 
          While enveloped by the continuous gravitational field it is the innate degeneracy and associated `self-interactions' of the underlying matter field components that are proposed to \textit{hold} such propagating wave-packet states coherently together under equation~\ref{GRQT}, in a stable manner consistent the property of equation~\ref{phbarku}.

        That is, the spreading of the wave-packet due to the mismatch between the group velocity and phase velocity of the Fourier components according to a dispersion relation 
        $\omega = \omega(\bk)$, as  might otherwise be anticipated for a single matter field component in isolation, could then be avoided (\cite{QGrav} subsection~6.2).   
         In quantum mechanics the dispersion of a propagating wave-packet can be interpreted as a spreading of the probability distribution for locating the particle in a measurement.
      Here, the degenerate self-interactions will again have a counterpart in the `self-energy' Feynman diagrams for initial and final states in QFT and the associated apparent cloud of `virtual' states accompanying
       such `renormalised' real propagating particles (see also~\cite{Unifi} section~11.3, figures~11.9 and 11.12(b)).

       That the new framework might plausibly reproduce the successes of QFT in the flat spacetime limit can be surmised on noting that the theory entails general properties such locality, Lorentz invariance and unitarity that are essentially equivalent to the general constraints employed by the $S$-matrix bootstrap as reviewed in section~\ref{boot2}. 
       The consistency requirements of unitarity and analyticity are here implied through the connection with complex transition amplitude structures in equations~\ref{pddddm} and \ref{uprob}.
       The significant \textit{additional} feature of the new theory in comparison with the $S$-matrix bootstrap as employed for strong interactions and depicted in figure~\ref{bubble}, 
        as noted after equation~\ref{uprob} with reference to figure~\ref{sinter},  
       is the ability to also   account for the detailed structure of the Standard Model elementary particle content, Lagrangian and interaction terms.
     With the full exceptional Lie group symmetries  of generalised proper time broken over the local external 4-dimensional spacetime substructure the Standard Model particle content may be reproduced as reviewed in the previous subsection.  \mbox{As described for equations~\ref{extgpt}--\ref{asmlag}} the resulting implied field redescription  constraints $\Rcon$ then play the role of interaction Hamiltonian
   $\Hint$   terms  in the corresponding QFT Standard Model Lagrangian, incorporating also the relevant coupling constants and mass parameters.
      
     The $S$-matrix bootstrap general constraints alone, as applied for the bubble diagrams of figure~\ref{bubble}, tightly constrain the nature of strong interactions and cross-section calculations for particle states. Given then  the additional input of the explicit physics to fully replace the Standard Model Lagrangian, as incorporated into processes such as depicted in figure~\ref{sinter}, there is a  reasonable plausibility of fully reproducing the successes of QFT as applied in the particle physics laboratory environment and for the Standard Model in particular. Essentially largely the  \textit{same information} at the simplest and  most elementary level of matter is input from both the left and right-hand side of equation~\ref{pddddm}, and hence the respective calculations may well converge, albeit with the possibility of `new physics' deriving from the fundamental basis in generalised proper time on the left-hand side as reviewed in the previous subsection.

     Any low-energy particle theory consistent with the basic principles of special relativity and quantum mechanics will inevitably look like a quantum field theory,
     with QFT in this sense already considered a low-energy approximation to a more fundamental theory
          (\cite{WeinQ} volume~I, chapter~1 opening). 
      Considered as a flat spacetime limiting case from the new theory, the calculations of QFT as an effective theory have nevertheless proved extremely successful in themselves. This success may then be largely down to the QFT formalism itself providing a consistent amalgamation of highly constraining properties,  incorporating a suitable Lagrangian, as motivated by empirical observations, \textit{as well as} the $S$-matrix bootstrap constraints which are essentially the broad principles of special relativity and quantum mechanics. As an indication of the significance of the $S$-matrix approach, we also noted
      in section~\ref{boot2} with reference to~\cite{White,BernDK,Alme} that explicit bootstrap techniques are still employed today to assist QCD calculations.

       A characteristic feature of QFT calculations with a complex transition amplitude, $\mcM_{fi}$ on the right-hand side of equation~\ref{pddddm}, is the \textit{interference} that can occur between the contributions of various intermediate processes. For given initial particle states $i$, as new types of possible intermediate states or interactions are added, this can result in a \textit{decrease} in the likelihood for specific final state outcomes $f$ or in the total cross-section itself. From the perspective of a degeneracy count probability,  $P_{fi}\propto D_+D_-$ coming from the left-hand side of equation~\ref{pddddm}, it might seem counter-intuitive that \textit{any} additional possible field exchanges could cause any cross-section to decrease. However, regardless of any such additions the total probability $\sum_f P_{fi} = 1$ will \textit{necessarily} be preserved, and hence while \textit{some} outcomes (such as possibly the case of `forward scattering', that is $i=f$) will increase in relative probability,  other eventualities (including sometimes the total cross-section for a non-trivial outcome) will inevitably indeed decrease (see also \cite{Unifi} pages 350--351 discussion).
       
        For this theory, once the full form of generalised proper time and the symmetry breaking structure is determined all possible intermediate matter field states, deriving from equations~\ref{sbreak} and \ref{gbreak}, and all possible $\Rcon$ field interactions, associated with equations~\ref{sconst} and \ref{slide}, will be given once and for all. No further ad  hoc additions would be possible, unlike the case for model building based on a QFT Lagrangian.  On following the connections through equation~\ref{pddddm} from left-to-right all the implications of the complex transition amplitude calculation and equation~\ref{uprob} in modelling the conservation of probability should be accounted for, including a role for apparent  `interference' effects when considering the relation between the various contributions to the degeneracy count.

      A similar situation will carry over for the non-relativistic limit of quantum mechanics. In the standard formalism the evolution of an individual particle state is modelled by a wavefunction $\Psi(\bx,t)$, as a function of 3-dimensional space coordinates $\bx$ and an absolute time parameter $t$,  evolving according to the Schr\"{o}dinger equation as punctuated by `collapse' events corresponding to `measurements'. This construction is central to the quantum mechanical description of the archetypal double-slit experiment 
       (see for example \cite{QGrav} figure~1 and discussion).  
     In the present theory, for the same experimental setup,  a single particle  state will be divided into two  propagating wave-packet contributions, one passing through each slit  (\cite{QGrav} figure~5). 
     With particle states emitted and detected as `whole' wave-packets consistent with equation~\ref{phbarku},
      the recombining of the two  contributions fully in phase will correspond to maxima in the likelihood distribution on the detection screen, while any fully out of phase coincidence will correspond to probability minima. It then remains to account for the complete probability distribution as calculated to follow a $\vert  \Psi(\bx,t) \vert^2$ interference pattern in quantum mechanics, similarly as it remains to fully account for QFT calculations on the right-hand side of equation~\ref{pddddm}.

       In all laboratory experiments the geometry of a single extended continuous 4-dimensional  spacetime solution for equation~\ref{GRQT} is to be determined. Within this enveloping geometry of general relativity there is a continuous interface between microscopic particle or `quantum' systems and the macroscopic apparatus or `classical' systems, without any discontinuity or `collapse' events (\cite{QGrav} subsections~7.3 and 7.4). 
        The 4-dimensional spacetime boundary conditions for obtaining solutions for equation~\ref{GRQT} include the geometry of the macroscopic apparatus and the potential for particle interactions as associated with particle sources and detectors, such as the source and detection screen in the double-slit experiment.

      These boundary conditions have implications for the possible measurements that can be made, just as the Copenhagen school of Niels Bohr considered that the choice of apparatus setup determined which features of a system could be observed and in particular which of a pair of quantities could be measured, in the original interpretation of quantum mechanics, with this intrinsic `complementarity' exemplified by wave-particle duality.
      For example for the present theory, that the electron state has electric charge means it has the \textit{potential} to undergo electromagnetic interactions. If a detector is placed at one of the slits 
      in the double-slit experiment
       that would imply a different set of boundary conditions and a  \textit{different} set of possible solutions for equation~\ref{GRQT}, such as with   the  position of an electron recorded at the slit, and the interference pattern on the final   screen  would not appear.

         Any `spooky action at a distance', that is any seeming non-locality or even `retrocausality', is then sown into the full 4-dimensional spacetime geometric solution, including the case for `EPR'-type or even `delayed choice' experiments. 
         In classical general relativity itself large-scale solutions are full extended 4-dimensional spacetime geometries as might be determined with the aid of spacelike boundary conditions imposed in the past \textit{or} in the future of the spacetime region. Here with quantum particle systems treated as laboratory-scale solutions for general relativity incorporating the most elementary and simplest level of matter, past and future boundary conditions are again equally valid, albeit now with an intrinsic element of indeterminacy entering through the underlying degeneracy in equation~\ref{GRQT} in identifying full 4-dimensional spacetime solutions.

         It is then a question of calculating the relative likelihood for a consistent web of particle interactions, linking for example source and detector events, for such a solution. The overall event probability in an EPR-type experiment will then depend upon the consistency of the preparation of an apparent `initial'   entangled state of for example a pair of electrons together with an apparent `later' detection of their respective spin states, treated as a full 4-dimensional spacetime solution for equation~\ref{GRQT}.
         Action might appear `spooky' from \textit{our} perspective since we have to \textit{wait} until the final measurement and then may neglect its retrocausal influence on this solution. If we were to try to make an earlier observation \textit{that} new measurement would again change the boundary conditions and the nature of the 4-dimensional spacetime solution, similarly as described above for the double-slit experiment.   
         
         This account of events is consistent with quantum mechanics in the sense that the choice of measurement setup restricts the range of possible outcomes, and also in the sense that the retrocausal influence and intrinsic local indeterminacy imply  the possibility of violating Bell's inequalities for the correlations observed in EPR-type experiment, quite unlike the case for hidden-variable theories.   
         The full 4-dimensional web of interactions under the umbrella of general relativity is also consistent with causality in that no signal can propagate faster than light. That all empirical phenomena, including quantum   processes, have been found to satisfy this constraint  is itself  evidence for how the principles of  relativity have surveillance over even the microscopic  quantum world.
       This observation contributes to the motivation for adopting the perspective of prioritising general relativity over quantum mechanics as suggested in the opening of this subsection.

      With the physical structure of the gravitational field, key to understanding these microscopic processes, not playing any role in standard quantum mechanics, such phenomena may seem `spooky' when modelled by a wavefunction and its apparent sudden collapse.
      To continue the analogy discussed shortly after equation~\ref{GRQT}, on neglecting any reference to the geological bed and banks a river \textit{could} be described in terms of the water alone, requiring the invention of a strange guiding force to account for the bends and meanders and general evolution of the water.
      In non-relativistic particle experiments, on neglecting the curved geometry of the 4-dimensional spacetime the invention of a wavefunction  $\Psi(\bx,t)$, representing our best knowledge of the state, is employed to guide the likelihood calculations, with seemingly strange and counter-intuitive implications. Since the gravitational field is far too small to detect directly in such laboratory experiments, this has historically inevitably been the way quantum theory developed.

      To fully understand quantum phenomena the full 4-dimensional spacetime `block' perspective of solutions for equation~\ref{GRQT} is required. However, we as `creatures of time', in perceiving the universe as revealed to us through a `dynamical' temporal flow, construct wavefunctions to help navigate through a course of observations into the future.
      As noted earlier, the apparent need for a range of quantisation methods and the ongoing debate over a range of interpretations concerning the foundations and applications  is an indication of the incompleteness of quantum theory. 
      In particular, the key missing element is gravity.

      In the new theory gravity is not only amalgamated with quantum phenomena but also classical general relativity \textit{takes the upper hand}, through equation~\ref{GRQT}, while QFT and quantum mechanics emerge in the special relativistic and non-relativistic limits respectively (as proposed in the heading of this subsection).
         We can also note that all of the technical difficulties associated with attempts to `quantise gravity', that follow from assuming quantum theory to have the upper hand, are naturally avoided.
          This is then in stark contrast with most approaches to `quantum gravity' that seek to apply an adapted set of quantum theory postulates directly to the gravitational field or spacetime geometry itself 
          (many of which have a long history~\cite{RoveQG}).
          
           Perhaps the most significant reason for not wishing to meddle  with the foundations of quantum theory itself is the enormous body of computational success that has been achieved with the  QFT and quantum mechanics of particle states based directly on calculations with complex amplitudes in accounting for all the corresponding \mbox{non-gravitational} elementary empirical observations. However, as we have also discussed in this section, a large part of this success may derive from the fact that such computations, which apply on a limited scale, are heavily constrained by requirements of  consistency.

      Here with gravitation taking priority, in the context of a theory based upon generalised proper time, not only are general relativity and quantum theory amalgamated but there is also a unifying notion of probability. Similarly as classical probabilities are associated with the relative `number of ways' things can happen \textit{in} spacetime, `quantum' probabilities are now associated with the relative `number of ways' that 4-dimensional spacetime can \textit{itself} be constructed, with a vast number of possible extended solutions for equation~\ref{GRQT} given the local matter field degeneracy underlying the external geometry.

      A main goal of this subsection has been to argue how, while identifying the \textit{origin} of quantum phenomena, this reconceptualisation might fully reproduce all the quantitative calculational successes of QFT and quantum mechanics. 
        This follows since the theory incorporates  the self-consistency constraints employed in the $S$-matrix bootstrap approach \textit{together} with an explanation for the source of explicit  Lagrangian terms for the Standard Model of particle physics underlying these calculations.

       This approach to `gravitising quantum theory' leads through the QFT calculation limit to an account of typical quantum mechanical  processes such as observed in the double-slit experiment.
        The `measurement problem'
     is   consistently addressed by giving an explicit account of `what is actually physically happening' in these quantum phenomena experiments without reference to a wavefunction and without the need for an `interpretation'. 
       
      The theory then simultaneously addresses the three outstanding issues of amalgamating general relativity with quantum theory, providing a firm foundation for QFT calculations, and accounting for the long-standing conceptual difficulties with quantum mechanics itself. In giving a detailed account of laboratory phenomena these aspects of the theory may also  prove testable, for example in the area concerning the apparent interface of `quantum' and `classical' systems. Some of the features of the conceptual framework described above will also be relevant on the cosmological scale, as will be discussed in the following subsection.


\subsection{The Dark Sector and Cosmological Evolution}
\label{boot33}

   In constructing the continuous extended 4-dimensional spacetime of general relativity as a solution for
   equation~\ref{GRQT} the considerable ambiguity inherent in the local degeneracy of the underlying matter field content and in piecing these local elements together provides the source of characteristically indeterminate quantum phenomena, as described in the previous subsection.
      This is in contrast with the local form for generalised proper time itself in equation~\ref{salpha} which provides very little room for ambiguity. However, while in subsection~\ref{boot31} we reviewed the unique sequence of mathematical forms for proper time, with a $\hG = \esi$, $\ese$ and potentially $\ee$ symmetry as  leading to structures of the Standard Model of particle physics,  this does \textit{not} express \textbf{\textit{the}} unique possibility for generalising local proper time intervals beyond the 4-dimensional spacetime form of equation~\ref{sxxxx} as consistent with equation~\ref{salpha}. 

   Indeed, in place of the above progression in forms, as described immediately after figure~\ref{sgener},  the standard form for proper time with $\sltc$ symmetry preserving the determinant of 
   $2 \times 2$ complex Hermitian matrices, describing the local Lorentz symmetry of 4-dimensional  spacetime, could instead be augmented to an $\slpc$ symmetry preserving the determinant of
     $p \times p$ complex Hermitian matrices for any integer $p>2$,  as also consistent with the
     $p^{\mathrm{th}}$-order
      homogeneous polynomial form of equation~\ref{salpha}. 
     On extracting the original local 4-dimensional spacetime base, as described for equations~\ref{sbreak} and \ref{gbreak} this now directly generates the symmetry breaking structure listed in table~\ref{ppdet} (as described in detail for  \cite{Basis}  table~2 and \cite{DEner} table~4).

\def\rai{+0.2ex}
\begin{table}[htbp]
\centering
\begin{tabular}{|l|ccccc|c|}
 \hline
  \raisebox{-0.5ex}{${p^2\;}$} \!\!\!\!\!\!\!
   {\mbox{\raisebox{+0.0ex}{\LARGE{$\diagdown$}}}} \!\!\!\!\!\!
      \mbox{\raisebox{+0.7ex}{$\slpc \!\, \supset\!\!\,$}} 
	    & \raisebox{\rai}{Lorentz} &
  \raisebox{\rai}{$\times\!\!\!\!\!\!$}
	    & \raisebox{\rai}{$\slptc$}   & 
	 \raisebox{\rai}{$\!\!\!\!\!\!\!\!\times\,\,$}
	    & \raisebox{\rai}{$\!\!\!\uod$}   &        
	    \raisebox{\rai}{matter} \\
 \hline
     $\;\, 4 \qquad$ & 4-vector & & $\bdx_4$  invariant & & $\!\! 0$ &    (external) 				
	\\
  $\, k^2\;\, (k=p-2)$ & scalar  & &
   $\bdx_{k^2} \to  S_k\:\! \bdx_{k^2}\:\! S^{\dagger}_k$ 
     & & $\!\! 0$ &   dark  $\bphiD$
	\\
  $4k$ & Weyl  & & 
       $\bdx_{4k} \to S_k\:\! \bdx_{4k}$
     & & $\!\! 1$ &   dark    $\bpsiD$	
	  \\
   \hline
  \end{tabular}
  \caption{\setb   
    Given the full $\hG = \slpc$ symmetry acting on the $n=p^2$ components of  
   the generalised form for proper time $(\delta s)^p = \det(\bxn)$,  with
      $\bxn \in \hpc$, the above symmetry breaking structure results from 
     the projection of the local external  4-dimensional spacetime subcomponents $\bdx_4$ as described for equations~\protect\ref{sbreak} and \protect\ref{gbreak}.
   The basis for
  matter states obtained includes $\bphiD$:  a set of $k^2$  (with $k=p-2$)
   scalars under the external Lorentz symmetry  transforming as the matrix $\bdx_{k^2} \in \hkc$ under  internal 
    $S_k \in\slptc$ actions and neutral under $\uod$,  together with  $\bpsiD$:  a set of \mbox{$k$ 2-component} complex  Weyl spinors transforming under the standard representation of $\slptc$ acting on
        the $k\times 2$ complex matrix $\bdx_{4k}$  and  charged under the internal $\uod$.
       These structures provide the basis for a substantial dark sector, as denoted by the 
        `$D$' subscripts and explained in the text.
        }
\label{ppdet}
\end{table}   

   With the internal gauge symmetry $G = \slptc \times \uod$ being \textit{independent} of that identified for the exceptional Lie group Standard Model branch of generalised proper time described in subsection~\ref{boot31} the matter fields $\bphiD(x)$  and $\bpsiD(x)$ deriving from table~\ref{ppdet} are necessarily `dark'. The only interaction between this dark sector and the visible sector of matter is through the gravitation of classical general relativity as associated with the curvature of the common external 4-dimensional spacetime base, with both sectors contributing through equation~\ref{GRQT}.

    A significant new feature for the branch of generalised proper time in table~\ref{ppdet} 
   is the \textit{non-compact} nature of the 
    internal gauge group $\slptc$   for \mbox{$k=p-2\ge 2$}. 
    This implies there will be gauge bosons associated with the positive energy and positive pressure contribution of  a gauge field component $\Apls(x)$ \textit{as well as} gauge bosons associated with the \textit{negative} energy and \textit{negative} pressure contribution of a 
     gauge field component $\Amin(x)$, as can hence be produced together \textit{out of the vacuum} via the 
    self-interactions of this non-Abelian gauge symmetry.
     While the vacuum is filled with a raging sea of mutually created and annihilating $\Apls$ and $\Amin$ gauge bosons the equilibrium state can be gravitationally completely benign on the macroscopic scale owing to the symmetry between the positive energy and pressure terms and their negative counterparts (\cite{Basis} figure~2 and equations~17 and 18).

     However, a contribution from the positive energy and pressure of  matter states 
     $\Mplu = \{\bphiD, \bpsiD, \AmaxD \}$  from table~\ref{ppdet}
     (with the gauge bosons $\AmaxD$  associated with the \mbox{internal $\uod$}),
        as also produced in the vacuum through mutual creation and annihilation interactions
     and chains of interactions 
       involving the $\Amin$ states,  will perturb this symmetry.
    With the equation of state now incorporating an asymmetry, as the collective $\{\Apls,\Mplu\}$ states act as the counterpart to the $\Amin$ contribution, it is then possible to obtain an extremely small, but net \textit{positive} energy density for the vacuum $\rvac > 0$, with $\vert  \rAMp \vert > \vert \rAm \vert$, together with an extremely small, but net \textit{negative} pressure $\pvac < 0$, with  effective contributions
   $\vert  \pAMp \vert < \vert \pAm \vert$, consistent with the net equation of state (\cite{Basis} figure~3 and equations~19 and 20):
\begin{equation}
 \label{weqos}
    \pvac \, = \,  \wvac \rvac  \quad \mbox{with} \quad \wvac = -1
\end{equation}  
      Such a value for the equation of state parameter $\wvac$ corresponds to the Einstein equation~\ref{Eineq} for the vacuum case in general relativity with an effective cosmological constant $\Lambda$ term  appended to the left-hand side:
 \begin{equation}
   \label{Einvac}
       G^{\mu\nu} + \Lambda g^{\mu\nu} = 0
 \end{equation}

   This equation can be interpreted in the context of equation~\ref{GRQT} with
    the $\Lambda g^{\mu\nu}(x)$ term transferred to the right-hand side as a specific form for $T^{\mu\nu}(x)$ as
  the macroscopic effect of the interplay of the underlying  $\{\Apls,\Mplu,\Amin\}$ field contributions to the central $f^{\mu\nu}(x)$ part of equation~\ref{GRQT}.  
   Such an energy-momentum term with  effective uniform positive energy density $\rvac$
    and negative pressure $\pvac$ such that $\wvac=-1$ corresponds to a particular case for `dark energy', which more generally requires an equation of state parameter $w<-\frac{1}{3}$ to generate a large-scale
     cosmic acceleration.
     
     Obtaining a
          tiny residual dark energy effect, due to the
     \textit{almost} complete cancellation in the gravitational impact of the  positive and negative energy
density and pressure contributions, is analogous to the mechanism responsible for the 
     Earth's magnetic field, as a small residual effect deriving from the \textit{almost} complete cancellation
      in the electromagnetic impact  of the positive and negative charge constituents of the planet. 
 In both cases the residual phenomenon results
from an asymmetry in the properties of the plus and minus components. While
the density of $\pm$ energy states in the dark vacuum may be much higher than
that of the $\pm$ charge states in the Earth, the fact that gravity is $\sim \!\! 10^{40}$ times
weaker than electromagnetism will be a factor in the dark energy residual effect
only being observable as an accelerating expansion on the cosmological scale.
   
    A key factor in obtaining such small values for $\rvac>0$ and $\pvac<0$ in 
    equation~\ref{weqos}, and hence
     also for $\Lambda>0$ in equation~\ref{Einvac}, is that the production of the $\Mplu$ states may be relatively very much suppressed (due to kinematic or other reasons as listed in 
  \cite{Basis} shortly after  equation 21 therein), and hence only provide a very minor perturbation from an otherwise
   symmetric $\{\Apls,\Amin\}$ vacuum state.  
    It is the $\Amin$ states of the non-compact gauge sector that provide the crucial input of
   negative pressure as an essential  property for dark energy, whether in the effective form of a cosmological constant $\Lambda>0$ or a more general case with some deviation from
    $w=-1$ permitted.
   With all $\{\Apls,\Mplu,\Amin\}$ contributions constantly being replenished out of the vacuum, even as the universe expands, the vacuum equilibrium state is essentially uniform in space and time as also required for dark energy.
   
   Unlike the various models that \textit{postulate} new fields or interactions contrived to account for the present era cosmic acceleration, the candidate for dark energy presented here has the significant advantage of a basis in a \textit{fundamental} theory, as founded upon generalised proper time. 
   The characteristic features of darkness, ultra-low net positive energy density and associated negative pressure, and spacetime uniformity all \textit{follow directly} from this foundation as summarised above.
 We also note that while this is \textit{not} a scalar field theory, unlike the typical case for dark energy models as in part motivated by simplicity, scalar objects quadratic in the non-Abelian gauge field strength tensor, essentially based on
    the $FF$  terms of geometric origin discussed for equation~\ref{sconst} in the previous subsection, are central to this construction (see also \cite{DEner} section~5).

   With the only particle-like interactions being internally restricted to the fields deriving from table~\ref{ppdet} this dark energy vacuum state is also \textit{stable}, in that there is no mutual creation of negative energy $\Amin$ states coupled to the familiar positive energy states of the Standard Model -- associated with the alternative branch of generalised proper time as summarised in subsection~\ref{boot31}. Such a coupling would allow the vacuum to decay, from our visible sector perspective, as a potentially catastrophic instability.  
  Due to this apparent danger, while the branch of generalised proper time of table~\ref{ppdet} was already considered for
    (\cite{TimeE} equations~86 and 87) it was then suspected of implying a possible inconsistency, and  cautiously avoided as unphysical similarly as for negative energy states in most other theories. 
    
      On the issue of stability
    it is important to note then that, as described in the previous subsection, in this framework gravity is \textit{not} quantised and in particular there are \textit{no} `graviton' states. Owing to their universal nature, such hypothetical gravitons would allow particle-like exchanges between the particle states of the dark and visible sectors, as would hence generate an instability given any dark energy candidate with a negative energy component. This can be problematic for example for `phantom' dark energy models (\cite{DEner} figure~2(a)  discussion), but is not an issue here due to the classical general relativity nature of the gravitational interaction associated with equation~\ref{GRQT}.
    Hence a non-compact non-Abelian gauge sector \textit{can} have consistent physical consequences in the context of this theory.

    There is an apparent \textit{symmetry} between the  Standard Model and dark energy sectors in that they are generated through parallel branches of generalised proper time and  with the energy density of both dominated by non-Abelian gauge components, namely colour $\suth_c$ and dark $\slptc$  respectively. However, the above discussion around equations~\ref{weqos} and \ref{Einvac} would seem to imply an apparent \textit{asymmetry} in the very early universe for positive cosmic time $t_c \to 0$, with the visible Standard Model sector characterised by extremely high energy density
     $\rSM \to \infty \vert_{t_c \to 0}$ while the corresponding vacuum energy of the dark sector
     $\rDS \equiv \rvac \vert_{t_c \to 0}$,
      from the temporal uniformity of equation~\ref{weqos},  remains ultra-low.  A more symmetric picture might suggest a more even balance in the partition of the initial energy density with
       $\rSM \simeq \rDS \vert_{t_c \to 0}$, to within an order of magnitude or so, at that earliest epoch. 
 
   If, similarly as for the visible sector, it were also the case that  $\rDS \to \infty\vert_{t_c\to 0}$, or approached an arbitrarily high value in the extremely early universe, while still  temporarily maintaining consistency with the vacuum case of equations~\ref{weqos} and \ref{Einvac}, that would tend to describe a very early \textit{inflationary} epoch.
    For inflationary models  the initial rapid exponential expansion is typically completed by of
     order a mere $t_c \simeq  10^{-33}$ seconds
  and followed by a phase transition, through which the high potential energy density of the inflaton field is converted into radiation and the creation of Standard Model states  in a thermalised `reheated' stage
   initiating the hot Big Bang  (see for example \cite{Tsuj}).

   Given the lack of interaction between the dark and visible sectors in the present theory an analogous phase transition might instead convert such an extremely high initial vacuum energy density $\rDS$
   into that of a broader 
    dark sector, potentially including a dark matter component with positive energy density $\rDM$. 
     With a non-vacuum equation of state, and
    also a non-negative pressure, such a dark matter component might then form such that $\rSM(t_c) \simeq \rDM(t_c)$ are of the same order and  evolve in parallel after the early inflationary era as a function of cosmic time.
   
   Given that the  $\Mplu$ states break the symmetry in energy density and pressure of the
     $\Apls$ the $\Amin$ components, a primordial universe with both a high density of states and relatively large production and contribution of the  $\Mplu$ states might provide suitable conditions to drive such an
      inflationary epoch consistent with equations~\ref{weqos} and \ref{Einvac}.
       A `phase transition', marking the end of inflation, might then correspond to a major suppression of 
   $\Mplu$ state production and annihilation rates.
   Such a phase transition could involve an initial local 4-dimensional spacetime projection out of equation~\ref{lpvn} for cosmic time $t_c \to 0$ with $\vert \bv_4(x) \vert \to 0$ and 4-vector  $\bv_4(x) \simeq 0$,  associated with the $\bxf$ components in equation~\ref{sbreak}, that rapidly stabilises to a finite vacuum
    value  $\vert \bv_4(x) \vert = h_0$ as alluded to after equations~\ref{sconst}--\ref{asmlag}, similarly as described for the  visible matter sector in (\cite{Unifi} section~13.2, see figures~13.3 and 13.5 therein), as here applied in parallel for the dark sector of table~\ref{ppdet}. 
               Such a transition would change the physical properties of the $\Mplu$ states, including their
                   interaction rates,
                and might then leave
    a large part of the existing positive energy excess 
    in the  $\{\Apls,\Mplu\}$ contribution \textit{frozen out} to form the dark matter sector.

      The positive energy contribution from the $\Apls$ states is associated with the \textit{compact} subgroup $\sukd \subset \slptc$ and hence such a non-vacuum state could generate a `hidden QCD' candidate for dark matter, dominated by the $\{\Apls,\Mplu\}$ states deriving from table~\ref{ppdet} (\cite{Basis} section~4, \cite{DEner} section~7). Whether as dark nuggets and glueballs or of a more diffusive nature, this describes a  candidate for cold dark matter (CDM)  that is most probably not  highly self-interacting, in the sense of forming a numerically relatively low density of massive states, as required  given empirical observations of galactic properties. 
      
      As regions of the dark matter eventually gravitationally collapse a diffusive relation might be established between this non-vacuum dark matter and a residual dark energy vacuum component.
       This latter component 
       consisting of a background 
      equilibrium of  $\{\Apls,\Mplu,\Amin\}$ states, now with only a very minor $\Mplu$ contribution similarly as originally
       described for equations~\ref{weqos} and \ref{Einvac} with   ultra-low uniform and constant energy density
        $\rDE = \rvac>0$,  would everywhere fill the vast voids between the web-like structures of dark matter and could account for the present day net cosmic acceleration. This overall scheme might then be compatible with the standard $\Lambda$CDM cosmological model, potentially together with an early inflationary era, albeit with some room for deviations that might assist in accounting for the various tensions between 
        $\Lambda$CDM  and cosmological observations~\cite{Abda}.
        
         The dark sector in this framework may also have further structure due to the existence of yet further mathematically permitted branches of generalised proper time, in addition to that of 
         table~\ref{ppdet} (with the case of a higher-dimensional \textit{quadratic} form for proper time also discussed in \cite{Basis,DEner,Dsect}). 
              The parallel generation of the Standard Model of particle physics sector and 
              such a multi-component
               dark sector, together with the diffusive relation between dark matter and dark energy          
               within specific dark sector branches, might also help address the various cosmic coincidences
               (see for example~\cite{Velt}) 
                concerning the present day average cosmic densities of visible matter, dark matter and dark energy (with their respective net contributions of 5\%, 26\% and 69\% as consistent with an overall critical density of a spatially flat universe,~\cite{PDG22} section~2).
                The presumably high entropy of the
uniform and ubiquitous dark energy component, increasing in proportion to the
volume of the universe, may also be significant with regards to the second law
of thermodynamics and the evolution of the entropy of matter states across all
sectors from the earliest cosmic epoch.

        While hence potentially having significant implications for galactic and cosmological observations, we note that the possibility of any laboratory detection of the above dark matter candidate would seem unlikely. This is due to the apparent lack of \textit{any} particle-like interactions between the visible and dark sectors, with independent gauge forces acting on each and only a classical gravitational interaction linking them via the common external spacetime geometry of equation~\ref{GRQT}.
     For a laboratory detection  a weak particle-like interaction between  Standard Model states and dark matter states would be required in a manner that does not jeopardise the vacuum stability of the dark energy component. There also remains the possibility that new physics could be identified beyond the Standard Model but within the visible branch itself, as considered in the two previous subsections, that could contribute to `dark matter' and still in principle  be detectable in the laboratory  (see also  \cite{Basis} discussion towards end of section~4, \cite{Dsect} section~6). 
     
      The dark energy vacuum described above consists of a seething ocean of real positive and negative energy 
    states  $\{\Apls,\Mplu,\Amin\}$ as mutually created and annihilating, leaving a net ultra-low vacuum dark energy density $\rDE=\rvac$ as described for equations~\ref{weqos} and \ref{Einvac}. 
   With gravity taking priority through equation~\ref{GRQT} energy-momentum is necessarily everywhere conserved, in the sense of compatibility with the contracted Bianchi identity as described after equation~\ref{GRQT}, as an overarching constraint upon all microscopic particle interaction phenomena, and in particular  those underlying the macroscopic vacuum state of equation~\ref{Einvac}.
   
    This is in stark contrast to the standard QFT vacuum calculation, which is associated with the fleeting creation and annihilation of virtual particle and antiparticle pairs, all of positive energy, and which notoriously predicts a vacuum energy density around $10^{120}$ times too large to be appropriate for the present dark energy era.
    The creation of the virtual pairs does \textit{not} locally conserve energy-momentum, but their apparent existence with energy of order $\Delta E$ is seemingly permitted for a brief time $\Delta t$  by quantum mechanics according to the `Heisenberg uncertainty principle' as representing the minimal uncertainty in energy and time via the relation:
 \begin{equation}
    \label{HUPE}
          \Delta E \Delta t \simeq \hbar
 \end{equation}

    While QFT cross-section calculations  have been extremely successful when tested against experiments,
      `virtual states' are  merely  an interpretation of mathematical elements in those calculations as linking energy-momentum conserving interactions between \textit{real} particle states, with questions surrounding the meaning of such virtual particles remaining unresolved as alluded to in the previous subsection.
      For the standard QFT \textit{vacuum} virtual states are apparently the \textit{only thing going on}, and if their `existence' is questioned then the whole nature of the QFT vacuum might be brought into doubt. 
      The spectacularly unsuccessful prediction of the dark energy density from the QFT vacuum calculation itself strongly hints at the possible incompleteness of the theory as also noted in the previous subsection.

       We also note that while the QFT vacuum with virtual states has had some apparent success in accounting for the Casimir effect, as associated with a drop in vacuum energy density
        as some vacuum fluctuation virtual modes of the electromagnetic quantum field are excluded by the boundary conditions between two very close macroscopic plates, which feel a corresponding
         Casimir force,  there are other proposed explanations for such an observed effect
          (such as a van der Waals kind of force~\cite{Jaffe,Niko}).
               Further, the apparent interactions between vacuum fluctuations  and for example orbital electrons in the hydrogen atom, as associated with  the `Lamb shift', or the 
             shielding effects of virtual pair creation, as associated with the phenomenon of energy-dependent    `running coupling', 
             all   concern elements of calculations connected with the properties of
              \textit{real} particle states and their `higher-order corrections' or `self-interactions'.
          Essentially it is not possible to directly probe the vacuum without employing
           a non-vacuum probe, which will be subject to physical effects of its own.

    The dark energy vacuum in the present theory is composed of the
     real states $\{\Apls,\Mplu,\Amin\}$, deriving from the branch of generalised proper time described for table~\ref{ppdet}, and while avoiding the problems associated with the QFT vacuum should have no discernable impact on laboratory observations in only interacting with the visible sector in a classical gravitational manner.  Further, the influence of gravity on the vacuum state itself will be negligible 
      in a terrestrial setting. However, 
    one environment that involves both a strong gravitational field and the vacuum state is the vicinity  of a black hole. In the present theory the continuous mutual creation out of the vacuum of    real $\{\Apls,\Mplu,\Amin\}$ states could play a significant role at the horizon of a black hole, with a corresponding partitioning of positive and negative energy states that might relate to the theoretical phenomenon of black hole evaporation. Hence this may be an area of calculation to further probe the  mathematical consistency of the present theory, while again
     here there is no involvement  of `virtual states' as might be associated with equation~\ref{HUPE} in QFT.

      Indeed the `uncertainty relations' of equations~\ref{HUP} and \ref{HUPE} 
       are here \textit{derived} for 
      real particles, such as composing  the initial and final states in figure~\ref{sinter} and  described by propagating wave-packet functions in the gravitational and matter fields. 
      As described following equation~\ref{phbarku} the basic Fourier analysis
       of such a  wave-packet
     shows that the product of spread in angular frequency $\Delta \omega$ with the width in the temporal structure $\Delta t$ satisfies the relation $\Delta w \Delta t \simeq 1$, which together with the time component of 
      equation~\ref{phbarku}, that is the de Broglie relation $E=\hbar \omega$,  \textit{implies} equation~\ref{HUPE}.
     Since the quantum relation of equation~\ref {phbarku} is here itself  \textit{derived} for real particle states propagating between interactions, the uncertainty relations \textit{only apply for such states}.

      Hence here there is no conception of `virtual particle' states, neither as intermediating between real particle interactions nor as created in a vacuum; rather  
      the fleeting production of virtual pairs out of the vacuum would rely on 
      taking equation~\ref{HUPE} out of context.    
    The meaning of the relations of equations~\ref{HUP} and \ref{HUPE}  is more concerned with what properties  can be \textit{specified} for real states, rather than what can \textit{happen}, similarly as for quantum theory limit in this application for real particle states. In particular, in no sense is there any implication of a possible violation of 
    energy-momentum conservation.

    The  uncertainty principle of quantum mechanics is sometimes employed in a heuristic argument stating that \textit{anything} coupled to a quantum system must also be quantised as would also then apply to gravity~\cite{EppHa,Adelm}, with the uncertainty in the position of any mass 
    presumed to imply
     a quantum uncertainty in the corresponding gravitational field (see also \cite{Hugg} section~3, \cite{Kief} section~1).
      However, that not only assumes a particular interpretation of the uncertainty principle but moreover involves a somewhat circular argument since it essentially takes the 
      uncertainty relations
      to be immutable and universal in the first place.

       Here the consistent framework for `gravitising quantum theory', as reviewed in the previous subsection, sees all of QFT and quantum mechanics as derived as a limiting case and  applying only to non-gravitational fields, and hence manifestly \textit{not} universal. Here gravity, which \textit{does} have a universal character via the  constraint of equation~\ref{GRQT} as applied in deriving equation~\ref{phbarku}, is central in particular to deriving in turn the `uncertainty relations' of equations~\ref{HUP} and \ref{HUPE}, as applies to all real particle states propagating as wave-packets. The uncertainty relations are therefore not valid as  providing a basis for either the standard QFT vacuum or  an argument implying the necessity of `quantising general relativity'. 
       
       For the present theory on introducing Planck's constant $\hbar$ as described for equation~\ref{phbarku}, as incorporated in turn into equations~\ref{HUP} and \ref{HUPE}, together with the speed of light $c$ and also the gravitational constant $\kappa$ in equation~\ref{Eineq}, the Planck scale as associated with the Planck length of order $10^{-35}\,$m can still be uniquely identified. 
       However, here with gravity \textit{not} quantised this neither denotes the scale at which `quantum gravity' should become manifest nor at which anything physically significant should emerge in the structure of spacetime itself. On the other hand the uncertainty relations of equations~\ref{HUP} and \ref{HUPE}  may still indicate a \textit{limit} on how far structure can be physically \textit{probed}, owing to the possible formation of a black hole in any practical attempt to reach the Planck scale. 
       
       With gravity, based on the external spacetime curvature, essentially a different kind of force to those with quantum properties, as based upon internal gauge symmetries, there is no \textit{a priori} reason to expect a comparable interaction strength, as alluded to in the opening discussion of 
        subsection~\ref{boot32}. For the visible matter branch of generalised proper time the Standard Model internal gauge symmetries themselves derive directly  from the symmetry breaking projection of the local 4-dimensional spacetime form through equations~\ref{sbreak} and \ref{gbreak}  which, as described in subsection~\ref{boot31}, also incorporates  the electroweak symmetry breaking element. There is then only \textit{one} symmetry breaking mechanism and a \textit{single} fundamental  scale as 
         set by the stabilised value of  $\vert \bv_4(x) \vert = h_0$, which is essentially that of the electroweak or Higgs mass scale, and with no further `Higgs' or equivalent needed at a `Grand Unification' scale. 
         
         With neither a physically significant Planck scale nor Grand Unification scale, in the sense described above, there is then no issue with any `hierarchy problem', as might be associated with the many orders of magnitude separating either of these scales from the  electroweak scale. Correspondingly, no new physics is required to maintain such a hierarchy, as would typically involve protecting the Higgs mass against the radiative instability that would ensue if exposed to those much higher mass scales (see for example~\cite{ArkDD,HasLM}).

       With classical general relativity prevailing there are  no graviton states in the present theory, as noted earlier in this subsection.
       This is contrary to the common assumption that gravitons exist as based upon the presumption that gravity should be quantised. 
        The cosmological bootstrap, as reviewed in section~\ref{boot2}, suggests that the existence of spin-2 gravitons, and the associated interactions during an inflationary era, might plausibly leave an imprint in 3-point and 4-point spatial correlations in the distribution of galaxies as might be observed from our present day perspective. 
        This might then provide one possible handle to demonstrate the \textit{lack} of gravitons and absence of any quantisation of gravity, noting there is currently no empirical evidence to suggest gravity should be quantised.

        There may also be another more pronounced impact for the new theory arising from the dark sector of generalised proper time described in this subsection. Local fluctuations in the spontaneous creation of matter field $\{\Apls,\Mplu,\Amin\}$ states   in the  primordial inflationary epoch, as expressed through equation~\ref{GRQT}, would perturb the uniform vacuum geometry of
         equation~\ref{Einvac} with microscopic inhomogeneities, associated with local energy density variations,  imprinted 
          in the very early universe. As  initially frozen into the spacetime geometry such minor perturbations would continue to grow in scale and depth, after the phase transition to an expanding universe dominated by a   dark matter  $\{\Apls,\Mplu\}$  component, under gravitational evolution as seeding the eventual formation of visible matter galactic structures in a manner potentially with observable consequences today.

          The initial primordial states   $\{\Apls,\Mplu,\Amin\}$ generated,  with the associated energy density and geometry perturbations,  could be stretched apart by the inflationary expansion in what might be again termed an effective `cosmological collider' as reviewed in section~\ref{boot2}. Given the non-Abelian gauge symmetry involved, such processes would be analogous to the production of quark and gluon jets in a terrestrial  HEP experiment, with a corresponding equivalent of `hadronic strings' connecting the initial  $\{\Apls,\Mplu,\Amin\}$ states similarly as for the process of `hadronisation'. An imprint of the initial local energy density variations in the rapidly growing background sea of $\{\Apls,\Mplu,\Amin\}$ states could still be left in the large-scale geometry.
          
          Through the cubic and quartic self-coupling of the non-Abelian gauge theory, two or three positive energy $\Apls$ states could be generated in one primordial interaction, generating 2-point and 3-point correlations, with contributions also from $\Mplu$ state production to these and higher order-correlations, as might be observable through the consequent structure of the   CMB or galactic distributions.
          The possibility of  `dark QCD strings' connecting these initial states, as          
           alluded to above, could have implications for
           the structure of the cosmic web as observed over the 
           largest scales of the universe, incorporating 
             clusters and superclusters of galaxies
           and even `Great Wall' features  (see for example~\cite{SheDi,HorBa,EinLi,LibWe}).  
              
             Further, unlike the inflationary production of real particle-antiparticle pairs, originating spontaneously as virtual states as permitted according to equation~\ref{HUPE} and all of positive energy as for the primordial fluctuations reviewed in section~\ref{boot2}, here the direct mutual production
              of real $\{\Apls,\Mplu\}$  and $\Amin$ states, of positive \textit{and} negative energy,   out of the primordial vacuum will generate regions of spacetime with \textit{both}  a positive energy density excess and also \textit{deficit} respectively.
             As a new feature, the primordial production of $\Amin$ states might then `seed'   regions of spacetime with a geometry more likely to lead to `voids' as seen in the CMB or galactic distributions.
             
              Through the creation of two or more $\Amin$ states in a single primordial event correlations between voids themselves might be identified, as well as 2-point and higher-point correlations between  over-dense regions and the voids as might be observable in the CMB or galactic surveys.        
              Indeed, in addition to regions of excess  energy density, ongoing studies probe cosmological models through observations of the dynamics and distribution of voids and `supervoids'. Similarly, predictive features for the new theory with such a negative energy contribution might hence be testable, and potentially help account for any particularly large or deep voids or even apparent anomalies such as the  CMB `Cold Spot', which appears to be of primordial origin (see for 
              example~\cite{BosWe,PisaS,HamaP,Kovac,SchuH,OwusF}).
              
         It was suggested in (\cite{DEner} section~6) that local fluctuations in the present era of ultra-low dark energy density through  $\{\Apls,\Mplu,\Amin\}$     vacuum-plasma creation and annihilation events, as associated with fluctuations in the spacetime geometry on the microscopic   scale through equation~\ref{GRQT}, might in principle be probed by any accumulative influence upon visible particles travelling great distances through spacetime. 
         Here then we are suggesting how fluctuations in the far larger primordial energy density through
          $\{\Apls,\Mplu,\Amin\}$   state production could be magnified through inflation to far greater cosmic length scales and mapped into the large-scale structure of the universe as probed by correlations and other observables associated with the CMB and galactic surveys.  
         Whether or not there might be a role for `asymptotic freedom' to aid such explicit calculations involving this `hidden QCD'  non-compact non-Abelian gauge sector at an extremely high absolute energy scale,
           techniques of the `cosmological bootstrap', as reviewed in section~\ref{boot2}, might also play a simplifying role.
         
         The present theory, as described here throughout section~\ref{boot3}, is \textit{not} itself a bootstrap model, but rather can be compared and contrasted with the cosmological bootstrap of figure~\ref{cosmo} as described below for figure~\ref{rules}.
          
  \vspace{-1pt}   
         
\begin{figure}[htbp]  
\centering
\leavevmode
\includegraphics[width=14.4cm]{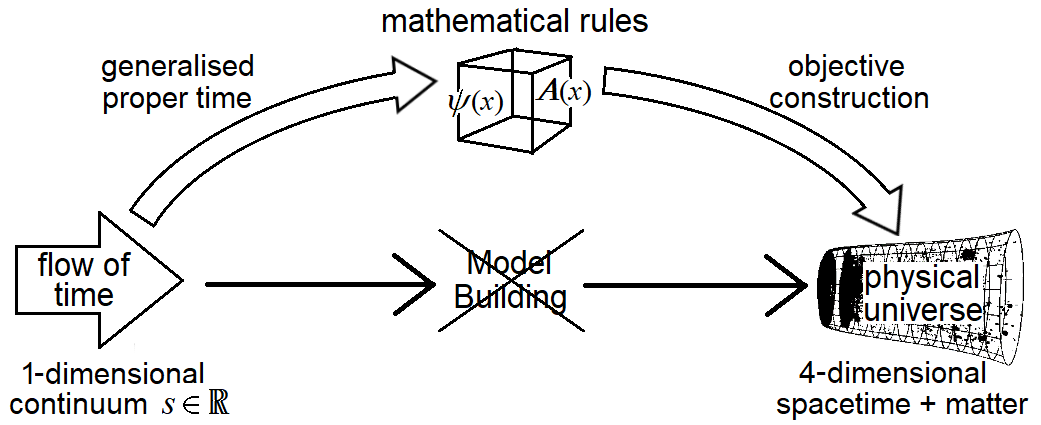}
\vspace{-24pt}    
\caption{\setb
   A schematic of the strategy employed for generalised proper time, exhibiting a common feature with the cosmological bootstrap of figure~\protect\ref{cosmo} in the avoidance of specific models featuring \textit{postulated} fields or interactions. However, while for the bootstrap approaches of section~\protect\ref{boot2} the results follow from basic principles and constraints alone, here the physics derives from \textit{explicit calculation} utilising the basic mathematical rules implied in constructing a 4-dimensional spacetime universe through generalised proper time alone, as described in these three subsections. Given the simplicity of the continuous flow of time as the basic entity, these mathematical rules are nevertheless themselves highly constraining.  
   }
\label{rules}
\end{figure}

    The physical universe constructed through the mathematical rules deriving from generalised proper time as represented in figure~\ref{rules}, from the infinitesimal local level through the extended laboratory environment and on to the global cosmological scale, exhibits properties throughout that connect with known empirical structures. These features include a basis for the Standard Model of particle physics and beyond, as reviewed in subsection~\ref{boot31}, a framework for amalgamating general relativity with quantum phenomena, as summarised in subsection~\ref{boot32}, and a possible means of accounting for the full evolutionary history of the cosmos as broadly consistent with standard inflationary and  $\Lambda$CDM  models, as described in this subsection. 
    
     With all of these properties deriving from generalised proper time, which depends only upon the existence of the continuous flow of time of equation~\ref{srrr} as represented on the left-hand side of figure~\ref{rules}, the question of the source for `time' itself might naturally be raised, particularly given the remarkable simplicity of this foundation.
      While this theory is not considered a bootstrap model as such, the identification of a possible origin of time \textit{will} inevitably lead to a system with a manifestly `bootstrap' character, and with universal scope, as we describe in the following section.


\section{The Universal Bootstrap}
\label{boot4}

\subsection{Chronogenesis: The Interrelation of Time and Physics}
\label{boot41}
   
    For the theory described in this paper the elementary structures of matter, in both the Standard Model of particle physics and the dark sector, together with a framework for amalgamating general relativity with quantum theory, all derive from intervals of the flow of time in equation~\ref{sint} via the generalised form of equation~\ref{salpha}, as reviewed in the previous section.
      This approach might then be termed a theory of `chronogenesis' -- meaning the generation of all physical structures from the continuum of time of equation~\ref{srrr} alone (as suggested in \cite{DEner} section~8).  
       Having provided an account of the origin of matter in 4-dimensional spacetime as collectively deriving from and supported by the basic entity of time alone, this leaves the origin of time itself, although of a fundamental and simple nature, as a seeming loose end of the theory to be addressed.  
       
       Empirically there is no proposal here to build an experiment or make physical observations to `detect time' as this basic entity, in the manner evidence for a new particle or interaction might be sought.  
       While physical predictions of the theory can be tested through empirical observations in particle physics and cosmology, as consequences of constructing the theory from time alone, the flow of time itself is a concept that we are \textit{intimately} familiar with, without the aid of any external apparatus.
        Indeed the term `chronogenesis' has another application, as read in the opposite sense to that above, meaning the generation of our subjective sense of a flow of time from a
         physical basis in neuroscience~\cite{chwww}, as we review here.
        
        Here we are interested in subjective, or mental time, concerning in particular `the duration which is perceived both as present  and as extended in time', as can be termed the `specious present' which,
         as proposed by William James in 1890, is `the prototype of all conceived times'       
         (\cite{Poid} section~4).  
        The subjective passage of time is marked by a continually changing experience in
         the short time span of the  present moment 
      of which  we are intimately and incessantly aware. This subjective present  hence stands     in contrast with the
       objective present or moment in time, as might be associated with a static durationless instant.
      
       Compared with other perceptions, that of time is seemingly unique in that it is not associated with a sensory system, indeed with the passing of time and temporal duration still experienced through patterns of thought even without sensory input. Rather our subjective sense of time is generated in the brain by
        the apparent integration of a range of internal states, regardless of any contribution from external sense data. It is then a matter for neuroscience to investigate how  neural systems 
  of the brain can be structured    with this capacity to   underlie our subjective experience of the flow of time.
        
        By the 1980s experiments had been conducted to detect a link between physical neural activity and the subjective experience of thoughts evolving in time towards an awareness of intention to act~\cite{Lib1}.  
        The apparatus involved sensors placed on the scalp of subjects to record electrical activity of the brain, providing evidence for cerebral activity (the `readiness potential') that precedes the
        subject's conscious awareness  to act by around 300$\,$ms. This demonstration of a correlation between the timing at an objective neural level and a subjective experience level, together with subsequent studies and interpretation, led to significant debate over the role of `conscious will' in performing  voluntary actions~\cite{Lib2}.

        More recently, advances in medical imaging techniques such as fMRI (functional magnetic resonance imaging) have allowed non-invasive probing of more extensive brain structure, including regions of particular interest for the subjective experience of time. 
        The precuneus is anatomically buried in the interhemispheric fissure in the posterior region of the
         medial surface of the parietal cortex (\cite{Cavan} figure~1, with the back of the brain towards the left in that figure). While this region had been relatively unexplored due to its inaccessible hidden 
        position in the cerebral cortex, the strategic location and wide-spread connections underlie the significance of the precuneus in association with a variety of functions.
        
        In particular it has been demonstrated that the precuneus plays a central role in the neural basis for our subjective experience of the present, that is the `now' of our time perception~\cite{Tang}. In fact precuneus activation is found to be significant in association with subjective feelings of the past, present and future, as evoked by speech
         stimuli, with the response up to three times greater for the experience of `presentness', in a manner that is independent of the language used.
         The fact that the strongest response of the precuneus is to presentness is presumably important for survival, with a clear evolutionary advantage in focussing upon an accurate representation of the current  situation of the individual and state of affairs of the external world.
         Further, the response to the `future' comes second, as is again reasonable given the need to make plans for upcoming events, while response to the `past' comes third, perhaps since `it is no use crying over spilt milk' (\cite{Tang} figure~3D and discussion).

         Given the resulting map of the past, present, and future of mental time over the precuneus region of the cerebral cortex the further aim is to determine the basis, in terms of
           structural connectivity and neural activity,
           for our subjective sense of a    \textit{flow} of time~\cite{chwww}. While the present `now' is most closely identified
           with this subjective flow of time, the past and the future provide the perspective for this directed progression.    Further planned research projects (\cite{chwww} B01) include the study
           of how making a decision maps into the 
           dynamics of neural activity representing the future options for the subject, particularly when the choices  seem of little relative value as for example in acting to grab  a bottle or a cup first when both are needed.
        These studies may indicate how such ongoing brain activity, as a basis for a decision making strategy, might contribute to generating our subjective perception of a flow of time.
        
        Other areas of the brain also play a  significant role in constructing our  conception of time. For example there are rich functional and anatomical connections between the precuneus and hippocampal 
         network, with the former important for retrieving memories from the latter. Memory itself is concerned not only with the past, but also with a re-experience of the past in the present.
         The hippocampus itself contains neurons called `time cells' 
          which fire at specific times and provide a `time stamping' of events as required for their
           temporal ordering. The hippocampus then plays a key role in the formation of
            `episodic memories' and storage of the associated unified representation of
             the `what', `when' and `where' of experience~\cite{Tsao}.
                  Episodic memory, as sequentially ordered in time, is the store of previous events as personally experienced by an individual, in contrast to `semantic memory', that is the more general
            non-autobiographical knowledge of the world~\cite{Cavan}. With the precuneus implicated in episodic memory retrieval and sense of the past, this can provide a background contributing to our subjective sense of a flow of time in the present. 
             
      The role of memory and network structure in the generation of subjective time can also be studied with artificial neural networks and machine learning, aiming to model the behaviour of a human brain.
                   Such an adaptive neural network has been implemented in a system  termed a 
               `Mind Time Machine' (MTM)~(\cite{chwww} B01, \cite{Ikeg,Ikeg2}).
                   The MTM is
                      coupled to an open environment through an arrangement of video cameras generating mutual feedback loops,
               providing multiple sensory inputs and creating unrelated episodic memories as stored in a `latent space'. 
        The system's memory development and self-organising internal time structure has then been investigated, as influenced in part by work of Libet.
                  One aim is to prove how  in a living biological system a single  subjective time axis thread can emerge by consistently combining a vast number of  episodic memories.
           The possible implementation  through deep learning protocols in an artificial intelligence system such as an MTM,  capable of maintaining its own identity and autonomous behaviour over a
            sustained period of time,
                  would suggest the possibility of a computational, and in turn mathematical, account of the generation of time. In particular, an ambition for such a project would be to extract the fundamental basis for the subjective \textit{flow} of time.

            In terms of human neuroanatomy the aim is to map out the biological neural network and 
       systems of information transfer between regions of the brain that generate \textit{our} subjective perception of the flow of time. This includes both the grey matter of the cerebral cortex on the brain surface, as responsible for higher-level brain function and with the precuneus considered a central component, as well as the interior white matter, as providing further transmission pathways in the broader network of
        structural connectivity.
        
        Such studies of the brain architecture can be performed using fMRI while the subject is performing various mental tasks. When neurons are active their firing is fuelled by an increase in the local rate of blood flow and its oxygen level. Small changes  in these can be measured by fMRI which is sensitive to the magnetic properties of  iron-containing haemoglobin, which in turn depend upon the degree of oxygenation.  
        Brain activity associated with performing tasks can then be recorded through this Blood Oxygen Level Dependent (BOLD) imaging.   
         The brain representation underlying human cognition can be described with a 
         voxel-based model, where `voxels' are essentially 3-dimensional graphic pixels that can be used to visualise anatomical data. 
            The aim of such analysis is to provide a voxel-based map of brain activity as associated with for example visual, auditory or linguistic information~\cite{Nakai}, as can be applied to determine the function and connectivity of specific brain regions such as the precuneus. 
       
        Functional imaging also shows how cerebral blood flow and metabolism varies across different regions while the subject is in a resting state, suspended from performing tasks, highlighting baseline processes.
        This organised baseline state, or the `default mode of brain function' corresponds to neural activity of the `default mode network' (DMN).   Within the DMN the precuneus is of particular interest in showing the \textit{highest} resting metabolic activity among the `hot spots'~\cite{Cavan}.
         As the core node or hub of the DMN the precuneus is engaged in the continuous processing of internal stores of information, rather than being directly concerned with external sensory input.
         This activity includes the retrieval or processing of episodic memories, conscious representation in the form of mental imagery, spontaneous thoughts and the manipulation of ideas for problem solving and future planning -- that is, the overall contemplative thoughts of resting consciousness against a background of minimal engagement in external sensory or motor activity.
               While the various functions can be associated with various adjacent and overlapping areas of the 
         precuneus region there is no evidence for interhemispheric specialisation for the precuneus, unlike the functional left-right asymmetry identified for many other regions of the cerebral cortex~\cite{Cavan}.

          These observations hence converge upon the precuneus, with its multiple functions and rich connectivity, as playing a central role in the neural network serving the internal  processes of
           reflective self-awareness and conscious  experience. The wider `posterior hot zone' of conscious activity in the parietal lobe  is mapped out in (\cite{Koch} figure `Footprint of Experience', with back of the brain towards the right), subsuming the precuneus      
         (\cite{Cavan} figure~1, shown from the opposite side of the brain).
         In the context of the present concern these observations support the identification of the precuneus as central to the irreducible conscious baseline of internal thought processes and the associated subjective experience of a continuous flow of time.
         
         Observations with fMRI have moreover demonstrated a central role for the precuneus in integrating
         multiple neural systems, including not only for
         conscious self-reflection and subjective time but also  for visuospatial imagery and the internal
         manipulation of mental pictures as alluded to above, all in a mutually interrelated manner~\cite{Cavan,Peer}.
         In particular,
           perception of time has a close connection with
          our \mbox{3-dimensional} spatial perception, which subjectively takes the form of a continuous flow and is structurally \textit{not} frame based as for the images of a movie film or typical machine vision.
         There is ongoing work to elucidate the correlation between space and time processing and the nature of our spatiotemporal  cognition~\cite{chwww}.
         
         The role of the precuneus in our sense of self, time and space is crucial for our sense of
          `orientation', that is the tuning between the subject and the internal representation of the external world with regards to people, events and places~\cite{Peer}.  
            This self-projection or orientation with respect to reference frames across different domains of the multifaceted self relies more broadly on the shared processing system of the DMN, with the precuneus as the central hub hence key to our overall sense of \textit{engagement} in the physical world.
            In this subsection we are mainly concerned with the generation of the subjective flow of time, or `chronogenesis', while spatial perception will play a role in the following subsection and self-consciousness more explicitly in subsection~\ref{boot43}.
                The aim of the above  neuroscience   review  of the evidence for the hardwiring of our perception of time in  brain systems has  been to show how chronogenesis, 
           the generation of a  subjective flowing temporal continuum as a structure of the mind,
                is fully open to  both  empirical and theoretical scientific analysis.

      Objective neural structures and activity should not be directly \textit{equated}  with this temporal continuum, but rather they \textit{accommodate} the subjective experience of a continuous flow of time. Hence while it is not possible to directly `point' to this subjective flow of time in the physical brain, \textit{time is in the world} in the sense of having a supporting physical basis, similarly as for example a subjective experience of the colour `green'.    
      However, unlike visual impressions, sounds, tastes and so on, which are all at their most vivid 
      \textit{while being experienced} through sensory input from the \textit{external} world, the subjective feeling of progression in time is \textit{internally hardwired} and largely independent of sensory data, as we have reviewed above.  We do not need to `look at a clock' to feel a sense of time passing.
      Further, unlike sensations such as `green', the \textit{qualitative} nature of this
       subjective perception of time as the flow of a  gapless \textit{continuum},
        without any notion of discreteness,   has a direct \textit{quantitative} correspondence in the gapless  real continuum described by the real numbers, permitting a \textit{mathematical} analysis
        of this conception of time in itself.

      That is, as will be discussed further in subsection~\ref{boot43}, as a simple continuous progression the \textit{subjective} flow of time is identical, that is structurally isomorphic, to the \textit{objective} flow of time  as modelled by the real continuum  $s\in \rrr$  of equation~\ref{srrr} and utilised as the basis for the present physical theory.
         This apparent equivalence, hinging upon the simplicity of this continuum, then suggests an \textit{origin for time} itself in the subjective temporal flow generated by structures of the mind as the basis for the objective flow of time through which \textit{all} physical entities in the world are perceived, and with which those entities are intimately connected through physical laws.
          In turn figure~\ref{rules} at the end of the previous subsection might hence be augmented as depicted in figure~\ref{mind} below, with 
           the underlying source of time  taken to be this subjective one, as indicated by the 
         first link on the bottom-left of figure~\ref{mind}, with a supporting physical basis in neuroscience.

     The conception and properties of the continuum, as expressed through the  mathematical correspondence with equation~\ref{srrr}, implies that      
          any finite interval or moment of duration of the flow of time can be analysed down to infinitesimal intervals. This hence provides the link to a mathematical structure that contains intervals of time at the infinitesimal level of equation~\ref{sint} and on to further mathematical analysis via the direct arithmetic equality of the generalised form in equation~\ref{salpha}. The mathematical possibilities this expression \textit{intrinsically} entails implies that time itself \textit{can} provide the basis for the 4-dimensional spacetime geometry and its matter content, as described for equations~\ref{sbreak} and \ref{gbreak}, that we observe. 
         Noting that \textit{everything} that we perceive or observe in the physical world is indeed experienced \textit{in time}, there is a compelling case for the world to actually \textit{be constructed} this way at the most elementary level.
         From this perspective, moving from left to right across the top half  of 
          figure~\ref{mind}, time is not something in the world but rather \textit{the world is in time}.
         
\pagebreak         
         
 \begin{figure}[htbp]  
\centering
\leavevmode
\includegraphics[width=14.4cm]{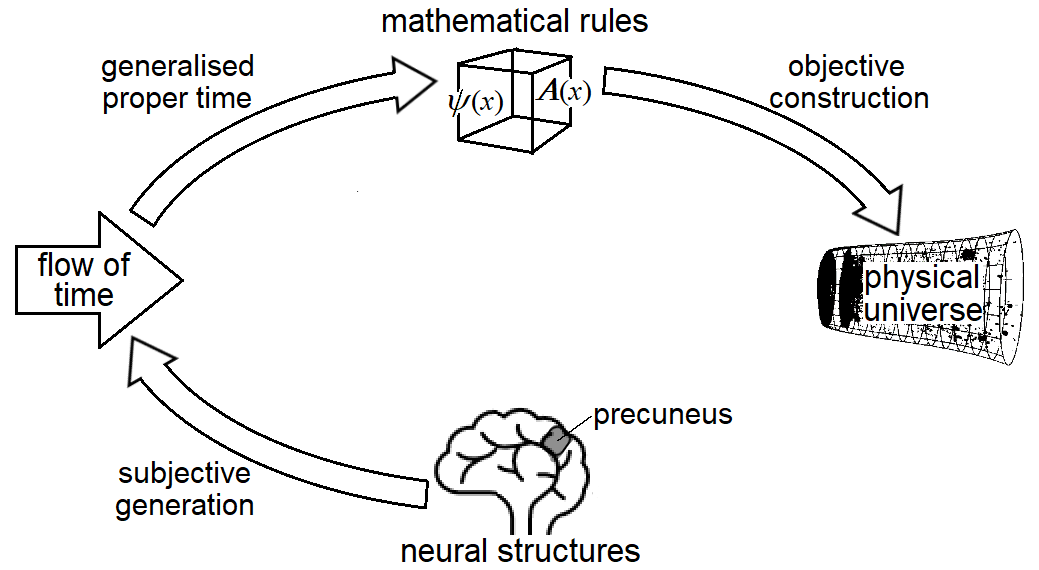}
\caption{\setb
      For the theory based upon generalised proper time physical structures of the universe derive directly from
       the arithmetic substructures implicit in the flow of time alone, as reviewed in section~\protect\ref{boot3} and represented in figures~\protect\ref{sgener} and \protect\ref{rules}. In turn a basis for a continuous temporal flow can itself be identified as an intrinsic subjective aspect of the mind emerging from neural structures in the brain
       as reviewed in this subsection. The precuneus, located in a central interhemispheric region of the cerebral cortex, plays a key role in this `chronogenesis'~\protect\cite{chwww,Tang}.
   }
\label{mind}
\end{figure}            
         
         Having identified a basis for time in the neural structures at the base of figure~\ref{mind}, and noted the interrelation between time and the physical world as described above, the further step towards the `universal bootstrap' is inevitable, as we expound in the following subsection.


\subsection{An All-inclusive Self-sufficient Bootstrap System}
\label{boot42}

     In this theory the principles for constructing the physical universe are determined by the mathematical rules deriving directly from the generalised form for proper time, as an inherent arithmetic property of the flow of time, with no further entity required to be added or postulated. Given the simplicity of the continuous flow of time as the starting point of the theory, the possibilities for the properties of matter and laws of physics that can be derived are necessarily highly constrained and restricted, sharing that common feature with existing bootstrap models as noted for figure~\ref{rules}.
     
     The mathematical rules overseeing the construction of the entire 4-dimensional universe, seen as a `block' whole solution for equation~\ref{GRQT} and as represented on the right-hand side of figure~\ref{rules}, are of course not themselves specifically centred at any particular spatial or temporal location in the cosmos.
       On the other hand,
      as proposed in the previous subsection the source of the flow of time, and hence the resulting mathematical rules, can be equated with the subjective sense of a temporal continuum supported by the physical architecture hardwired into the neural structure of an organic brain, as depicted in  figure~\ref{mind}. Such neural entities \textit{are} physically embedded in the material universe at  specific locations in space and time, amongst the general empirical matter content of the world.
      This further, inevitable, link implies the possibility of augmenting  figure~\ref{mind} such that the
       \textit{circuit closes} as depicted in figure~\ref{cycle}.

\vspace{2pt}
       
\begin{figure}[htbp]  
\centering
\leavevmode
\includegraphics[width=14.4cm]{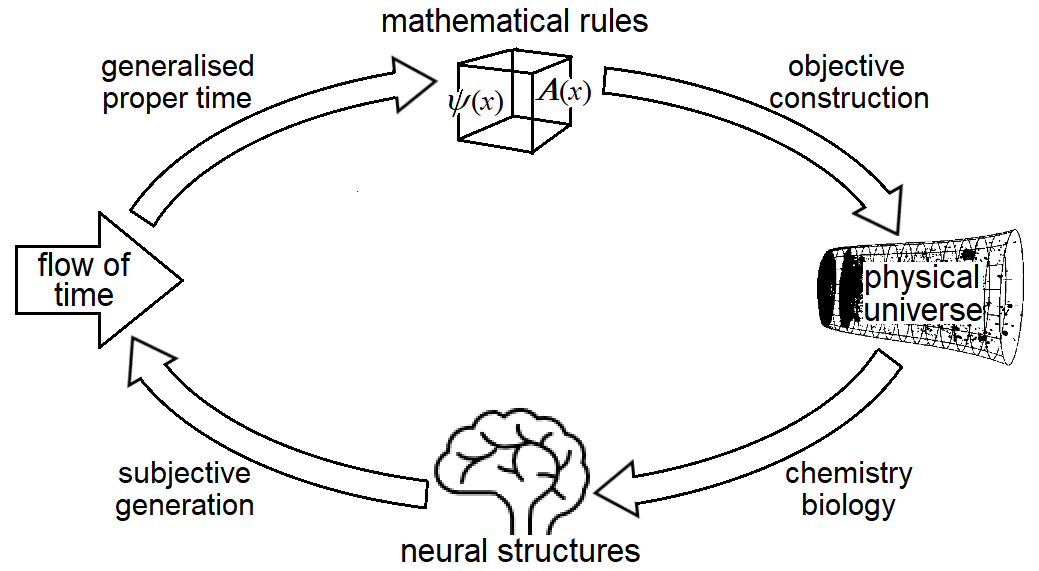}
\caption{\setb
      While the theory of generalised proper time describes how the physical world can be constructed from mathematical rules implicit in the flow of time, the temporal continuum itself can have a subjective origin supported by neural structures embedded within the physical universe itself.
      The circuit of the `universal bootstrap' can then be seen to close in a coherent, consistent and fully self-contained manner.
   }
\label{cycle}
\end{figure}                 
       
       The universal bootstrap thus constructed in figure~\ref{cycle} has a close analogy with the image of an individual `picking himself up by his own bootstraps', as alluded to in the opening of section~\ref{boot1}. Both the conceptual cycle of figure~\ref{cycle} and the talented Baron von M\"{u}nchhausen exhibit a self-contained and self-sustaining system capable of floating free, detached from any external support.
       
       We emphasise that the full circuit of the universal bootstrap in  figure~\ref{cycle} is at every stage fully scientific and amenable to both theoretical and experimental reasoning and analysis.
       Beginning on the left-hand side of figure~\ref{cycle} the theory takes the continuum of time with $s\in \rrr$ of equation~\ref{srrr} as the basic entity, then via the intrinsic arithmetic structures of generalised proper time employs the implied mathematical rules as indicated at the top of the figure for constructing the physical universe depicted on the right-hand side. This leads directly to the empirical structures of the Standard Model of particle physics, with gravity consistently amalgamated with a quantum theory limit, and  also a source of dark energy and dark matter components as described in section~\ref{boot3}.

        The further physical sciences of atomic and molecular structure and through to macroscopic entities with their familiar chemical and biological properties, accompany particle physics on the laboratory scale and astrophysical structures up to the cosmological scale to provide the extensive scientific understanding of the physical universe on the right-hand side of figure~\ref{cycle}.
        A path through contemporary biology, neuroscience and psychology, and associated elements of the physical universe  as represented in the link to the lower part of figure~\ref{cycle}, then leads to the generation of a subjective sense of the flow of time as modelled by the continuum $s\in \rrr$ and back to the left-hand side.
        
         Although not all links are in areas traditionally covered by physics, at all stages around the cycle the description is wholly scientific, as is then  the full picture of the universal bootstrap. 
      We note in particular that  through the detour into neuroscience, time, as the basic entity of the physical theory, is itself
         `bootstrapped' within this self-contained system, along with all of the physical structure
          it supports.

       In order for the circuit to be complete a well-defined, rational and unequivocal account must be given for   each link around the cycle. That the link on the left-hand side of figure~\ref{cycle} should be definitive and clear strongly suggests that the physical theory of matter in 4-dimensional spacetime \textit{must} indeed be constructed solely from the flow of time $s\in\rrr$ of equation~\ref{srrr} alone with no additional basic entities. This is fully consistent with the perspective that has been adopted in developing the theory of generalised proper time, and has been key to identifying the empirical successes described in section~\ref{boot3}.
       
        However, the one further input that is required is a \textit{mechanism} to break the symmetry of the general form for proper time of equation~\ref{salpha}, through the extraction in equations~\ref{sbreak} and \ref{gbreak} of a distinguished local 4-dimensional spacetime form incorporating a 3-dimensional spatial component, as key to generating the mathematical rules for the physical construction of the universe.
   The subjective nature of our temporal perception also motivates consideration of the degree of subjectivity, and implications, of our 3-dimensional spatial perception of the world. 
   That there is such a subjective spatial element, that is also linked with subjective time,  was alluded to in the previous subsection.

    In fact the realisation of an intrinsic and necessary role for the mind in constructing our 3-dimensional spatial perception of the world has a long history~(see for example \cite{Berkl}). 
    Our visual perspective at any moment is essentially rooted to a single location, with anything viewed along a common line of sight projecting to this one point, implying that spatial distances can not strictly be perceived by our sense of sight alone (\cite{Berkl} paragraph~II).
  Binocular disparity, the geometric leverage gained from having a pair of eyes, does not 
   affect this conclusion since in all cases judgement of distance relies on experience
      (\cite{Berkl} paragraphs~XII--XX). Rather, the primary and immediate objects
       `apprehended by the Eye' should be distinguished from the secondary objects generated by the
        `intervention of the Mind'  (\cite{Berkl} paragraphs~XLIII--LI).
        
        As Berkeley noted, while the primary sensory input can suggest the idea of distance to our thoughts, it is the secondary object that makes the stronger and more vivid impression on us.
         However, since it is difficult to discriminate between these two things
         we habitually, and mistakenly, attribute to the immediate, primary, objects of sight what in fact belongs to the mediate, secondary, objects of sight.  Such historical references  demonstrate, even without the modern day observations of neuroscience, that through reflection and self-examination of thought this distinction can still be identified and that while our sense of sight provides input the subjective element in mind is still needed to interpret vision in terms of 3-dimensional spatial perception of the world.
         In the modern day we are also aware of this necessity on a constant basis: on viewing
          stationary or moving images on a flat 2-dimensional screen we subjectively \textit{perceive}
           the perspective and depth of a fully 3-dimensional scene as constructed through the 
            internal workings and  neural activity  of our brain structure.
            
            While for Berkeley the secondary reconstruction of the depth in a field of view
            was a capability that could be learnt through experience,  more recent analysis suggests that significant aspects of depth perception are innate and essentially implanted in the mind from the outset
            (see for example~\cite{Bruce} chapter~7). Given the 2-dimensional retinal image detected by the eyes there are a range of cues, including binocular disparity, motion parallax, obstruction, perspective and shading, which are in general individually  ambiguous. These cues are initially separately processed in the brain and then integrated into a single stable and unambiguous scene as neural mechanisms analyse and represent this optical information to generate our fully 3-dimensional spatial perception.
   Studies with fMRI can then indicate the areas of the brain where the initial processing and where this secondary integration takes place.

       The primary visual cortex of the brain does \textit{not} directly contribute to visual
        \textit{experiences}, similarly as for other primary sensory processing. Rather, the  information is passed to the `posterior hot zone'  where conscious perception is generated (\cite{Koch}, see also the figure therein as discussed in the previous subsection). 
         This also raises the neuroanatomy  question concerning what exactly it is about the architecture of cortical regions that do or do not directly contribute  to subjective content (an issue we return to in the following subsection). 
         
         Homing in more specifically our visuospatial perception has been connected with neural activity in the precuneus,  as alluded to in subsection~\ref{boot41}.
         We also reviewed there how the precuneus  serves a central role for subjective time,
        and noted  how our visual perception is inextricably linked with perception in time, both in terms of the ongoing flow of visual perceptions in the specious present and in terms of the visual imagery associated with episodic memory recall.
   As a key region associated with  visuospatial information processing and internal imagery the precuneus can be considered the  `mind's eye' \cite{Cavan}.
     In addition to its role in more abstract mental imagery tasks, the precuneus is part of the neural system 
   guiding body movement behaviour given also the  spatial relations required for such a function. 
    These neural structures are then central to both  temporal  and spatial perception and these  aspects of our orientation and mental engagement in the external world, as also reviewed in the previous subsection.

   As proposed there in leading to figure~\ref{mind}, here we are identifying the continuum of the subjective flow  of time with the objective progression in time as parametrised by $s\in \rrr$ in equation~\ref{srrr}. Through this mathematical correspondence,  on approaching the infinitesimal  limit as permitted by the continuum property,  intervals 
      of time can then be expressed in the general form of equation~\ref{salpha}.
   This general form for `proper time' for $p\ge 2$ in equation~\ref{salpha}  itself intrinsically incorporates substructures with a quadratic form as can be associated with a local spatial geometry.
   Through a further structural isomorphism  this quadratic mathematical  form then
   provides our further innate  predisposition to perceive  3-dimensional spatial forms with something to \textit{attach itself  to}. Hence a key role for subjective space, as well as subjective time, can be identified for the construction of the universal bootstrap.
   
   \pagebreak
 
    The extraction of a local 3-dimensional spatial form, as part of a 4-dimensional spacetime component, implies the symmetry breaking and direct derivation of the elementary structures of matter from the residual components over 4-dimensional spacetime as described for equations~\ref{sbreak} and \ref{gbreak}. By comparison also with figure~\ref{sgener} we can express the symmetry breaking of generalised proper time in the schematic form:
 \begin{equation}    
 \label{subrk}          
        \underbrace{  s \in \rrr  }_{\mbox{\small{time}}}   
          \;   \to  \;      \underbrace{  \delta s \in \rrr  }_{\mbox{\small{interval}}}   
          \;   \to  \!\! \!\!\!   \underbrace{   (\delta s)^p  }_{\mbox{\small{generalised}}}  \!\!\!\!\!
             \to  \;
            (\delta s)^p /  \underbrace{  (\delta s)^2 }_{\mbox{\small{space}}}                                                                
             \;   \to \;      
                  \underbrace{ \{\delta x^a\}\; \mbox{residual} }_{\mbox{\small{matter}}}                                                                
\end{equation}  
     
     The symmetry breaking structure implied at the fourth stage above is essential for the construction of the entire physical world and its matter content identified at the fifth stage.
     Here we are proposing then that the \textit{mechanism} for the symmetry breaking extraction of the spacetime components may itself have a  subjective basis. That is, the fact that the brain is structured to perceive the world in time \textit{and} space may provide the source of both generalised proper time \textit{and} its symmetry breaking, and hence the origin of the mathematical rules at the top of the universal bootstrap of figure~\ref{cycle}, which itself might then be augmented as depicted in figure~\ref{cycles}.

\vspace{2pt}
     
 \begin{figure}[htbp]  
\centering
\leavevmode
\includegraphics[width=14.4cm]{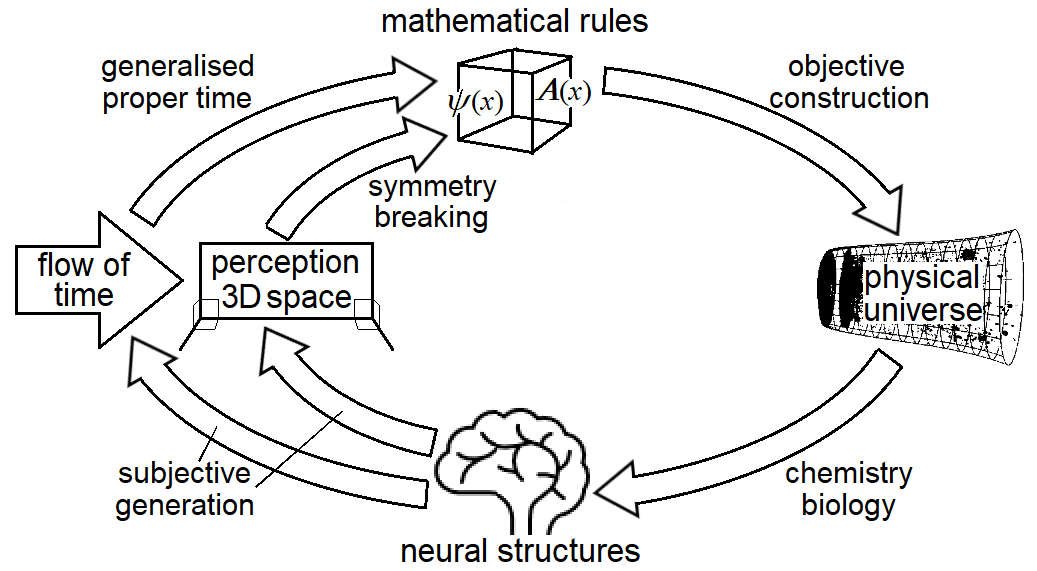}
\vspace{-24pt}    
\caption{\setb
     The significance of neural structures may not only lie in the generation of subjective time, closing the circuit as described for figure~\protect\ref{cycle}, but also the associated subjective perception of 3-dimensional space, as required to impose the symmetry breaking that generates the mathematical rules that determine the structure of the physical world within the universal bootstrap system. 
     It is not so much that perception of a moment in time generates the full objective continuum $s\in \rrr$, or analogously for space, but
     rather  anything that \textit{can be} perceived is necessarily perceived \textit{through} time and space, as the \textit{forms} of subjective impressions hardwired into the brain, and \textit{that  alone} is sufficient  to generate the laws governing the physical world as observed. 
   }
\label{cycles}
\end{figure}

The physical universe, on the right-hand side of figure~\ref{cycles},  is then necessarily constructed within the constraints of the mathematical rules in a manner that accommodates such a physical brain. 
  The limited form of our perception of a world within these constraints of time and space is analogous to the way we might view a \mbox{3-dimensional} image as encoded within the confines of
   a particular 2-dimensional autostereogram figure, except in the former case the physical form of the observer is inextricably embedded within the `image' as identified with the objective  physical world itself
   (see also \cite{Tflow} section~5).
       
       We can perceive 3-dimensional depth in a flat 2-dimensional photograph or autostereogram  when 
       it  \textit{isn't really there} but rather only indirectly represented. On the other hand the 3-dimensional depth we can perceive as associated with the quadratic form at the fourth stage of equation~\ref{subrk} \textit{really is there} since this algebraic mathematical form has properties directly \textit{isomorphically equivalent}
        to the geometric structure of \mbox{3-dimensional} space. In this case the  `image' generated is robust and stable since it corresponds to the extended physical world within which \textit{we ourselves} are irreducibly physically integrated through the universal bootstrap.

        To further the above analogy, the extended 3-dimensional image generated by an autostereogram can be perceived without focussing any attention upon the local pixel components, the geometric properties of which are essential in constructing the depth observed in the image. 
       Similarly our  subjective 3-dimensional perception supervenes on the overall construction of the physical world, without the possibility of focussing upon the local infinitesimal elements in equation~\ref{subrk}, the quadratic forms of which are essential in underlying the construction of the extended 4-dimensional spacetime solution for equation~\ref{GRQT}.
      While these 
         infinitesimal quadratic elements of spacetime (pictured in \cite{Tflow} figures~4 and 5) permit the construction of  the extended physical world (\cite{Tflow} figures~7 and 8) they are not only of course far too small to directly perceive but also meld into the smooth 4-dimensional spacetime continuum.

       In terms of the physical properties of the extended world constructed in this manner,
        the simplicity of  emphasising time  alone as the basic entity
       together with the extraction of a 3-dimensional spatial form
        is very constraining.
       These constraints, on constructing the theory from infinitesimal elements of time  as described for equations~\ref{sint}--\ref{gbreak} and figure~\ref{sgener}, imply locality, causality and Lorentz invariance as noted in subsection~\ref{boot32} (regarding causality in particular, an analogy and contrast is made with the construction in causal set theory  in \cite{Tflow} section~3 and figures~4--8 therein as alluded to here in leading to equation~\ref{GRQT} and again above).
      With the conservation of probability modelled by complex amplitudes, as described for equations~\ref{pddddm} and \ref{uprob}, there is also an implication of unitarity and analyticity,  for calculations to be mathematically well-defined, as also noted in subsection~\ref{boot32}. Hence the source of this set of constraints, as also featuring as key inputs for the $S$-matrix bootstrap reviewed in section~\ref{boot2} for figure~\ref{bubble}, is itself \textit{contained within} the universal  bootstrap of figures~\ref{cycle}
       and \ref{cycles}  \textit{without} needing to be imposed from the outside.

       Hence the $S$-matrix bootstrap is in this sense \textit{subsumed} within the universal bootstrap, with the latter theory also providing the source of specific matter states and interactions as described for equations~\ref{extgpt}--\ref{slide} and figure~\ref{sinter} as arising directly from the symmetry breaking of generalised proper time. These matter states and interactions replace the role of a Lagrangian and account for the action principle as employed in a standard field theory such as that of the Standard Model of particle physics.
      Properties of the Standard Model particle multiplet structure itself are here identified in the local symmetry breaking pattern for the $\esi$, $\ese$ and potentially $\ee$ branch of generalised proper time as reviewed in subsection~\ref{boot31}.
       Given the overall set of constraints we argued in subsection~\ref{boot32} how a quantum theory limit, ultimately deriving from equation~\ref{GRQT} as consistent with general relativity, could reproduce the successes of standard QFT on noting the similarity of the respective inputs constraining each end of equation~\ref{pddddm}.   Alternative possible branches for generalised proper time also provide a direct and explicit source of a dark sector as described in subsection~\ref{boot33}.

       With the constraints on particle interactions augmented by assumptions concerning the symmetry of the large-scale structure of the universe as evolving from the earliest times, methods analogous to those of the cosmological bootstrap, also reviewed in section~\ref{boot2}, 
        might  also be incorporated within the top half of figures~\ref{cycle} and \ref{cycles} as a means of simplifying some calculations relating to the cosmological scale,  without requiring a full computation of the microscopic details.
           However, while the cosmological bootstrap deliberately bypasses specific models, as described for figure~\ref{cosmo}, here the actual particle and interaction content of the theory can in principle be derived in detail explicitly for both the visible and dark sectors from the symmetry breaking  of generalised proper time and the implied specific mathematical rules as indicated in figure~\ref{rules}.
           For example primordial fluctuations in the creation of dark sector $\{\Apls,\Mplu,\Amin\}$ states could leave  a characteristic  imprint in the  CMB as well as in the spatial distribution  of galaxies and the voids between as observable in the present day, as proposed and described before figure~\ref{rules} in subsection~\ref{boot33}.
        Similarly, with gravity not being quantised in this theory,
           the lack of any spin-2 graviton states, in particular in the extremely early universe, may have implications for these cosmic observables.

   While the pragmatic techniques of the $S$-matrix and cosmological bootstraps, with their closely related sets of constraints associated with the mathematics of amplitude calculations, can be applied as general heuristic tools, they do \textit{not} represent fundamental theories.
      Reaching beyond the $S$-matrix and cosmological bootstraps, in being \textit{fully} self-sufficient with all entities and all constraints supported internally as described for the big picture of figure~\ref{cycle}, and its proposed augmentation to figure~\ref{cycles},  the universal bootstrap
      not only realises a fundamental theory but also 
       offers an account for  \textit{why anything exists at all}.
       That is, the theory  provides a means of addressing the question of `why there is something rather than nothing', in a manner that is fully scientific and seems not possible for most other physical theories, 
        as we shall consider further in section~\ref{boot5}.

       The universal bootstrap of figure~\ref{cycle} is also predictive and testable, at least in the sense of being falsifiable, in reinforcing the notion that all of the structures of the physical universe must derive from the mathematical rules  implied in basing the theory on the sole entity of the continuum of time. The completed circuit of figure~\ref{cycle}, emphasising this starting point for a theory of generalised proper time in the flow of time alone as required  to close the circuit,  plays then not just an abstract conceptual role but has concrete physical consequences. As well as reinforcing the physics described in section~\ref{boot3}, through the overall self-consistency and self-contained nature of the  system of figures~\ref{cycle} and \ref{cycles}, the universal bootstrap implies further considerable explanatory power in itself  as we describe in the following subsection.

\pagebreak

\subsection{Physical Implications and Explanatory Power}
\label{boot43}

        Through the self-contained circuit of the `universal bootstrap' of figure~\ref{cycle} this theory
extends into areas 
that most other theories cannot reach. However, it is not \textit{essential} to construct such a cyclic system  as an inevitable consequence for the approach of generalised proper time. The link in the bottom-left of figure~\ref{cycle} (and similarly for the augmented circuit of figure~\ref{cycles}) could be broken, hence exposing \textit{three} loose ends, as pictured in figure~\ref{cycleb}. This structure and the questions it raises map closely onto the corresponding issues for other proposed unification schemes and most theories in general as discussed below.
        
\begin{figure}[htbp]  
\centering
\leavevmode
\includegraphics[width=14.4cm]{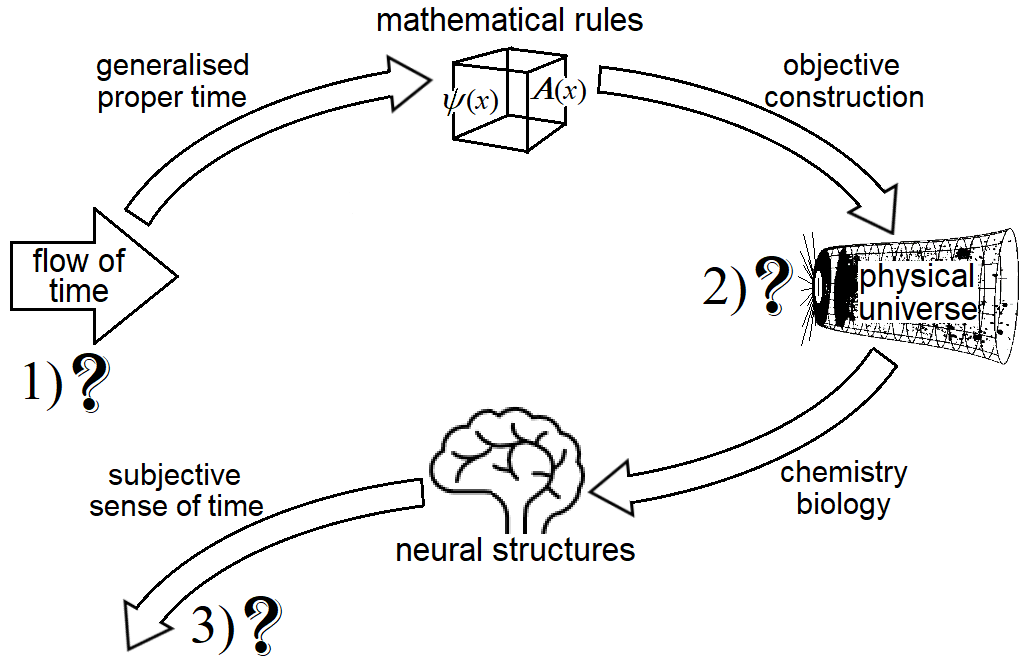}
\caption{\setb
      On breaking open the circuit in figure~\protect\ref{cycle} three loose ends are exposed, concerning 1) 
      the source of `time' as the basic element of the theory, 2) the origin of the Big Bang and the initial conditions of the early universe and 3) the seemingly hugely improbable generation of the complex structures required to support sentience, how this is possible at all, and its possible role if any in the cosmos.
   }
\label{cycleb}
\end{figure}     
      
\vspace{-3pt}         

         As summarised in figure~\ref{cycleb} the significant loose ends entail:  
\vspace{-1pt}         
\begin{itemize}
\item[1)]{The need to postulate the ordered flow of `time' as the basic entity, with this temporal continuum parametrised by $s\in\rrr$ in equation~\ref{srrr} as the basis for the mathematical rules and  structure of a physical theory  through equations~\ref{sint}--\ref{gbreak}.}
\vspace{-1pt}
\item[2)]{Even given the established laws of physics, the question of why the physical universe itself should apparently have come into existence `in the first place' around 13.8 billion years ago  and what determines the specific initial conditions.}
\vspace{-15pt}
\item[3)]{Given a physical universe, the desire to account for the possibility and significance of  the
complex neural systems with the 
 seemingly highly specialised capabilities of supporting self-reflection and observing the physical universe itself. }
\end{itemize}

       Essentially all existing theoretical unification schemes have these three loose ends, or an equivalent in terms of a particular set of postulates in the place of `1)'. The nature of each loose end appears  miraculous in itself and raises seemingly intractable questions in any attempt to find an explanation. However, for the present theory a possible means of comprehending a simultaneous  solution for all three is provided by the universal bootstrap. On linking the two loose ends on the left-hand side of figure~\ref{cycleb}, connected as suggested for figure~\ref{mind} in subsection~\ref{boot41} 
       as incorporated into figure~\ref{cycle} in subsection~\ref{boot42}
          and accounting for `1)',  the physical universe on  the right-hand side of the closed circuit is seen as a complete 4-dimensional spacetime `block' \textit{within} the overall universal bootstrap system.  The Big Bang and the very early universe is then just a special  \textit{region} of this full spacetime solution,
          as constructed within the implied mathematical rules and physical laws,
        with this shift in perspective addressing loose end `2)'. In particular, the apparent `initial conditions' are 
           \textit{required} to have the appropriate features  in order that the necessary complex brain structures can exist within the physical universe block for the purpose of generating subjective temporal experience capable of closing the circuit, as the primary role for `3)'.

       Hence the above \textit{three} seemingly intractable problems summarised in figure~\ref{cycleb} and the above list can be transferred to \textit{one} question concerning the possibility of closing the circuit of the universal bootstrap. Hinging upon the identification of the subjective flow of time with an objective flow of time on the left-hand side of figure~\ref{cycle}, through their equivalent and simple structure as modelled by the continuous progression of real numbers  $s\in\rrr$, and supplemented by an associated subjective predisposition to perceive the world in 3-dimensional space as described for figure~\ref{cycles}, a plausible argument can at least be framed to address this single question as we have discussed in the previous subsection. 
   While addressing the three major open questions of figure~\ref{cycleb} 
      the universal bootstrap,  having a fully scientific basis as also described in the previous subsection, carries with it a series of further  implications as will be described below, starting from the left-hand side of figure~\ref{cycle} and essentially working around in a clockwise sense.
      
      A possible objection to the proposal of joining up loose ends `3)' and `1)' by employing the perceived flow of time as the source of objective time parametrised by $s\in \rrr$ in equation~\ref{srrr}, and hence as the underlying basis of the physical theory, might be the apparent \textit{variable rate} of the subjective flow of time. There are several examples of this including the familiar phenomenon of feeling time to pass more slowly during periods of boredom while passing quickly on happier occasions. 
      Further, in the present moment a compression of perceived time regularly accompanies voluntary eye movements, or saccades, giving rise to the `stopped clock illusion' in which the second hand temporarily appears frozen on first glance. Such `chronostasis' is caused by a compensating mechanism of the brain retrospectively filling in the perceptual gap due to eye movement, effectively projecting the new visual field back in time by a noticeable fraction of a second~\cite{YHagg}.

         As well as dependence on changing sensory input, an apparent slowing of time in the present moment can also be perceived while preparing for an action.  This might occur 
         when a rapid motor response is required, such as in playing a fast moving sport or experiencing
          a sudden physical danger. In such circumstances accelerated visual information processing maximises the opportunity for corrective motor action and directly generates a perceived slowing of time~\cite{NHagu}.
         Metabolic and processing rates in areas of the brain such as the precuneus can be studied to determine the neural mechanisms associated with the subjective sense of how quickly time is passing.
         
         Mental time as constructed by the brain appears then distinct to physical time as measured by an instrument in that the former changes rate depending upon the internal brain state and external environment conditions and does not seem to coincide with the flow of time in the broader physical world.
          The fact that our perception of time is seemingly unstable in this manner, as modulated by ongoing tasks and less \textit{precise} than a mechanical clock, might then cast doubt on the employment of subjective time as the basis for equation~\ref{srrr} and the theory of generalised proper time.

      We note though that in physics to some extent we can adopt abstract mathematical structures to represent what we \textit{choose} them to represent, wherever that is found to be useful. On employing the real continuum $s \in \rrr$ here we are \textit{not} interested in \textit{actual} ``real numbers'' such as $s=1.5,\pi,7.83,\ldots$ etc. 
      (or in the rational numbers or even integers with which `time' is typically measured out 
       as for example in counting $1,2,3,\ldots$ seconds).
      Rather, we are interested in  subjective time in the specious present as experienced as a \textit{continuum} 
        as can be isomorphically linked with \textit{that} property of the mathematical real 
          continuum. Hence we are taking $s \in \rrr$ to represent what we \textit{mean} by the gapless continuum of progression in time as we experience it, for the purpose of allowing a detailed  analysis of this continuum property in mathematically precise terms. 
          In place of a \textit{formal} mathematical definition the initial motivation for employing the real number system is the immediate \textit{intuitive} concept of the continuum, with $s \in \rrr$ employed to express \textit{this} property which can then be rigorously mathematically explored.

          Time is ultimately not `measured' or `counted' by the numbers $s \in \rrr$ but rather by \textit{events} in the physical world, which itself  is here \textit{generated} using the continuum property of $s \in \rrr$  on employing the
          mathematical analysis
              summarised in equation~\ref{subrk}, implying the mathematical rules underlying the construction of the physical universe as represented in figures~\ref{cycle} and \ref{cycles}.
              With the full physical theory constructed upon this basis via generalised proper time the objective measured rate of `ticking' physical clocks is `fixed' by the necessary physical laws of such simple systems as carried along in the objective `inertia of the world'. It is then not surprising that the subjective rate of time can apparently vary, as noticed relative to simple clocks, since brain processes are far more complex and less rigidly repetitive. 
              
              This variable rate is also then not problematic, since the subjective flow of time can still be employed in connecting the loose ends on the left-hand side of figure~\ref{cycleb} and closing the circuit of the universal bootstrap as described for figures~\ref{mind} and \ref{cycle}, as long as it exhibits the crucial continuum property. We can also choose to associate actual numerical values of $s\in \rrr$  with a physical measure of objective proper time, such as indicated by a standard clock, utilising the freedom in the scale of real numbers, effectively calibrated as described for (\cite{Unifi} equation~13.3) and as we have generally presumed.
              
              There is an historical connection here with Hermann Weyl's view that mathematical knowledge might be founded in the intuition or insight presented to the mind, with in particular our immediate intuition of the flow of time and of motion capturing the essence of the mathematical notion of the continuous real number line, as he expressed in \textit{Das Kontinuum} (see for example~\cite{BellK} section~3).
              The question Weyl considered is whether such a mathematical construction provides a suitable representation of the temporal and physical continuum as it is actually experienced, with the focus upon the fundamental continuum of the immediately given experience of flowing time at every moment rather than underlying physical real world processes.
              
             This subjective continuum does not dissolve into `points'; that is, exact time-points or space-points, or any such `atomic' elements of duration or extension, are not given in experience. 
           Weyl noted that such points have no independent existence, but rather in the physical world they are only encountered in terms of values of a coordinate system. 
           Since definite real numbers, such as particular values for $s\in \rrr$, have no correspondence in terms of a direct intuition of temporal points, there is a potential discrepancy between the mathematical conception of real numbers and experienced time. Indeed, with the intuited continuum of temporal flow not grasped mathematically as a set of individual points or discrete `stages' there is a seemingly unbridgeable chasm between the two concepts.
           
           This led Weyl to repudiate any atomistic theory of the continuum, as influenced by the work of
            Luitzen Brouwer (\cite{BellK} section~3.2),  bridging this chasm by endowing the real number line with the perpetual fluidity of a `true' continuum, as compatible with the primordial intuition of time taken as an indecomposable whole that cannot be divided into separate fragments or a set of points. These developments foreshadowed the later emergence of smooth infinitesimal analysis, with the real numbers not generated from discrete elements but rather expressing this indecomposability of the intuited temporal continuum in a formal mathematical framework.
            By encapsulating the notion of infinitesimal quantities this also supports the development of physical theories that employ an analysis of infinitesimal elements. 
            
            It is precisely this possibility that is exploited in the present theory via the direct analysis of infinitesimal elements of time $\delta s \in \rrr$ in equation~\ref{sint} as expressed in the generalised form for proper time in equation~\ref{salpha}. Through the extraction of the local infinitesimal elements of 
            3-dimensional space a 4-dimensional spacetime continuum can be constructed, with the residual components providing the matter content, again as summarised in equation~\ref{subrk}. The resulting mathematical structures, indicated in the top part of figures~\ref{cycle} and \ref{cycles} then underlie a full unified physical theory, which, as based upon the properties of infinitesimal elements as representing space, time, and matter, is also motivated in a similar spirit as the unified field theory sought by Weyl (see for example~\cite{BellK} section~4.1).
           
              It is the regularity of the resulting physical laws, the `inertia of the world' alluded to above, that creates the \textit{illusion} of a fully objective physical world, out there apparently independent of \textit{our} existence and seemingly detached from any necessary subjective elements.
              However, there is no fully objective external physical world, and our innate perception in time and space does not require structures in such a pre-existing external world to \textit{latch on to}. Rather, the entire external physical world can be constructed \textit{out of} the subjective forms of perception in time and space alone, as based upon equation~\ref{subrk} and  within the implied constraints, which indeed ultimately determine the  physical laws and ensure their regularity and reliability.

      With respect to the `big picture' of figures~\ref{cycle} and \ref{cycles} the full 4-dimensional spacetime extent of the physical universe on the right-hand side represents `one thing' as a single link in the chain of the universal bootstrap, as alluded to above for figure~\ref{cycleb} in addressing loose end `2)'. 
      This emphasises the nature of equation~\ref{GRQT} as utilised to determine the entire geometric solution of a full-scale 4-dimensional spacetime as a single coherent entity. This entity is constructed \textit{entirely from} the flow of time, as entering into the right-hand side of figure~\ref{cycle} through the implied mathematical rules at the top of the figure, as incorporated into the internal structure of equation~\ref{GRQT}. The `block' picture, as central to the general relativistic worldview, is hence coherently amalgamated with the `dynamic' view with a central role for progression in time, as underlying the local evolution of the apparent states in the quantum theory limit, which itself here derives from the local degeneracy in the matter field
      composition of spacetime as described for 
       equations~\ref{GRQT}--\ref{phbarku}.

      This construction, as also described following equation~\ref{GRQT} with reference to a `river' analogy, avoids any `problem of time' and might rather be considered to represent a `theory of time' (\cite{Tflow} section~7). The position of the physical universe in figure~\ref{cycle} then resonates closely with the approach of amalgamating general relativity with quantum phenomena  described in subsection~\ref{boot32}, with the construction solely from elements of time key to this
       unification. 
      The local degeneracy of  matter field states underlying a common local external 4-dimensional spacetime geometry implies a vast number of possible solutions for constructing such a physical universe and correspondingly a vast number of ways to complete the circuit of the universal bootstrap in figure~\ref{cycle}. These `many solutions' for distinct physical universes are analogous to the
       bifurcating `many worlds'  proposed in one interpretation of quantum mechanics.
       We note though that here the `many solutions'  conception is \textit{not} an \textit{interpretation}, but rather represents the intrinsic nature of the theory.

       With a full 4-dimensional spacetime solution for equation~\ref{GRQT} constructed from elements of time alone, as derived from our innate subjective time perception,  we as `creatures of time' necessarily explore the universe in a  temporal sense. In being unable to detect the minute  local warping of the spacetime geometry that irreducibly accompanies elementary particle and quantum events, such as represented by the contour 
        sketched  in figure~\ref{sinter},   and not knowing `which' \mbox{4-dimensional} spacetime solution we are embedded within, we can never be sure `where we are going' at the microscopic level. Hence, for such laboratory experiments, it seems the best we can do is to invent the mathematics of wavefunctions and their collapse  to make  probabilistic predictions, albeit with inevitable  counter-intuitive implications and ongoing interpretational questions. 
      These issues can be resolved when quantum mechanics is seen as a pragmatic limit obtained on taking our temporal perspective from within a complete \mbox{4-dimensional} world,  which itself is fully compatible with a single general relativistic spacetime solution in the context of the present theory.
      This is also consistent with the observation, noted in the opening of subsection~\ref{boot32}, that quantum phenomena are of a limited scale and scope and only make sense \textit{within} a classically defined framework, which is here provided by the external world of classical general relativity.

    As discussed towards the end of subsection~\ref{boot32}, with reference to well-known quantum mechanical laboratory phenomena such as seen in double-slit or EPR experiments, an element of apparent `retrocausality' can be identified. This is implied in consistently constructing a full  4-dimensional spacetime extended solution for equation~\ref{GRQT} describing the external geometry and internal matter content. 
      From the context of the physical universe as a whole, and the right-hand side of figure~\ref{cycle}, there is a further significant degree of retrocausality from the cosmological perspective of our own 4-dimensional spacetime location, with the occurrence of the Big Bang and nature of the initial conditions apparently being conducive for our physical existence here and now in the universe. This relates in particular to the connection between the loose ends `2)' and `3)' described for figure~\ref{cycleb} above. 
      
       That is, for the necessary neural structures, for the link in the universal bootstrap at the bottom of figure~\ref{cycle},  to exist in the physical universe very special initial conditions for the 4-dimensional spacetime solution are required in order to create a physical environment suitable for such highly complex organic structures to have evolved within the universe within the constraints of laws of physics
       deriving from the mathematical rules at the top of figure~\ref{cycle}.
        Such initial conditions are typically postulated or contrived in many cosmological models. For the present theory the universe does not exist \textit{because} of a `special event' 13.8 billion years ago in the Big Bang, rather it exists within \textit{a particular instantiation} of the universal bootstrap \textit{since}
                full solutions for the circuit of figure~\ref{cycle} are possible. This solution demands the inclusion of sentient neural structures within the context of an apparent cosmological evolution, hence implying a seemingly retrocausal aspect, constraining the `initial conditions' of the universe, from our perspective.
       
       The notion of `retrocausality', whether for small-scale laboratory experiments or large-scale cosmological evolution, is itself relative to an apparent `arrow of time' linking causally connected events in the physical universe. In the present theory with the source of time, going into the left-hand side of figure~\ref{cycle}, being of a subjective origin a further implication for physics can be immediately noted. It is the generation of the temporal continuum  through the continuous \textit{directed} progression of subjective thought processes itself that \textit{drives} the apparent \textit{arrow} of time. More precisely, time does not have an `arrow' but rather \textit{is this continuous progression} (see also discussion in \cite{Tflow} section~6).
       
       This subjective causal progression, modelled by the continuum $s \in \rrr$ of equation~\ref{srrr} and mapped via the intermediate local quadratic form of equation~\ref{sxxxx}  into the  local  light cone structure of an extended 4-dimensional spacetime, supports the causal flow of the matter content, as deriving from the general form for proper time of equation~\ref{salpha} and objectively carried along in the physical world (\cite{Tflow} figures~7 and 8).
          This psychological causal asymmetry in the temporal continuum then takes precedence over other proposals for the possible source of an arrow of time, such as the increase in entropy. From the present perspective the second law of thermodynamics, and the non-decreasing evolution in the entropy of a system, is rather more passively an inevitable objective statistical law that applies \textit{in} time.

       In principle if the physical and thermodynamic environment were suitable, even without special initial conditions or an organised biological evolution, a momentary  `Boltzmann Brain' (see for example~\cite{Linde2,Page}) might potentially generate a subjective sense of time to complete a circuit
       of a universal bootstrap 
        similarly as for figure~\ref{cycle}. 
          (The possibility of an ordered state spontaneously appearing out of disorder illustrates why entropy cannot drive an arrow of time).  
       For a  very long-lived or temporally open-ended universe, such as for
       eternal inflation models or even  the
        standard $\Lambda$CDM  cosmology, it is difficult to assess the likelihood  of events of  almost vanishingly low probability to   occur given potentially infinite opportunity. This is the case for Boltzmann Brains, 
        arising from random fluctuations in thermal radiation or even more fleetingly from a vacuum state,
        accommodating the properties of conscious experience analogous to that of the 
        `ordinary observers' we presume ourselves to be.
        
        It is sometimes suggested that occurrences of such a configuration of matter in a high-entropy thermal equilibrium bath randomly generating such intelligent observers,  together with planets and galaxies, without passing through the familiar stages of the Big Bang and cosmic evolution, could be infinite per finite comoving   volume of the cosmos given infinite time. Further, the extended physical form and environment for such a 
        Boltzmann Brain observer may not be required since an intelligent mind might be generated with an 
          apparent  `knowledge'  merely simulating these elaborate surroundings and even 
            an imagined evolutionary past.  Such states would then vastly outnumber the finite production of ordinary observers in the same comoving spatial volume of the cosmos as actually evolving from an initial low-entropy Big Bang within a limited time frame. 
            
            Several arguments explaining why `we' can't be Boltzmann Brains have then been constructed, typically involving the nature of what is entailed in being an `intelligent observer' (see for example~\cite{Gott, Carr1}).
              For one thing,  even a `sentient' Boltzmann Brain would typically experience  far less coherent impressions than `we' are generally accustomed to.
    In the context of the present theory, in order to complete the circuit of the universe bootstrap 
    with a Boltzmann Brain, substituted into the bottom of figures~\ref{cycle} and \ref{cycles} in place of a neural structure, such an entity would be required to actively \textit{generate} an internal subjective sense of time and innate 3-dimensional spatial perception as well as have a sophisticated level of sentient \textit{engagement} in the external physical world.  
    These features, concerning the subjective temporal aspect in particular, may require a sustained 
    \textit{process} rather than a fleeting \textit{state}.
    
    While Boltzmann Brains might be problematic in the context of some physical models of the microscopic and cosmological scales, they do still require the underlying support of the physical world and may not be compatible with the constraints of an ultimate physical theory. In particular it may be that the physical laws arising from generalised proper time suppress, or do not allow, the generation of Boltzmann Brains,
   but do
    permit the evolution of ordinary observers with
       the properties of stable neural structures as incorporated into figures~\ref{cycle} and \ref{cycles} as required to complete the circuit of the universal bootstrap.

    With the large-scale cosmological picture for the present theory closely compatible 
    with the $\Lambda$CDM model, as reviewed in subsection~\ref{boot33}, there is no eternally repetitive or steady state but rather an exponentially expanding future evolution of the universe with ordinary matter becoming ever more diluted. This would then ever further diminish the likelihood of a thermal fluctuation generating a Boltzmann Brain structure. Also the physical properties of the dark energy vacuum state, composed of fluctuating dark sector
     $\{\Apls,\Mplu,\Amin\}$  fields states, are very different to that of the visible matter substrate of ordinary brains, and again unlikely to be capable of supporting Boltzmann Brain production.

        This then raises the question concerning exactly \textit{what kind of} physical structures and subjective elements of engagement in the world \textit{are} required to complete the circuit of the universal bootstrap through the bottom link in figure~\ref{cycle}. On the theoretical side speculation concerning a role for G\"{o}del's theorems, or a related or analogous mathematical structure, was considered in (\cite{Unifi} section~14.1).
        We there sought the mathematical correlate of the subjective experience of a momentarily  `\mbox{undecidable} proposition' such as   ``shall I pick up the pen or the pencil?'',  which has a close parallel in the studies of neural activity   for an individual choosing whether to
         first ``grab a bottle or a cup'', alluded to here in subsection~\ref{boot41} (with reference to \cite{chwww}). 
             Such decision making processes are here considered to contribute to our experience of a temporal progression and the generation of time itself, as an essential link in closing the circuit of figure~\ref{cycle}.

         Investigation into the mathematical properties needed to model this generation of a subjective sense of time might then in principle provide input to neuroscience in terms of identifying a parallel correspondence of these properties in a physical brain. Such a mathematical analysis might also  connect with the  artificial intelligence simulations reviewed in subsection~\ref{boot41} with reference to a Mind Time Machine~\cite{Ikeg,Ikeg2}.
   As described for figure~\ref{cycles} a further property required to construct the universal bootstrap   may be an innate predisposition to perceive the world in 3-dimensional space. 
   As alluded to in subsection~\ref{boot42} human visual perception is inextricably linked to our perception of time~\cite{chwww}, with the precuneus region of the brain playing a key role not 
      only in our perception of time but also in visuospatial imagery (see for example \cite{Cavan,Peer}). 
   This raises the further question of precisely \textit{how} such a subjective framing of spatial perception is generated by the  brain and might even be  implemented in the architecture of a machine.

            Given then the primacy of temporal and spatial perception, as properties of the brain, the  implications for neuroscience concern the  understanding of the overall structure of the brain and  the interpretation of its function and purpose. 
      In this manner a reciprocal interconnection between neuroscience and the theory of generalised proper time might be established, contributing to the interpretation of neural structure and activity in the former while further clarifying the conceptual foundations of the latter, with perception in time and space playing a key role for both. 
      This interplay between the physical theory based upon generalised proper time
      and the functions of neural structures  is key to completing the universal bootstrap
       of figures~\ref{cycle} and \ref{cycles}.

     An historically grounded mainstream  approach to neuroscience can be termed the `outside-in' perspective, based on first identifying preconceived intuitive cognitive concepts concordant with our experience and \textit{then} exploring how these map on to structures in the brain. From this perspective the brain's task is to perceive and represent the outside world, process the information and respond accordingly.  
      The alternative `inside-out' strategy~\cite{Buzsa} argues that the pre-existing connectivity and
       self-organised dynamics of the brain \textit{constrains how} it views the world, while generating actions and predicting the consequences  in order to make sense of the world. The aim is to investigate this 
       self-sustaining neuronal network architecture and its properties  and
        \textit{then} associate these with meaningful cognitive function
      to understand how such brain operations generate our cognitive faculties.
      
      This change in perspective, in beginning at the neuronal level, is particularly significant
       for questions relating to timing or temporal ordering (see also~\cite{Schar}).
       While motion and change can be directly experienced from optic or tactile flow, time itself cannot be assumed to be a pre-existing category the passage of which is detected by the brain.
            In fact the inside-out strategy may be key for the pertinent concepts of both space and time, for which we do not possess  any specific sensors and cannot track down a corresponding dedicated neural mechanism by employing an outside-in approach (\cite{Buzsa} chapter 10, \cite{BuzsL}).
            Spatial extent and temporal flow are not `re-presentations' of real entities out there in the world,  rather they are  two prominent concepts invented by the human brain from an   inside-out perspective as mental constructions that are essential for everything we do in the world.
       
       Any neuronal assembly network that can support a self-sustained sequential activation can potentially correlate with a succession of events or track the passage of time. Such an activation can be compared with an external clock but does not faithfully emulate physical clock time, rather the subjective passage of time can be warped and manipulated as we reviewed earlier in this subsection.
        Such self-generated, sequentially evolving neuronal activity, moving perpetually, is the default state of most neural circuits, even when there is no external sensory input.
    
        Anatomically, both the parietal cortex and the hippocampal system accommodate neuronal sequence generators that continuously bridge the gap between events to be linked  (\cite{Buzsa} chapter 10, \cite{BuzsL}).  As we have discussed, the posterior parietal cortex, containing the precuneus, is a key substrate for egocentric representation and our temporal and spatial engagement in the world.
             It can  be argued that most of the attributes of time and space can be supported in the brain by sequential neuronal assemblies and trajectories. Further, it is likely that the sequential organising and ordering of events whether in temporal or spatial terms can be represented by the same structures, neurons and mechanisms. Indeed, the experience of time is closely linked with experience of space, as related for example in perceiving motion.
             
         With space and time not \textit{sensed} by the brain, structures of sequential neuronal activation do not correspond to a \textit{passive} representation of perceptions of space and time. Rather these neuronal assemblies underlie the \textit{construction} of relationships and the \textit{generation} of our conceptions of space and time from a brain-centred inside-out perspective~\cite{BuzsL}.  
         This model-building role of brain mechanisms can then in turn help clarify the nature of fundamental concepts such as space and time.   
         While   their \textit{origin} is not independent of our engagement in the physical world, the notions of space and time can be abstracted from our experience in the world and studied as geometric and mathematical entities in their own right. 
         In so far as these conceptions  are employed in physics, neuroscience may then also help address the associated conceptual issues  regarding the laws of physics as apply to the external world.
             
    A more typical approach to neuroscience  might concern a two-way correspondence between mind and matter, with `primary qualities' such as extension and duration \textit{input} to the brain via sensory data from objective physical structures while `secondary qualities' such as the subjective sense of green, although dependent on sensory information, are essentially projected from a construction of the mind onto the object concerned. Taking the inside-out approach literally the primary qualities of extension in space and the flow of time might \textit{also} be seen as constructions of the mind and projected outwards as constraints on the construction of the \mbox{physical world itself}. With the underlying brain structures embedded in the physical world, these projections lead to the link in the bottom-left of the universal bootstrap in 
figures~\ref{cycle}~and~\ref{cycles}.
    
     That is, for the present theory and the universal bootstrap the place of the neural structures in figure~\ref{cycles} is essentially that of an extreme or limiting case of the 
     inside-out approach when it comes to the origin of space and time.
    While it is well established that the brain must internally build a model of the external world in order to navigate in that world, here we are essentially proposing  that ultimately the  `model' constructed \textit{is} the physical world itself.
    It is perhaps already remarkable that the brain is able to  build a representation 
        \textit{of} the physical world;  it is  one further step to propose that this is not \textit{just
         a representation} but rather the world itself can be \textit{constructed} through the constraints imposed by the subjective forms of space and time originating in the mind.
     Since the brain's model of the world and the external physical world are then not two different things, this provides a unifying picture directly linking our `internal' conception with the world `out there'.

      While the present theory of generalised proper time is embedded in the top half of figures~\ref{cycle} and \ref{cycles} and neural structures are embedded in the bottom half, both are essential in closing the circuit of  the universal bootstrap as a whole. 
        In terms of the link between objective physics and subjective experiences, in connecting matter and mind the central brain region of the precuneus is in this theory the analogue of the pineal gland  as the seat of the interaction between  body and soul in the $17^{\mathrm{th}}$ century worldview  of Descartes.
      Here the subjective forms of time and space, accommodated by the neural structures associated with the precuneus, through the theory of generalised proper time generate the objective structures of matter in spacetime, including the precuneus itself.

       It is not so much that we gather sensory input and \textit{then} interpret what we see
        in terms of time and space, rather the physical world must \textit{always} and \textit{everywhere} conform to these subjective forms of time and space. Constructing the world from the infinitesimal level through generalised proper time as described for equations~\ref{sint}--\ref{GRQT}, and summarised for equation~\ref{subrk},  ensures conformity with these forms on all scales. Forms of perception in time and space are \textit{all that is needed} to construct the physical world itself. This approach is then  reminiscent of the \textit{esse est percipi}, `to be is to be perceived', philosophy of Berkeley, except here a well-defined physical theory and the fundamental laws of physics are \textit{derived} from these constraints alone.
        
        Time and space are not pre-existing entities  `out there', as an independent background to the matter content of the physical world,   that we perceive and learn about \textit{from experience}.
        Rather they are  subjective \textit{forms of experience},  with the subjective flow of time as the \textit{origin} of the objective progression of time as mathematically represented by the continuum of real numbers $s\in\rrr$.
  Through analysis of infinitesimal intervals of time  $\delta s\in\rrr$, as permitted by the continuum property, the generalised form     $(\delta s)^p$  incorporates both a quadratic substructure
   as a substratum for our  innate predisposition to see a world in space and also residual components as  the \textit{source} of its matter content. 
         In one sense this conception  is akin  to the
         \mbox{\textit{a priori}}  forms of time and space proposed by Kant in being \textit{prior}  to any experience in the physical world, although here there is an explicit and key \textit{source}
        of these temporal and spatial forms  
          rooted in neural structures \textit{within} that world itself.
         
           The mathematical structures obtained from these forms for time and space in this manner, as summarised for equation~\ref{subrk}, provide the basis for the laws of physics as described in detail in section~\ref{boot3}, with the full physical theory encompassing all the structures of matter in the world, including that of the physical brain.
    With time, accompanied by the closely related generation of space, 
      ultimately sharing a common origin in 
      the structures of  sequential neuronal activity described above
       and  as experienced together, these neural elements provide a critical link in the universal bootstrap of figure~\ref{cycles}.

     According to this worldview   brain structures accommodating these subjective perceptual forms  are not just a chance feature in the universe but an irreducible link required for the whole system, and anything in it, to be realised.
     Indeed, from the perspective of the universal bootstrap the \textit{primary role} of neural structures is to generate the subjective forms of time and space to facilitate the closure of the circuit in figures~\ref{cycle} and \ref{cycles}.
          A more conventional interpretation might be that the prime function of a brain is to optimise the survival prospects of the associated individual in the context of evolutionary biology.
        Such functions, that provide a survival advantage in a given habitat, are here considered to fall into place in providing a secondary supporting role as consistent with, and following in the wake of,  the prime directive.  
         The universal bootstrap provides a bigger picture in which there is a prime and necessary role for the brain in generating subjective perception  in time and space.

        We also note that
        evolutionary survival does not seem to necessarily imply sentience, which is perhaps most evident in the case of  plants in completely lacking any neural structures. 
         Even in humans the cerebellum, or `little brain', at the base of the rear of the brain, contains four times more neurons than the rest of the brain combined but does not seem to play a direct role
          in subjective consciousness.  The neural circuitry of the cerebellum is extremely uniform with
          feed-forward, largely independent, parallel modules controlling motor and cognitive systems. 
         While performing key functions for survival and containing the most dense structure of neurons, the apparent lack of conscious awareness associated with the cerebellum may relate to the lack of 
          complex feedback loops~\cite{Koch}. This may in turn provide a clue for the question  of what  exactly  is required for sentience to arise.

          It is indeed a major unresolved question for neuroscience  to comprehend the scientific relation between the objective physical world and our subjective experience of sentience, and to define the precise nature of the latter. Whether special neurons or specific connections in certain brain regions are required to generate subjective awareness is an open issue, although as noted in the previous two subsections with reference to~\cite{Koch}  the `posterior hot zone' is known to be the central area of interest (see also~\cite{Koch2}).
  Further, several fundamental theories of consciousness have been posited, with two outlined below
  (\cite{Koch}, \cite{KHoel} section~1 and references therein).
   Given the intrinsically  subjective aspect of the problem and  the means accessible to investigate such proposals there is also an open question of the extent to which theories of consciousness might be 
        satisfactorily tested (see for example~\cite{KHoel,DoerSH}).
  
   In the `Global Neuronal Workspace' (GNW) framework consciousness emerges from a sudden self-amplifying process leading to a brain-scale pattern of neuronal activity over dispersed networks with long-range interconnections. On creating this global access to neuronal systems consciousness acts as a 
   gateway  mediating from  sensory input through to consequent memory recall or action execution,
   broadcasting the contents of this spotlight of attention  to a 
      multitude of cognitive brain systems. The GNW framework is well suited to computational modelling and in principle could  be implemented on a computer. 
      
    On the other hand for  `Integrated Information Theory' (IIT), motivated by the huge variety of experiences and the unity of each, consciousness corresponds to the capacity of a system to integrate information, as generated by a complex of elements. With the integration corresponding to the unified structure of a specific experience the integrated information cannot be reduced to independent components. The \textit{quantity} of consciousness is proportional to the amount of causally effective information that can be integrated, suggesting a graded level of consciousness as might apply for infants or animals. In fact for IIT a  level  
       of internal conscious experience arises for any complex interconnected mechanism encoding such an integration, but must be built into the structure of the system and  cannot be simply
       `computed'.

      Ultimately the architecture of living systems and neural structures must conform to the laws of physics, just as for any other system of matter. The physical theory developed in this paper deals primarily with the elementary structure of matter on the microscopic scale and implications for the cosmological scale, as traditionally associated with the fields of particle physics and cosmology respectively.
       A third intermediate scale of significant interest is that at which highly organised complex structures of matter can be found as a basis for emergent phenomena such as subjective experience. A physical theory is then still required to explain under what circumstances such sentience can arise in any such physical system.

        Here we note that such experience is inextricably connected with a subjective sense of the flow of time in the ever-present moment `now' as identified with the specious present.   
        The question of how the brain generates this sense of a continuous temporal flow is an active field of neuroscience, as reviewed in subsection~\ref{boot41}, and points to the precuneus region of the brain as playing a central role. As also noted there, the precuneus is anatomically the hub of the default mode network as located in the posterior hot zone, as consistent with the notion that subjective time is central to conscious experience.  
        
        While the question of \textit{how} neural mechanisms can generate subjective time and consciousness are a matter for neuroscience, the present physical theory does offer an account of \textit{the necessity} for the mind to generate time. As an irreducible link in the universal bootstrap of figure~\ref{cycle} subjective time plays a pivotal role which is hence required for \textit{anything}, and in particular the physical world at large, to exist.  
         Since such \textit{subjective} elements, incorporating time as well as  space perception in figure~\ref{cycles}, are essential to close the circuit of the universal bootstrap it could be suggested that any complex system of matter capable in itself of closing the loop in such a manner has \textit{sentience}.  As an alternative to the GNW and IIT proposals reviewed above this might then itself form the basis of a theory of consciousness, defined in terms of its role in the context of the universal bootstrap.  Through the mutually interdependent interplay of the universal bootstrap the subjective elements of consciousness are just as fundamental as the objective elements of physical laws.
         
         As a corollary of this discussion, for either a `Boltzmann Brain' or artificial intelligence system to be sentient  would presumably then also require internal subjective powers of perception in time and space    and a sufficient level of engagement in the world   as substituted into the bottom part of figure~\ref{cycles} to achieve the closing of the universal bootstrap.
    As alluded to earlier this might prove impossible for a Boltzmann Brain given the limitations of the laws of physics and the random nature of such an entity.
     However, in requiring more than ever-increasing raw computing and processing power alone, 
      such a perspective might  provide insight into the practical possibility of building  a machine, whether as an extension of a Mind Time Machine or more generally, capable of internal sentient experience through a stand-alone  ability to close the circuit of the universal bootstrap.

       The whole system of figure~\ref{cycles}, with  subjective time and space having a source in physical brain structures as in turn embedded in the physical universe which originates through the constraints of the forms of subjective time and space,    revolves around an apparent `circular argument'. However, this is a `circular argument' that sustains itself from within in a positive self-consistent manner. This interplay of `mind and matter', between the subjective elements associated with sentient experience and the objective structure of the brain embedded in the broader physical world,  in providing mutually supporting roles with neither taking priority over the other, also has a fully scientific basis as we have described.
      The overall system is also rich in philosophical connections as we have partially alluded to above.
     
     Other theories and models typically have a foundation in `matter' alone, as based upon a set of \textit{postulates}, leaving the \textit{origin} of the structure of matter and the issue of `why there is something rather than nothing' as a seemingly intractable open question, as associated with the first loose end described for figure~\ref{cycleb} in the opening of this subsection. 
     Indeed, paraphrasing all three loose ends, from the purely materialist viewpoint the fact that matter exists \textit{and} that the physical universe was created `in the first place' \textit{and} that it has led to the evolution of sentient beings is indeed truly miraculous.

      A yet further question from the perspective of the standard view of the physical universe, as a detached entity in itself on the right-hand side of figure~\ref{cycleb},  might concern the apparently  astronomically unlikely probability for it to be `now', as well as `here',  for the seeming benefit of any particular sentient individual in the context  of an apparently objective and potentially infinite in space and eternal in time 4-dimensional spacetime cosmology. That a chance Big Bang event followed by a period of cosmological and biological evolution, as permitted by postulated physical laws,    should happen to  \textit{result in} such a sentient individual, with a particular genetic heritage situated within a particular environment,  would seem unreasonably fortuitous.

      However, on giving equal priority to the `mind' perspective,
       as is the case on placing the physical universe itself \textit{within} the context of the universal bootstrap in figure~\ref{cycles}, any sentient being will necessarily subjectively encounter the world from  a situation `now' and `here' as an essential characterising feature of \textit{that} experience (similar in spirit to the `life is absurd' conclusion of  existentialist philosophers, although perhaps less pessimistic). With this subjective engagement in the world required to close the circuit of figure~\ref{cycles}, and for anything to exist at all,  the locations of such `spacetime events' associated with a sentient observer are hence in this sense of greater significance than the rest of the 4-dimensional spacetime cosmos.   
       While in principle only one such sentient `event' is required to close the circuit, the conditions of a physical universe required for its generation  will likely generate very many (\cite{Unifi} figures~14.6 and 14.7  and discussion).
       
       The retrocausal  elements discussed earlier in this subsection,  from the scale of laboratory quantum physics experiments to cosmological initial conditions and evolution, also apply for the biological evolution and life history leading to any given experience of any particular sentient individual.
       The full 4-dimensional spacetime solution for a physical universe on the right-hand side of 
        figure~\ref{cycles} also naturally encompasses  the physical neural structures underlying any
         subjective temporal experience. In turn the apparent `retrocausal' phenomenon of physical
          cerebral activity preceding the conscious will to act by of order a second, as initially observed by Libet as reviewed near the opening of subsection~\ref{boot41}, is consistently subsumed within the system of the universal bootstrap. That is, the universal bootstrap opens up capacity 
      for the compatibility of freewill and the laws of physics from the dynamic interplay of the subjective and objective elements it entails (\cite{Unifi} section~14.2).

     This focus upon the role of sentient beings in the physical world, with the associated subjective elements providing an essential link in the overall functioning of the universal bootstrap in figure~\ref{cycles}, is highly `anti-Copernican'. 
     On the other hand the 4-dimensional spacetime physical universe is still vast and our location within it, other than the basic anthropic selection of habitability, is unprivileged from the  broader perspective of large-scale homogeneity and the cosmological principle.  
     A degree of anthropic selection freedom may also be available as associated with the details of the symmetry breaking of generalised proper time, including the stable vacuum value for $\vert \bvf(x) \vert = h_0$ discussed in section~\ref{boot3}, as might account for the apparent fine-tuning of physical parameters required for the  support of carbon-based life forms
       (see for example~\cite{Tflow} section~5).  Nevertheless, 
     there are still presumably vastly `many solutions' for \mbox{4-dimensional} spacetime worlds, as consistent with the local degeneracy in solutions for equation~\ref{GRQT} as noted earlier in this subsection, accommodating the subjective elements required to close the loop of the universal bootstrap in figure~\ref{cycles}.
     
     However, most other theories \textit{only} have the Copernican elements, with nothing special about \textit{our} presence in the vast or infinite universe, which even itself may be only one component of a  `multiverse' containing a potentially limitless number of other physical worlds that are essentially independent of ours and predominantly `lifeless'.
      For the approach described in this paper our presence in the universe is \textit{essential} to the question of the completeness of the theory, through the overarching universal bootstrap of figures~\ref{cycle} and \ref{cycles} as described in this section.
       We assess how other theories have broached the foundational issue of completeness  in the following section.


\section{Alternative Approaches to Foundational Questions}
\label{boot5}
\vspace{-1pt}

    In this section we make some comparisons with various points of view relating to the question of `why there is something rather than nothing' as adopted in other theories.  Examples that can be compared directly with the self-sustaining cycle of the universal bootstrap in figure~\ref{cycle}, as further reinforced in figure~\ref{cycles}, for the present theory are rare. However, there is one well-known significant case in the shape of John Wheeler's `participatory universe' 
    (see for example~\cite{Wheel1} figure~1, \cite{Wheel2} figure~1, \cite{WheelW},  \cite{MTZ} figure~3 and discussions therein) as we first assess  here.  

   Influenced by Bohr's philosophical views on quantum mechanics (associated with the Copenhagen school as alluded to in subsection~\ref{boot32}), and encouraged by empirical success in relation to Bell's theorem, Wheeler essentially turned the measurement problem on its head and took the implications of the act of measurement as the ultimate basis for any attempt to actually derive all of quantum theory itself. Such a basis could be summarised by the assertion that `no phenomenon is a phenomenon until it is a recorded phenomenon', emphasising the role of the observer~\cite{MTZ}.
   
   This perspective was motivated by Wheeler on noting that the possible range of future outcomes is determined by `participator' selection of what to measure, 
    as associated with the quantum complementarity in the choice of questions we can ask of the world
    through the setup of apparatus 
   as exemplified by  `delayed-choice' laboratory experiments.
   On extrapolating the consequences to the largest scale, he proposed that the universe as a whole evolves to generate `observer-participators' that in turn, through technology and telescopic instruments, observe the early universe thus imparting a `tangible reality' upon it. The resulting `self-excited circuit', or Wheeler's ``U'', is depicted in figure~\ref{WheelU}.

\begin{figure}[htbp]  
\centering
\leavevmode
\includegraphics[width=11.0cm]{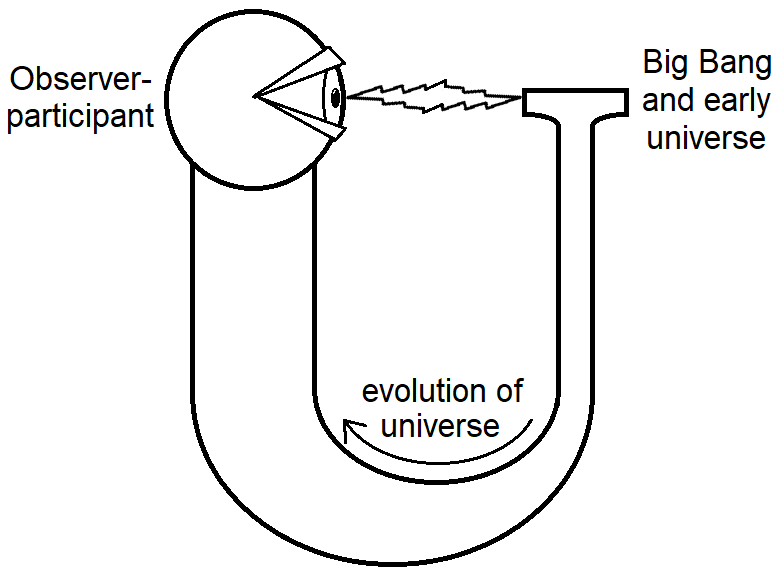}
\caption{\setb
   Depiction of Wheeler's ``U''. The self-excited universe proposed by Wheeler  is brought into being by acts of observer-participators whom are in turn brought into being through the evolutionary structure of the universe (\protect\cite{WheelW}, \protect\cite{MTZ} figure~3).
   }
\label{WheelU}
\end{figure}      

   In this picture the universe itself is hence created through a radical far-reaching act of `retrocausality', of a similar nature to that inferred in delayed-choice experiments on the laboratory scale, running counter to a more conventional notion of strict causality. 
 The actual laws of physics governing the evolution of matter and the universe are themselves proposed by Wheeler to derive statistically from a large number of \mbox{lawless} events, namely elementary acts of observer-participancy, effectively generating `law without law'~\cite{Wheel3}. He speculated that the basic building blocks of the universe could be information, correspondingly coining the phrase `it from bit', generating the properties of the universe that would be habitable for observer-participators~\cite{MTZ}.

  The elementary acts of observer-participators were also taken to transcend the concept of time itself.
  This `timelessness'  view of Wheeler was partly motivated by the presumed breakdown of spacetime, and   the continuum of time itself, on the Planck scale,
   with notions of `before' and `after' losing meaning for intervals less than of order
    $10^{-43}\,$ seconds~\cite{Wheel2}.
    If the concept of time fails on the Planck scale it would seemingly necessarily fail everywhere.
    The directed progression, or asymmetric arrow, in time was hence considered to be a secondary and derived concept. 
    We also note that while the incompleteness of quantum mechanics is inherited by the
    $S$-matrix bootstrap, the emergence of time is a perspective shared by the cosmological bootstrap, with both of these features discussed in section~\ref{boot2}.

     This secondary nature for time with respect to Wheeler's ``U'' in figure~\ref{WheelU} marks a major difference from the self-contained universal bootstrap of figure~\ref{cycle} for the present approach in which time plays  the primary and most essential role for the physical theory. For the new theory the continuum of time, and of 4-dimensional spacetime, holds down to arbitrarily small intervals. This is the case even below the Planck scale, with no quantisation of gravity or spacetime itself as
      reviewed in subsections~\ref{boot32}~and~\ref{boot33}.
     There are also several other significant differences as we discuss in the following.

     While for Wheeler's ``U'' the quantum mechanical act of measurement plays a primary and essential role in figure~\ref{WheelU}, here for the new approach  the entire formalism of quantum theory is a secondary aspect of the physical universe depicted on the right-hand side of figure~\ref{cycle}, derived in the appropriate  limit as applies in the  laboratory environment and without assuming \textit{any} quantum postulates as input, as also described in subsection~\ref{boot32}.
   For Wheeler the questions of `Why quantum?' and `Why existence?' are both important and intimately interrelated~\cite{Wheel1}, while for the new theory the `quantum' is just one feature within the physical world for which the existence mechanism is provided by the universal bootstrap in figure~\ref{cycle}.
   Further, for \mbox{Wheeler's ``U''} no account is given of the explicit forms of matter as empirically observed, while structures of the Standard Model and suitable dark sector candidates \textit{can} be directly derived  
   at the elementary symmetry breaking level for generalised proper time as reviewed in subsections~\ref{boot31} and \ref{boot33} and as represented by the mathematical rules in the top half of figure~\ref{cycle}.
   
   On the other hand, there are also notable similarities between the universal bootstrap of the present theory and the self-excited circuit of Wheeler,  with respectively figures~\ref{cycle} and \ref{WheelU} both exhibiting a closed cyclic structure.
   The inevitable need for some form of circuit, as opposed to a foundation in `tortoises all the way down',  is discussed for example in (\cite{Wheel1} page 313 -- section on `Why a Circuit Rather than a 
     Foundation?').
    A role for retrocausality in underlying the Big Bang and initial conditions of the early universe has also been described for each of the two circuits   compared here, in the previous subsection for figure~\ref{cycle} and in the present section for figure~\ref{WheelU}. In both cases there is a connection with the retrocausality implied in laboratory quantum systems, which for  Wheeler  is conceptually quite literal as
    noted after figure~\ref{WheelU}.
    
      For the present theory the apparatus and  phenomena of   
      the double-slit, EPR and delayed-choice experiments  for example are all enveloped under
      a single extended cosmological solution for 
     the \mbox{4-dimensional} spacetime geometry through equation~\ref{GRQT}, with
     elements of retrocausality in these laboratory environments   implied by the consistency of this solution considered \textit{objectively}. On the other hand the element of apparent retrocausality constraining the initial conditions on the cosmological scale is required to ensure the extended solution for the physical universe  accommodates  neural structures capable of supporting the  \textit{subjective} component needed to  complete the universal bootstrap via the bottom half of figure~\ref{cycle}.

   Perhaps most significantly, there is an irreducible role for a sentient observer and subjective elements in both of the  self-supporting systems of figures~\ref{cycle} and \ref{WheelU}. 
    For the participatory universe the apparent incompleteness of quantum mechanics, as encapsulated  by the measurement problem and as alluded to in the second paragraph of subsection~\ref{boot32},  motivates 
     emphasising the 
     distinction between the subjective `probe' and object `probed' (to the left and right respectively in figure~\ref{WheelU}) and their interplay as key to the system. For the universal bootstrap it is the distinct roles and mutual interplay of
    subjective `mind' and objective `matter' (towards the left and right in figure~\ref{cycle}) that sustains the complete system, as elaborated towards the end of the previous subsection.
    
   \pagebreak
    
     With reference also to figure~\ref{cycles} we have also discussed how the importance  for the universal bootstrap of subjective temporal \textit{and} spatial perception as the primary functions 
     supported by physical brain structure and activity could establish a significant interdisciplinary connection with contemporary neuroscience and psychology. Indeed such a link can be made for example with both  explicit studies of the precuneus region of the brain and the perspective of the `inside-out' strategy
     for neuroscience
      as reviewed in section~\ref{boot4}.

      Examples of this kind are rare in fundamental physics, although a parallel of this particular feature of the present theory can be seen in an application of the wavefunction collapse model proposed by Roger Penrose and collaborators~\cite{PenH}. 
   In that model for the `objective reduction' of the wavefunction the key role played by gravitation~\cite{Pen1} takes place on the scale of sets of `microtubules' within brain neurons. Computations in these neural structures, associated with `orchestrated objective reduction' of the wavefunction,
       are then proposed as the source of our sentient awareness. In turn 
      these biomolecular processes and the associated phenomenon of consciousness 
       then plays an inherent central role in the laws of the universe operating at the  micro-structure level.
       
       The present theory also adopts an approach of `gravitising quantum theory' (as noted near the opening and closing of subsection~\ref{boot32}), with gravity playing a central role.
       However, while the approach of Penrose deals with the 
        wavefunction and an explicit mechanism of reduction as a means of 
        providing a `completion' for quantum mechanics, here there are no such wavefunction 
        constructions
         but rather the entire formalism of quantum theory is obtained as a limiting case 
         subsumed under the umbrella of classical general relativity 
        as  described for equation~\ref{GRQT} in subsection~\ref{boot32}. 
    Further, here quantum theory itself is \textit{not directly} related to brain function, but is rather a property of the physical universe no more dependent upon a sentient observer than any other physical structure or observables.
   As well as this difference in the objective physics, for the present theory the subjective elements also have a more significant and global role in the possibility for the existence of the entire physical universe through the universal bootstrap of figures~\ref{cycle}  and \ref{cycles}.

    The role that the mind does play in the orchestrated wavefunction collapse for Penrose partially relates to  loose end `3)' of figure~\ref{cycleb} and the subsequent discussion, in terms of understanding the  significance of neural structures for the universe. 
     The same model of Penrose has also been placed within the context of a `conformal cyclic cosmology'~\cite{Pen2}, proposed to account for the special low entropy conditions of the \mbox{Big Bang} and the role of the second law of thermodynamics in the universe on the largest scale. Within that model the origin of `our universe' 13.8 billion years ago is seen as deriving from the ancient remnants of a \textit{previous} aeon, as part of a perpetually cyclic chain. 
     Such an approach partially addresses loose end `2)' in figure~\ref{cycleb}, by absorbing the event of the Big Bang at \textit{our} cosmic time $t_c=0$ within an infinite regression of earlier epochs and corresponding `universes',   `pushing back'  the ultimate origin question indefinitely.

    The same can be said for other cyclic cosmology approaches.
    There are in fact numerous `oscillating' or `bouncing' cosmological models that are motivated in a large part by the desire to avoid a Big Bang spacetime singularity at  $t_c=0$, typically  through the construction of a mechanism that prevents the cosmic scale factor from correspondingly vanishing
    (for example~\cite{Maeda,WanZ,CabLR}).
    There are also more symmetric cosmological frameworks with the universe effectively expanding in two opposing temporal directions from the same `Big Bang' state; with a corresponding bidirectional `arrow of time' emerging in an intrinsic non-entropic manner from a `timeless' basis  in complexity (\cite{Barb} figure~4), or with a `mirror universe' as an image of our own reflected in the Big Bang, acting as a spacelike mirror, as motivated in part by the simplicity of the  $t_c \simeq 0$ conditions and by the  apparent empirical consequences (\cite{BTuro} figure~2). 
    Neither  orchestrated objective reduction nor any of the above cosmologies  are `bootstrap' models. Further, such cosmological models incorporating an eternal past as well as an eternal future do not provide any input to loose end `1)' in figure~\ref{cycleb}, regarding the source of the \textit{foundations} for the theory itself.

    A self-contained cyclic structure \textit{is} described and pictured in (\cite{Pen} section~1.4 and figure~1.3) in which Penrose considers the interconnections between three different `worlds':
    Mathematical $\to$ Physical $\to$ Mental $\to$ back to Mathematical; with each world emerging or created in entirety from some of the elements of its predecessor in the circuit. The three connections between the three worlds represent three `mysteries', with Penrose suggesting that a major revolution in our understanding of the basis of physical reality is required, and will likely be needed before the nature of the mind and mental processes can be fully understood.
    While not representing a specific theory or model in itself, the links of such a circuit
        (\cite{Pen}  figure~1.3) closely map onto the stages of the universal bootstrap in 
        figure~\ref{cycles} here, which \textit{does} accommodate a fundamental unified physical theory as based on generalised proper time.
     A self-supporting system like the universal bootstrap is generally not attempted in other theories and  unification schemes, which typically
    do not propose a closed circuit analogous to figures~\ref{cycle} and \ref{cycles} but rather leave the question of `why there is something rather than nothing' completely open. 
   In the following we return to consider other views towards this most fundamental of foundational questions.

    One approach would be to adopt the view that \textit{all} science, including fundamental physics and unification schemes, is simply about \textit{modelling} empirical observations with a mathematical and descriptive formalism, such that  the issue of ultimate foundations is not even raised. In a related vein perhaps the most standard `answer' to the question concerning the source of the basic entities for any physical theory or scientific description of any aspect of the universe is that `we have to start somewhere'. That is, the foundation consists in a set of basic entities or structures, typically of a form of matter in space and time, that are \textit{posited} as `brute facts' about the world (see for example \cite{Carro}).
   This direct  postulation of the foundations of a theory has been very much the prevalent approach, and as a perfectly reasonable basis for almost all science has proved hugely successful. However,  for an ultimate unified framework in physics with the desired broad 
    explanatory power such a \textit{blind spot} in the need to state a set of brute facts at the very \textit{core} of the theory is unsatisfactory, 
     inevitably raising the question as to whether  more  can be done to establish a
        \textit{complete}  theory.
     
     In this context perhaps the most common partial justification for the basic elements of a physical theory   is in terms of a notion of `mathematical beauty'.
     The aesthetic appreciation of `beautiful theories' was for example an inspiration for the origins of string theory in the 1970s (\cite{Schw2} sections~5 and 6) where it continues to provide a significant motivation half a century later (see also the general discussion in~\cite{Struct} section~5).  
     This might involve taking the position that the mathematics used in physics, particularly at a fundamental level, \textit{should} be aesthetically beautiful, or even more robustly that some mathematics is so beautiful that it is compelled to exist in a physical form.
     The hope would then be for this standpoint to assist in identifying the basis for the       
     `Mathematical $\to$ Physical' link by focussing upon such a subset of the mathematical realm 
     as potentially significant for our physical world.
     
     The view might also be adopted that 
        \textit{all} self-consistent mathematical structures are in some sense realised in a `physical' existence, somewhat akin to the realm of `Platonic Forms',
        with an element of weak anthropic selection for a world supporting self-aware 
         substructures such as manifested in ours~\cite{Tegm1,Tegm2}. 
   Whether or not the particular `mathematical universe' within which we ourselves reside is considered more or less `beautiful' compared with others in a vast multiverse of possibilities, it is then  the self-consistency that underlies its existence.
   However, if \textit{all} such mathematical structures can be considered physical then, even with the anthropic selection effect, this clearly does not help greatly in pinning down the elementary theoretical basis of  \textit{our} universe.

    Neither the postulation of brute facts nor the appeal to mathematical beauty are strictly `scientific', attaining their justification at best as provisional or heuristic guides to theory construction.  
  The lack of a scientific answer to the question of `why there is something rather than nothing' offered by most unification schemes might risk rendering the question as presumed beyond science. This difficultly in providing an ultimate foundation could have even stigmatised the question itself with a somewhat `fringe' or `taboo' status, particularly given that many `unscientific' proposals \textit{have} been made to address the issue. However, it could be argued that it is actually unscientific to assume that any question relating to the basic functioning of the  physical world should necessarily be perpetually beyond our comprehension, 
  rather than at least attempting a scientific account.

   As described for figures~\ref{cycle} and \ref{cycles} throughout section~\ref{boot4} all of the links in the universal bootstrap are fully scientific in proposing an ultimate basis for the existence of the universe.
   The fact that it is possible to frame an argument in the context of a unified physical theory based upon generalised proper time, with a corresponding basis in the continuum of time, to account for the ultimate foundational question 
   through the system of the universal bootstrap 
   presents in itself a significant  advantage over most other unification schemes. 
   As an alternative to an appeal to brute facts, beauty or even `naturalness', or to any kind
    of \textit{mystical} elements, there is no 
    substitute for a \textit{rational} scientific argument subject to the demands of reason.

   Regardless of whether this speculation is ultimately on the correct path, and there is no insistence that this is necessarily the case, the theory stands on the extent to which the traditional goals of unification in physics are achieved as reviewed in section~\ref{boot3}. These successes
   in the ability to directly account for the elementary structure of the physical world,  
     corresponding to the development of the upper half of figure~\ref{cycleb}, themselves compare favourably with the other theories.
     On top of this the new theory provides a means of addressing the \textit{three} otherwise seemingly intractable loose ends of figure~\ref{cycleb} in the shape of the \textit{one} big picture of the universal bootstrap, for which the scientific case has been presented.
    This potential closing of  the circuit as described for figures~\ref{cycle} and \ref{cycles}, in a manner that is both not obviously flawed and open to further investigation,  reaches into an area for which  most other approaches offer no   account, and hence in the very least represents a significant bonus feature for the new theory.


\section{Conclusions}
\label{boot6}

       There is a perspective on the aims of theoretical physics, alluded to in the previous section, that the physical universe functions as a collection of disparate phenomena, involving various systems or different scales, that can be described at best by a broad collection of models, involving a range of mathematical frameworks. Each model will be individually motivated by the physical phenomena it is constructed to represent, with the ultimate goal, and fate, of physics to establish an optimal set of such models.
   The alternative view is that \textit{all} physical phenomena are governed by a single fundamental physical principle, with the above empirical models and associated phenomena serving as a provisional guide as to what {\bf {\textit{the}}} ultimate unifying physical theory might look like and what it should explain.
  This latter view is taken for the present work.
  
   One approach to uncovering such a unification scheme would be to start `from the outside' with a significant subset of existing phenomenological models and try to chip away to reveal common features, in an attempt to gradually expose the underlying unified theory.
   As employed for example in direct attempts to quantise gravity
   this perspective has not proved very successful. 
    Another approach would be to begin `from the inside' with an underlying conceptual motivation, that is ideally simple, unique and conservative as consistent with the ambition of unification, and build up the theory from within aiming to assimilate the empirical models and associated phenomena with
    accumulative  ratchet-like progress.
   Setting out with a well-motivated basis in generalised proper time this latter strategy  has been adopted  in developing the present theory in one-by-one pursuing the main outstanding questions in fundamental particle physics, quantum theory, gravitation and cosmology. The accumulated successes achieved, as reviewed in section~\ref{boot3},  compound the justification for this approach to unification
   and provide an overall compelling case.
 
   Towards the end of his lecture on `Theory, Criticism and a Philosophy' at a conference in 1968~\cite{Bethe}, Heisenberg recalled an episode from his childhood in the early $20^{\mathrm{th}}$  century.
  In attaching the cover on a wooden box he was constructing to accommodate books and so on, the young Heisenberg hammered down the first nail hard, only to find that the cover had become misaligned with all  the banging. His handicraftsman grandfather, watching on, recommended removing the first nail and starting again -- this time gently tapping in each of the nails a little bit at a time until the cover was held squarely in place and aligned overall  before striking all the nails down more firmly into the wooden box.
  
  This then was Heisenberg's metaphor for how best to proceed in working on problems in theoretical physics. It is advice to have the big picture clear in mind \textit{before} hammering down too much to fix the theory with rigorous mathematics, which can distract attention away from ideas important for the physics. The necessary mathematical details can subsequently be filled in with complete rigour. 
   This advice seems as relevant today for tackling the outstanding questions of fundamental physics in the early $21^{\mathrm{st}}$  century, and as has been followed in developing  the theory of generalised proper time with the progress achieved summarised in table~\ref{nails}.
{
   \linespread{1.25} 
 \def\rai{+0.2ex}
\begin{table}[htbp]
\centering
\begin{tabular}{|l|l|}
 \hline
                \qquad  Key Questions in Particle Physics and Cosmology
                       &  References \\
 \hline           &  \vspace{-14pt}  \\
     $ \, \naild  \! \;\; $
                     Well-motivated conceptual foundation for unification 
                         &                  \cite{Struct}, \cite{Short}, \cite{Tflow}
	  \vspace{-3pt}  \\
  $  \!\! \nailc \,\, \;\; $   Standard Model symmetries and particle multiplets 
                         &    \cite{Unifi} 6--9, \cite{Novel},  \cite{TimeE}
	 \vspace{-1.6pt}  \\
 $  \!\naila\, \;\; $   Standard Model particle couplings and masses
                         &        \cite{QGrav} 5.1, 6.2, $[*]$~\ref{boot32}
	  \\
	 $ \nailb \;\; $   Higgs physics and electroweak symmetry breaking
                         &   \cite{Unifi}8.3, \cite{TimeE}4.3,   \cite{Gener}3,4   
          \\
       $ \,  \nailb \!\;\; $   The `hierarchy problem' and weakness of gravity
                         &   \cite{KKone}5.3, \cite{QGrav}7.2, $[*]$\ref{boot33}
          \\   
        $ \,\, \nailc \!\! \;\; $   Left-right asymmetry in weak interactions
                         &    \cite{Novel} 6,  \cite{Gener} 3.1   
        \vspace{-1.6pt}    \\
        $ \,\,\, \naila \!\!\! \;\; $   Neutrino and beyond the Standard Model physics
                         &    \cite{BSM} 6, \cite{Gener}  4  
                        \\     
      $\,\,\,\, \naila \!\!\!\! \;\; $   Origin of matter--antimatter asymmetry
                         &   \cite{Gener}  4  
                        \\     
        $ \,\,\,\,\, \naila  \!\!\!\!\!\;\; $   Strong $C\!P$ problem in quantum chromodynamics
                         &   \cite{QGrav}    7.2   
                        \\    
          $ \!\! \nailb \,\, \;\; $   Basis for Lagrangian and action principle
                         &    \cite{KKone},\cite{QGrav}5.1,\cite{Basis}5,$[*]$\ref{boot32}     
                        \\       
          $ \!\nailc \,\;\; $   Amalgamation of quantum theory with gravity
                         &  \cite{Unifi}11, \cite{QGrav}4,7.1,   $[*]$\ref{boot32}  
                        \\                            
           $ \nailc \;\; $  The `problem of time' for `quantum gravity'
                         &    \cite{QGrav} 7.1,  \cite{Tflow}  7
                        \\                                    
          $\, \naila \! \;\; $   Black hole singularity and information issues
                         &   \cite{QGrav} 2.3, 7.1
                        \\                            
          $\,\, \nailb \!\! \;\; $   Structure and properties of the vacuum state
                         &   \cite{DEner} 6, 7,  \cite{Basis} 4,    $[*]$~\ref{boot33}  
                        \\                                         
           $\,\,\, \nailb \!\! \! \;\; $   Foundation for QFT formalism and calculations
                         &  {\small \cite{Unifi}11.1-2,\cite{QGrav}5,7.2,$[*]$\ref{boot32}$\!\!\!$    }
                        \\     
        $\,\,\,\, \nailc\!\!\!\! \;\; $   The `measurement problem' in quantum mechanics
                         &    \cite{Unifi} 11.4, \cite{QGrav} 7.3, 7.4    
                 \vspace{-1.6pt}         \\                                                                                                                           
        $ \!\! \nailb\,\, \;\; $    Origin and nature of dark matter
                         &   \cite{Dsect}, \cite{Basis} 4     
                        \\ 
           $\! \nailb \, \;\; $   Origin and nature of dark energy
                         &   \cite{DEner}, \cite{Basis} 4,     $[*]$~\ref{boot33}     
                        \\       
                    $ \naila \;\; $   $\Lambda$CDM issues such as `Hubble tension'
                         &   \cite{DEner} 7      
                        \\           
        $\,\naila \! \;\; $   Cosmological constant and coincidence problems
                         &   \cite{DEner} 7, \cite{Basis} 4      
                        \\   
                    $\,\,\naila \!\! \;\; $   Early Universe `inflation' and structure formation
                         &   \cite{Unifi} 13.2,      $[*]$~\ref{boot33}    
                        \\        
         $\,\,\, \nailb\!\!\! \;\; $   Origin of Big Bang and initial universe conditions
                         &   \cite{Unifi} 14.3,        $[*]$~\ref{boot43}  
                        \\                                 
           $\,\,\,\, \nailb \!\!\!\! \;\; $  Fine-tuning of physical parameters for biological life
                         &   \cite{Unifi}$\,$13.3, \cite{Tflow}$\,$5, $ [*]\,$4.3
                        \\    
             $\,\,\,\,\, \nailb \!\!\!\!\! \;\; $  Driver of the apparent `arrow of time'
                         &   \cite{Unifi} 14.1,  \cite{Tflow}  6
                        \\                                               
          $\, \nailc \!\;\; $  Why there is something rather than nothing
                         &      \cite{Unifi}   14,  $[*]$
                 \vspace{-16pt}       \\        
                           &    \\                                     
   \hline
  \end{tabular}
  \caption{
     \setb  
      \setlength{\baselineskip}{1.28\baselineskip} 
      Within the big picture of a unification scheme based upon generalised proper time this theory has sought to connect with empirical phenomena and outstanding questions in most areas of fundamental physics. As an alternative to individually targeting each `nail' \raisebox{1pt}{$\nailz$} in isolation the approach has been collective, by analogy with the construction of `Heisenberg's box' as described in the text. In the first column the nails, some of which cover several issues, are shown `tapped in' roughly in proportion to how well the questions have been accounted for 
    by the theory, ranging from a provisional connection  $\naila$ to an advanced understanding
   ${\mbox{\raisebox{-1.5pt}{\LARGE{$\intercal$}}}
   {\mbox{\hspace{-9.352pt}\raisebox{-4pt}{$\color{white}\bigvee$}}}   \mbox{\hspace{-11pt}\raisebox{-1.5pt}{$-\!\!-$}} }$.      \vspace{-2pt}
     In the final column the 
     principal references with section numbers are listed in chronological order, 
     where each `$[*]$' indicates a significant further contribution in this present paper,  although 
       essentially all of the topics listed have been discussed to some degree herein.
        Within the proposed self-sufficient universal bootstrap system all of the above questions might then be addressed in a coherent and self-contained manner.
        }
\label{nails}
\end{table}   
}  

\pagebreak

   The basic motivation and conceptual foundation for generalised proper time, augmenting the local
    4-dimensional spacetime form, was summarised in subsection~\ref{boot31} where the direct connections  
 with the Standard Model of particle physics and indications of new physics in the Higgs and neutrino sectors were also reviewed. 
  On constructing an extended 4-dimensional spacetime a framework can be identified for amalgamating general relativity with quantum theory, with the latter proposed to derive in a suitable limit that reproduces and underlies QFT calculations, in a manner that can also address the measurement problem of quantum mechanics,  as described in subsection~\ref{boot32}. 
  Returning to the local level of 4-dimensional spacetime  a source for a dark sector, incorporating both dark energy and dark matter candidates as consistent with cosmological  evolution and outlined in subsection~\ref{boot33}, can be identified from the symmetry breaking structure of an alternative branch of generalised proper time itself, as might also provide the source for an initial inflationary epoch and the primordial basis for large-scale structure formation.
  
   Hence all of these major areas of physics, as summarised alongside the associated `nails' of
   `Heisenberg's box'  in table~\ref{nails}, are brought under the cover of a single unification scheme. In being based upon the continuum of time as the sole basic entity, the approach of generalised proper time is  consistent with the ambition of unification not only in having a simple basis but also in terms of the broad explanatory power in accommodating diverse empirical phenomena. 
   With regards to the question of whether mathematics in general is discovered or created, one view that  might be taken is that all mathematics is \textit{discovered} but the act of discovery is a \textit{creative} process. 
    The same could be said of theoretical physics, and in particular the quest for a unified theory, although, as advised by Heisenberg, with the creative process of discovery not necessarily initially \textit{led} by mathematical stringency. While the conceptual picture motivates the theory, rigorous mathematical structures then steer the course of its development, as seen for example in how the concept of generalised proper time leads inevitably to the symmetry groups $\esi, \ese$ and potentially $\ee$ and the connections established with the Standard Model.

   However, the above account of this unifying theory still leaves open three significant loose ends, as listed in and following figure~\ref{cycleb}, concerning the source of the basic entity of time itself, the origin of the Big Bang together with the initial conditions of the universe, and the improbable generation and role of sentient beings in the universe.
    Any purely `materialist' theory, based upon brute facts that may or may not involve beautiful mathematics as a founding motivation, also raises but does not answer a parallel set of questions, as discussed in subsection~\ref{boot43} and further in the previous section. In particular the foundational issue of `why there is something rather than nothing' can at best only be met with further regression to an ever deeper layer of material substratum. Such a substrate is generally postulated according to its pragmatic utility in accounting for the upper layers accessible to empirical observation, without providing any intrinsic explanation for its own foundation.

    As proposed in subsection~\ref{boot41}, a theory of generalised proper time as based on the continuum of time \textit{can itself} have a foundation in a subjective sense of time, arising from neural structures as studied in ongoing neuroscience research. With the corresponding physical brain structures embedded in turn in the  physical universe, as constructed through generalised proper time, this leads directly to the universal bootstrap of figures~\ref{cycle}  as a self-sufficient system capable of `levitating' free of \textit{any external} assumptions or constraints, as described in subsection~\ref{boot42}. As  discussed in subsection~\ref{boot43} this universal bootstrap framework ties up all three loose ends of figure~\ref{cycleb}, including the provision of a wider context for the origin of the early universe and a central role for sentient beings, while addressing the broader `why there is something rather than nothing' existence question. 
 
    Given the need to `step outside' the objective physical universe, on the right-hand side of 
    figure~\ref{cycle}, it may not be possible to find a good analogy for the big picture of the universal bootstrap \textit{within} the physical universe itself. However, a metaphorical correspondence can be identified 
     with some of the artwork of M.C.~Escher, in representing conceptions that also cannot be constructed or realised within the actual physical world.
    For example, the absence of a posited lowest level for the universal bootstrap  in figure~\ref{cycle} is closely analogous to the absence of a lowest step in the finite but endless staircase of
     M.C.~Escher's lithograph print \textit{Ascending and Descending} (1960).  
   In both cases the self-contained circuit can be followed around in either a clockwise `ascending'  or an anticlockwise `descending' sense without encountering any limit.
   
     Further,  the cycle in \textit{Print Gallery} (1956), with the observer \textit{of} a print in the gallery   embedded \textit{in} that same print, is reminiscent of the self-reflective element, and the interrelation between the subjective and objective, in figure~\ref{cycle}. 
     However, the role of the `observer' is more active in figure~\ref{cycle}, as also for the 
     augmented circuit in figure~\ref{cycles},  in \textit{generating} the subjective forms of time and space
     as the basis for the construction of the objective physical world itself.
     The above works of M.C.~Escher, as well as a number of other examples of `strange loops', 
     are discussed in
      (\cite{Hofst} figures~6 and 142 respectively), with such loops paralleling 
       the universal bootstrap as a self-consistent `circular argument' as described in subsection~\ref{boot43}.

    As a final example,  in the circuit depicted in  M.C.~Escher's \textit{Cycle} (1938)  a \mbox{3-dimensional} character runs down the stairs on the right-hand side and then evolves into a 2-dimensional tiled pattern which undergoes a simplifying transformation up the left-hand side before being
     reinterpreted as a 3-dimensional block in the building from which the original character emerges. In figure~\ref{cycle} the 4-dimensional physical universe on the right-hand side incorporates neural structures  capable of generating a purely 1-dimensional subjective sense of temporal flow on the left-hand side with an intrinsic arithmetic substructure that can be interpreted as accommodating the original 4-dimensional spacetime and its matter content.  The mathematical rules that determine  the structure of matter derive from the symmetry breaking of generalised proper time induced by the further subjective predisposition to
      perceive the world in 3-dimensional space as described for figure~\ref{cycles}.
 
  All stages of the universal bootstrap of figures~\ref{cycle} and  \ref{cycles} are `equally elementary' in the sense of a foundational basis for the whole system, but `time' is the most elementary entity in the sense of the simplest component and as the entry point for describing a theory of physical structure.
  Indeed, the big picture that this reveals  contains the theory of generalised proper time, in the top half of figures~\ref{cycle} and  \ref{cycles}, which does still exhibit the general character of a traditional physical theory in building upon an elementary basis to  derive empirical phenomena, albeit now reinforced through accommodation within the universal bootstrap.

    The approaches of the $S$-matrix and cosmological bootstraps reviewed in section~\ref{boot2}, are also effectively accommodated in the universal bootstrap as discussed in subsection~\ref{boot42}.
    The typical constraints \textit{imposed} for the  $S$-matrix  bootstrap, including causality, Lorentz invariance, locality and unitarity, here \textit{follow} directly from the simplicity of the basis in the continuous flow of time alone
      for a theory of the construction of matter in an extended 4-dimensional spacetime from elements of generalised proper time. 
      In addition to the these $S$-matrix constraints here specific matter fields and interaction terms also derive directly from the properties of this construction, as described for equations~\ref{salpha}--\ref{gbreak} in subsection~\ref{boot31} and equations~\ref{extgpt}--\ref{asmlag}
     in subsection~\ref{boot32}, in place of the need for a posited Lagrangian such as for the Standard Model of particle physics or for new phenomena beyond.

    The further constraints assumed for the cosmological bootstrap, concerning the symmetries of an early inflationary epoch,  might also be determined for the present theory as discussed  in subsection~\ref{boot33}.
    The nature of the Big Bang and initial conditions, associated with   loose end `2)' in 
    figure~\ref{cycleb}, is also constrained by the requirement of joining loose ends
     `1)' and `3)' as described in subsection~\ref{boot43}.
    In constructing an extended 4-dimensional spacetime universe, as a single coherent entity as part of the circuit in figure~\ref{cycle}, there are then significant `retrocausal' restrictions on the conditions of the very early universe.
   In completing the circuit the subjective sense of a continuous progression in time is not only the source of `time' but also of the apparent `arrow of time', with respect to which this retrocausality is oriented.
      With the continuous flow of time, as the basis for all the physics, \textit{enveloped within} the self-contained cycle of figure~\ref{cycle}, all of these constraints, including those of the 
       $S$-matrix and cosmological bootstraps,  as well as any specific models assimilated by the theory, are themselves implied and
      \textit{subsumed within} the universal bootstrap.

      The initial motivation for the physical theory itself involves a change of perspective
      or `gestalt shift' from the framework of extra dimensions of space to
      the employment of generalised proper time (see also \cite{Struct} section~4).
      As noted in the opening of subsection~\ref{boot31} (with reference to~\cite{Adams})  such a `subtractive solution', in initially removing \textit{all} spatial dimensions and starting from time alone, might easily be overlooked
      in favour of the manifestly `additive solution' of extra spatial dimensions. 
      While the ideal of unification is to strive for simplicity, there may also be an opposing tendency due to the related  phenomenon of
       `complexity bias' -- the predisposition when presented with a single holistic system to assume it to  consist of various complicated parts, as may require the authority of exceptional technical expertise.
      The development of complex theoretical structures may in turn require the investment of many people over many years and, regardless of how successful or otherwise, once established as a field in itself such academic inertia can be self-sustaining.
       Such an `existence bias' in seeing the status quo as more attractive or `correct', particularly when 
       coupled with longevity, is marked by a preference for  familiar but questionable assumptions  in favour  of a novel alternative that may be supported by good evidence but subjected to a higher standard of judgement~\cite{Eidel}.

      The conception of the universal bootstrap of figure~\ref{cycle}, with its interplay of subjective and objective elements,  perhaps runs up against all of these potential cognitive biases
       (which, in other contexts, can play a more positive role).
       Indeed,  the overall worldview concerning \textit{what} the physical universe is, by comparison with that of a standard purely materialist philosophy, demands a significant gestalt shift in itself in terms of a substantial  change in the overall perspective.
   The difficulty in addressing the fundamental question of `why there is something rather than nothing' was further discussed in section~\ref{boot5} through further comparison with other approaches and theories.
      The answer to this question in the present theory, through the closure of the circuit in figure~\ref{cycle}, requires this \textit{irreducible} interplay between the subjective mind and objective matter and hence does not even make sense from a solely  `materialist' standpoint.
  That is, on insisting upon a strictly materialist philosophy the universal bootstrap of figure~\ref{cycle} is clearly `wrong'; the question is then whether the required gestalt shift  to the alternative perspective from which it 
  \textit{does} make sense is permissible. 
      We emphasise that this does \textit{not} imply a shift towards a purely `idealist' philosophy, which would also raise seemingly insuperable  foundational questions, but rather towards a mutually supporting reciprocity between the idealist and materialist standpoints.

    Augmenting the subjective sense of time as combined with a subjective propensity for 3-dimensional spatial perception, again with a basis in neural structures, the origin of  the symmetry breaking of generalised proper time and the explicit mathematical rules for the construction of the physical universe can be fully accounted for, within the freedom for a degree of anthropic selection for some physical parameters as noted towards the end of subsection~\ref{boot43}. This further reinforces the universal bootstrap as described for figure~\ref{cycles}.   
  It is not just that we passively observe a physical universe within which we are physically embedded, rather the forms of time and space through which we are hardwired to perceive it are essentially \textit{sufficient in themselves} to determine the physical properties of that universe. That is the essential nature of the universal bootstrap, as can be summarised in the following three steps:
\vspace{+1pt}
\begin{itemize}
    \item[A)]{We are `hardwired' to subjectively perceive a world through the continuum of the flow of time and in the form of 3-dimensional space.}  
\vspace{1pt}
    \item[B)]{These subjective forms of time and space, via the direct arithmetic link with generalised proper time and the symmetry breaking extraction of the external spatial components, are  sufficient to construct the entire objective physical world.}
  \vspace{-13pt}  
    \item[C)]{This physical world contains neural structures capable of supporting sentience and  within which  we are objectively `hardwired'.}
\end{itemize}  

      The probing of this proposed system might be broken down into two parts. Firstly,
       independently of `our world' is the overall conception of this scheme coherent in a general and abstract sense, that is for `sentient entities' in any `physical world'? \mbox{If so, then} secondly, what is the evidence that our world might belong to such a system and how might this hypothesis be tested? 
       Given that, regarding the first part, 
        the universal bootstrap  is  logically coherent by construction,
         below we focus on the second question.
       
       The contention of point `A)' above is  supported by empirical research in neuroscience indicating that the brain really does \textit{generate} temporal and spatial forms of perception, as we have reviewed in detail in section~\ref{boot4} (with reference for example to~\cite{chwww,Peer,BuzsL}), consistent with the psychology of how we \textit{experience} time and space.
        Step `B)' concerns the main thesis of the theory of generalised proper time, with the physical world constructed as based on equations~\ref{sint}--\ref{gbreak} and~\ref{GRQT} as summarised for equation~\ref{subrk}, and as supported by the empirical successes described through section~\ref{boot3}.
         The third stage `C)' is seemingly the most implausible link in the circuit, in that inanimate physical matter should give rise to subjective awareness, and yet is perhaps the most indisputable aspect of our world in that we directly experience the consequences, even though we lack a full scientific understanding of how this is possible. 
         
         The universal bootstrap may itself shed some light on this latter point, since the closing of the circuit as described above may itself imply a need for sentience.
           That is, the closing of figure~\ref{cycle} in particular may be a sufficient criterion for 
            sentient awareness, given the seeming necessity of a stream of subjective time to accompany a stream of consciousness. It is a further question as to whether the augmentation to figure~\ref{cycles} is strictly necessary, as it seems for ourselves in our world, or whether an 
             alternative, with for example a different number of spatial dimensions or mechanism for symmetry breaking and correspondingly different physical laws, accommodating sentient beings in a very different world,    would be conceivable.
         
         Regarding the connection from `A)' to `B)' above,  it is not so much that we can think \textit{of} time, but rather we intrinsically and irreducibly think \textit{in} time as a necessary form for any `thoughts'. 
         The continuum property of this subjective flow of time is encapsulated by the continuum of real numbers providing the basis for mathematical structures that can accommodate the structure of space and matter as initially described at the local level for figure~\ref{sgener}. This provides the basic elements for the construction of  the entire physical universe, as we also perceive \textit{in} time and as incorporates the physical neural entities required in `C)' to close the circuit. It is this closing of the universal bootstrap that compels the structures it contains to `exist'.

   As a consequence of this gestalt shift away from a purely materialist worldview, the overall universal bootstrap then entails significant  explanatory power, in physics and other sciences, as elucidated in subsection~\ref{boot43}. 
   There is in particular an opportunity for a significant interplay between this theory of fundamental physics and the fields of both  neuroscience and artificial intelligence. The primacy of the subjective generation of time, as well as spatial perception, could play a role in further understanding the overall functioning of the brain, in particular on taking the `inside-out' perspective to the limit.  
  This  perspective 
   could also be significant for the question of whether and how a machine might become sentient.
   The question then is whether a sense of subjective time and space, and a sufficient level of engagement in the physical world, could be `hardwired', programmed or otherwise incorporated into the design of an artificial device such that it might `get off the ground' and be self-supporting in the manner of the above three `A,B,C)' steps and figure~\ref{cycles}.
   There is then rather  more to the potential explanatory power of the universal bootstrap, beyond particle physics and cosmology and the `nails' of table~\ref{nails}, with a series of pertinent philosophical questions also being illuminated as alluded to in  subsection~\ref{boot43}.

   It remains that the principal justification for the unification scheme of generalised proper time is in the ability to account for a series of major outstanding questions in physics as summarised in section~\ref{boot3} and  table~\ref{nails}.
   With all of the nails in that table requiring further taps of the hammer
  there are here also of course significant areas needing further work.
  Given the conceptual foundation in generalised proper time 
  one  branch of the theory is directed by  unique mathematical structures of exceptional Lie groups, with the need
  to determine the full $\ee$-type form for proper time to complete the correspondence with Standard Model particle states.
    Further work is also required to 
    fully establish the connection with a quantum theory limit consistent with general relativity and in deducing the detailed nature of the dark sector branches for this theory.
   The successes already achieved in pursuing these familiar goals with generalised proper time, now also supported by the proposal of the fully self-contained  universal bootstrap, make a strong case for an 
    ultimate    framework for   unification in physics.


{\setlength{\baselineskip}{0.89\baselineskip}

\par}



\par}

\end{document}